%% file: Main.tex
\documentclass[11pt]{article}
\input{packages}

\begin{document}
\title{Diffusion Distribution Model for Damage Mitigation\\in Scanning Transmission Electron Microscopy}
\author{
Amirafshar~Moshtaghpour\footnotemark[1]\footnote{Rosalind Franklin Institute, Harwell Science \& Innovation Campus, Didcot, OX11 0QS, UK.} \footnotemark[2]\footnote{Mechanical, Materials, \& Aerospace Engineering, University of Liverpool, Liverpool, L69 3GH, UK.}
\and 
Abner~Velazco-Torrejon\footnotemark[1]
\and
Daniel~Nicholls\footnotemark[3]\footnote{SenseAI Innovations Ltd., Brodie Tower, University of Liverpool, Liverpool, UK.
}
\and
Alex~W.~Robinson\footnotemark[3]
\and
Angus~I.~Kirkland\footnotemark[1] \footnotemark[4]\footnote{Department of Materials, University of Oxford, Oxford, OX2 6NN, UK.
\hfill\break \hphantom{aaa} Corresponding author: \url{amirafshar.moshtaghpour@rfi.ac.uk}}
\and
Nigel~D.~Browning\footnotemark[2] \footnotemark[3]}
\markboth{}%
{}
\date{}
\maketitle
%%%%%% End of Title %%%%%%
%%%%%% Start of Abstract %%%%%%
\begin{abstract} 
Despite the widespread use of Scanning Transmission Electron Microscopy (STEM) for observing the structure of materials at the atomic scale, a detailed understanding of some relevant electron beam damage mechanisms is limited. Recent reports suggest that certain types of damage can be modeled as a diffusion process and that the accumulation effects of this process must be kept low in order to reduce damage. We therefore develop an explicit mathematical formulation of spatiotemporal diffusion processes in STEM that take into account both instrument and sample parameters. Furthermore, our framework can aid the design of Diffusion Controlled Sampling (DCS) strategies using optimally selected probe positions in STEM, that constrain the cumulative diffusion distribution. Numerical simulations highlight the variability of the cumulative diffusion distribution for different experimental STEM configurations. These analytical and numerical frameworks can subsequently be used for careful design of 2- and 4-dimensional STEM experiments where beam damage is minimised.
\end{abstract}
\section{Introduction} \label{sec:intro} 
(Scanning) Transmission Electron Microscopy (S(TEM)) is a widely used tool for investigating complex structures at the atomic level~\cite{nellist1995resolution, james1999practical,zhang2018atomic}. This technique is now routine due to the development and implementation of spherical aberration correctors  ~\cite{Krivanek1999subA, Sawada47pmSTEM,haider1998electron},  improved electron sources~\cite{johnson2022near,pooch2018coherent,kuo2006noble} and direct electron detectors~\cite{faruqi2007direct,faruqi2015progress}. For STEM imaging, a high intensity coherent, convergent probe is scanned over a region of interest of a sample ~\cite{pennycook2011scanning, zuo2017advanced} which can lead to detrimental \textit{beam damage}~\cite{egerton2004radiation, jiang2015electron}. 

Beam damage can arise through various mechanisms. \textit{Knock-on} damage results from an electron-atom interaction, whereby the incident electron transfers kinetic energy to an atom in the sample~\cite{jenkins1982characterization,egerton2004radiation}. In this case, kinetic energy and momentum are conserved in the collision and an atom may be displaced from its equilibrium position or from its atomic site, if the transferred energy is higher than an atomic displacement energy~\cite{egerton2019radiation}.

\begin{figure}[t]
    \centering
    \includegraphics[width=1\columnwidth]{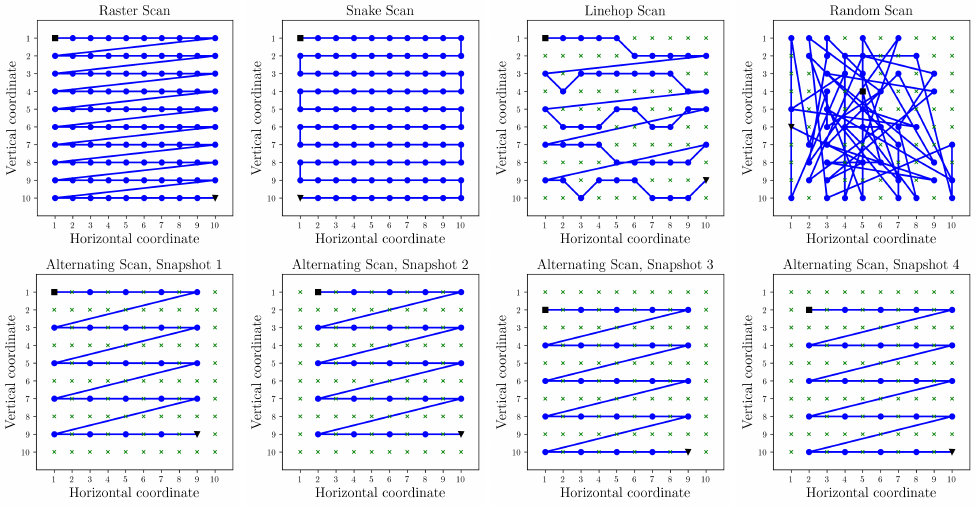}
    \caption{\textbf{Scanning probe trajectories on a $10 \times 10$ grid.} The first and last scanned probe positions are indicated by a black square and a triangle, respectively whereas the unscanned positions are shown by green crosses. Linehop and random scans are illustrated for 50\% sampling of probe positions. Four snapshots associated with an alternating scan of order 2 are shown in the bottom panel. As for raster and snake scans, the alternating approach scans the full FoV.}
    \label{fig:example_path}
\end{figure}

A second mechanism, predominately affecting insulators, known as \textit{radiolysis} describes the cleavage of chemical bonds within the structure of a sample~\cite{egerton2019radiation}. This occurs due to an inelastic electron-electron interaction, whereby the incident electron causes either excitation or ionization. If sufficient kinetic energy is transferred to a valence electron or an inner-shell electron, this can cause the generation of secondary electrons, unstable radicals or ions that subsequently result in the dissociation of chemical bonds and eventually induce defect formation or even amorphisation. If the time for electron-hole pair recombination is longer than a critical value, ions can migrate due to the Coulomb potential, which can be induced by the incident electron beam ~\cite{jiang2015electron,egerton2019radiation}. The secondary electrons that are released during this interaction can travel distances of the order of tens of nanometers, creating further damage away from the irradiated position~\cite{secondary_polys1999, secondary_litho2001}. These adverse effects can in principle be considered to be proportional to the electron fluence -- \ie the number of electrons delivered over a region of interest. However, due to the delocalised effect of radiolysis, regions of the sample that are
visited by the STEM probe at a later stage in the scanning sequence can be damaged when scanning earlier positions. In addition, because of the dynamic nature of radiolysis, due to the primary and secondary processes occurring on different timescales~\cite{EgerOutrundamage2015}, conventional raster scan as shown in Fig.~\ref{fig:example_path} can induce more damage if the scan step and dwell time favour a fast accumulation of the damage effects. However, although the physical process(es), \eg electrostatic charging, diffusion of radicals, ions, and/or heat, affecting a particular sample~\cite{egerton2004radiation} may not be clearly identified, these mechanisms all behave as a diffusion process that extends spatially in time. 

Emerging approaches for reducing  beam damage are based on controlling the probe trajectory for scanning the full Field of View (FoV). A recent report~\cite{interleaveSTEM2022} has shown damage reduction using an alternating scan (also known as interleaved or leapfrog scan) compared to raster scan for high resolution imaging of a zeolite sample. As shown in Fig.~\ref{fig:example_path}, in an alternating scan, all probe positions in a 2-D array of points are visited by skipping a fixed number of points in both horizontal and vertical directions on the grid. A previous report~\cite{randomSTEM2020} exploited a random scan strategy, selecting the next probe location at random for hyperspectral STEM imaging to reduce  emission instabilities in a cathodoluminescence experiment. In both examples, the same number of electrons are distributed over the scanned area as in the raster scan but adjacent probe positions are not used sequentially.

An alternative approach to mitigate beam damage is based on subsampling the probe positions. A high quality reconstruction is expected assuming that the image has an (approximately) sparse representation in a transform domain~\cite{CStheory2006}. Subsampled STEM modalities are often recognised as applications of the theory of Compressed Sensing (CS)~\cite{CStheory2006,candes2006robust} in the field of electron microscopy; subsequently, referred to as \textit{compressive STEM} for convenience. This subsampled acquisition allows for a larger average distance between neighbouring positions compared to a conventional full acquisition. Experimentally, subsampling of the probe positions can be achieved using a scan generator or a beam blanker ~\cite{kovarik2016implementing, beche2016development}. A scan generator is used to alter the trajectory of the probe such that only the desired positions within the FoV are sampled. A beam blanker, without changing the trajectory of the probe from a default raster scan, deflects the beam rapidly and selects locations
that are irradiated along the scan path.

Other approaches such as the fractionation of the fluence by multi-frame fast acquisitions have also shown damage reduction in contrast to a single acquisition, for a fixed total fluence in STEM spectrum imaging of a lead perovskite~\cite{multiframeSTEM2018}. Importantly, these findings support the diffusion-like behaviour of radiolysis and suggest that some of the damage processes of this mechanism are reversible.

As described earlier, the displacement of atoms by \textit{knock-on} involves a  simplistic mechanism, and a relatively large number of analytical and molecular dynamic models exist to describe and model that interaction~\cite{knockonmodels2018}. However, because of the complexity of the multiple processes triggered by radiolysis, there are a lack of models that describe this mechanism which has limited progress towards experimental strategies to mitigate its effects~\cite{EgerDose-rate1999}. A diffusion process, supported by experimental results, is an attractive concept that can be implemented as an analytical Fick's diffusion model~\cite{fick1855v, crank1979mathematics}.

\subsection{Background and outline}\label{subsec:intro_contributions}

Previous studies~\cite{nicholls2020minimising,jannis2022reducingpart2} have used diffusion models to model electron beam damage and its propagation through a sample during the image acquisition process. Nicholls \etal~\cite{nicholls2020minimising} first used this approach to provide insights as to why sub-sampled measurements, (as in compressive STEM), reduce beam damage. That work also demonstrated that linehop sampling \cite{kovarik2016implementing}; a sampling method designed to overcome scan coil hysteresis, is as effective as random sampling for reducing beam damage at low sampling rates (6.25\% of scanning probe positions). As shown in Fig.~\ref{fig:example_path}, linehop sampling involves constructing an image from a series of lanes, where each lane only samples one pixel from each column, and where the next pixel to be sampled is a neighbour to the previous column. This limits the range of motion during sampling whilst still allowing random perturbations, satisfying both the conditions for hysteresis from the scan coils and incoherence for data recovery using inpainting~\cite{CStheory2006}. 

Jannis \etal~\cite{jannis2022reducingpart2} have extended that work to explore experimental results obtained from alternating scan strategies in STEM. In this work, a 2-D diffusion process from a continuous point source is considered as the mediator of damage, and it was shown that by the implementation of a damage threshold the proposed model could explain the experimental results for a reduction in damage using this scan strategy compared to a raster scan. Since the detailed physics of the damage process may not be known, the concentration of a diffusing substance in the model is replaced by a  parameter defined as a ``state of the sample'', which is altered by the electron beam. Sample damage then only occurs when that parameter crosses a defined threshold. The total damage at a certain position is thus calculated by integrating a non-linear function, referred in this paper to as a damage activation function, over time.

The focus of this work is to further develop existing models~\cite{nicholls2020minimising,jannis2022reducingpart2} and to present a rigorous mathematical framework. Following the work by Jannis \etal~\cite{jannis2022reducingpart2}, we make use of a damage threshold, and include a damage diffusion distribution, or diffusion distribution for short, to replace the classical concentration of a diffusing substance.
 
This work provides the first explicit formulations of a damage induced diffusion process model in STEM by considering a realistic non-point source representation of the electron probe. We note that since our mathematical model supports an electron probe of arbitrary size, the findings of this paper can be directly applied to 4-D STEM scans~\cite{moshtaghpour2023exploring,robinson2023simultaneous,moshtaghpour2022towards}. This enables future studies of diffusion distribution \textit{(i)} in thick or three-dimensional (3-D) samples, \textit{(ii)} from pulsed (or modulated) electron beams, or \textit{(iii)} in a TEM geometry. 
A detailed summary of the key findings reported in this paper are:
\begin{itemize}
    \item Based on the Fick's second model~\cite{fick1855v}, we derive the spatio-temporal diffusion distribution of an instantaneous point source in a $d$-dimensional medium with an anisotropic diffusion coefficient (Eq.~\eqref{eq:general-pde-constant-d-solution-instantaneous-point}). This is a generalisation of the previous formulation in ~\cite{jannis2022reducingpart2}, from a two- to $d$-dimensional medium and from isotropic to anisotropic diffusion coefficients. 
    \item Using that solution and the principle of superposition we provide the diffusion distribution for an arbitrary-shaped source (See Eq.~\eqref{eq:general-pde-constant-d-solution}). 
    \item We extend that model to a continuous source and provide closed-form formulations for diffusion distributions for four special cases of continuous sources: \textit{(i)} point source, \textit{(ii)} square disc source, \textit{(iii)} circular disc source, and \textit{(iv)} Gaussian-shaped source.
    \item Our analyses of diffusion distribution in STEM assume that the 2-D Gaussian function is a close proxy to a more realistic airy disc~\cite{james1999practical} function. By computing the first and second derivatives of the diffusion distribution with respect to both space and time, we investigate the spatio-temporal behaviour of the diffusion distribution. Our findings in Lemmas~\ref{lem:stem_single_probe_derivatives_beamon}, \ref{lem:stem_single_probe_derivatives_beamoff}, \ref{lem:stem_single_probe_derivatives_t_beamon}, and \ref{lem:stem_single_probe_derivatives_t_beamoff} highlight the complexity of the diffusion process caused by activating an electron probe at a single location.
    \item We formulate in Section ~\ref{subsec:diffusion_in_stem} a Cumulative Diffusion Distribution (CDD) caused by activating an electron probe at multiple locations during STEM acquisitions. The CDD takes into account the diffusion distributions caused not only by the currently activated electron probe,
but also by every electron probe activated previously. 
    \item Our STEM diffusion model can be coupled to a damage threshold mechanism (Eq.~\eqref{eq:stem_did}). We propose two Diffusion Induced Damage (DID) parameters: \textit{(i)} the frequency of DID events, and \textit{(ii)} the intensity of DID events.  For each of these, we distinguish online \textit{vs.} offline damage observations. In the online case, the DID at a certain location is affected only until that location is
scanned; whereas in the offline case, DID changes at every location – regardless of whether it is scanned later on or not. Moreover, in Theorem \ref{thm:cs_stem}, we provide sufficient mathematical conditions, for which subsampling probe positions, as in compressive STEM, reduces the DID  compared to full sampling. That result supports the experimental findings in the literature reporting the advantageous of subsampling probe positions~\cite{nicholls2020minimising, jannis2022reducingpart2, nicholls2022compressive,nicholls2023potential} in reducing damage.
    \item We present extensive numerical simulations, in Section ~\ref{sec:numerical_results}, which illustrate the advantages of the proposed framework. 
    \item Finally, in Secs.~\ref{subsec:mask_design} and \ref{subsec:simulations_mask_design}, for given STEM acquisition parameters we have designed a Diffusion Controlled Sampling (DCS) strategy that selects as many probe positions as possible while ensuring that the global maximum of the CDD is constrained, or equivalently the DID is minimised. 
\end{itemize}

%%%%%% End of Introduction %%%%%%
%%%%%% Start of Mathematical Model of Diffusion %%%%%%
\section{A Mathematical Model for Diffusion}\label{sec:math-diffusion}
We base our analysis of diffusion distribution on the second Fick's model \cite[\textsection 1.2]{crank1979mathematics}, in which the diffusion distribution $\phi$ in a medium at a $d$-dimensional spatial location\footnote{In this paper, vectors and matrices are denoted by, respectively, boldface small and boldface capital letters.} $\bs r = [r_1,\cdots,r_d]^\top \in \bb R^d$ and at time $t>0$ is the solution of the following Partial Differential Equation (PDE) \cite[Eq. 1.5]{crank1979mathematics}, \ie
\begin{equation}\label{eq:general-pde-constant-d}
    {(\rm constant~diffusion~coefficient~model)}\hspace{1cm}
    \frac{\partial \phi(\bs r, t)}{\partial t} = \sum_{l=1}^d    D_l\frac{\partial^2\phi(\bs r, t)}{\partial r_l^2}.
\end{equation}
where $D_l>0$, in ${\rm m}^2\cdot {\rm s}^{-1}$, is the diffusion coefficient along the $l$-th coordinate. Equation \eqref{eq:general-pde-constant-d} assumes that the medium is homogeneous such that the diffusion coefficient does not vary from point to point. It also assumes that the diffusion coefficient is independent of the concentration of a diffusing substance $\phi$ as in, \eg the diffusion of organic vapours in high-polymer substances. In the remaining of this paper we assume that this diffusion is caused by an electron source interacting with an infinite medium.

For an instantaneous point electron source activated at a location $\bs r = \bs r_0$ and time $t = t_0$, the solution of the PDE in Eq.~\eqref{eq:general-pde-constant-d} is (Section~\ref{subsec:developing-instantaneous-point} gives the derivation)
\begin{equation}\label{eq:general-pde-constant-d-solution-instantaneous-point}
    \phi(\bs r,t) = \frac{Q_0}{\sqrt{|4\pi \bs D|(t-t_0)^{d}}}
    \exp\Big(-\frac{(\bs r - \bs r_0)^\top \bs D^{-1} (\bs r-\bs r_0)}{4(t-t_0)}\Big),
\end{equation}
where $Q_0$ is the initial rate of the diffusing species. and the diagonal matrix $\bs D \coloneqq {\rm diag}(D_1,\cdots,D_d) \in \bb R^{d\times d}$ contains the diffusion coefficients on its diagonal with $|\bs D|$ denoting the determinant of $\bs D$. Equation~\eqref{eq:general-pde-constant-d-solution-instantaneous-point} extends previous results, \cite[Eq. 2.6]{crank1979mathematics}, \cite[p. 150]{carslaw1906introduction}, and \cite{pattle1959diffusion} to a $d$-dimensional anisotropic diffusion model, which is equivalent to non-identical diffusion coefficients along different coordinates. 
\begin{remark}\label{remark:units_of_diffusion}
    In Eq.~\eqref{eq:general-pde-constant-d-solution-instantaneous-point} and subsequently, we assume that $Q_0$ is related to the electron beam current $I_0$ through a general function $\cl M: \bb R_{\ge 0} \mapsto \bb R_{\ge 0}$, \ie $Q_0 = \cl M(I_0)$. For the sake of convenience, we further assume the units of the rate of the diffusing species $Q_0$ to be $\qunit$ for an unspecified and arbitrary ``${\rm u}$'' here; hence, the unit of the diffusion distribution $\phi$ is $\diffunit$. This assumption, allows a comparison of different STEM scans with fixed beam current. By comparison the authors in \cite{jannis2022reducingpart2} considered that the electron probe deposits energy at every probe location and assumed $Q_0$ with units of ${\rm s}^{-1}$. However, we emphasise that we do not make any claim about the nature of the diffusive species in STEM. 
\end{remark}

Using the principle of superposition \cite[Eq. 3.5a]{crank1979mathematics}, the solution of the PDE in \eqref{eq:general-pde-constant-d} for a source with an arbitrary spatio-temporal activation profile $h(\bs r, t)$ can be written as;
\begin{equation}\label{eq:general-pde-constant-d-solution}
    \phi(\bs r,t) = \int_{0}^t \int_{\bb R^{d}} 
    h(\bs r', t')\frac{1}{ \sqrt{|4\pi\bs D|(t-t')^{d}}}
    \exp\Big(-\frac{(\bs r - \bs r')^\top \bs D^{-1} (\bs r-\bs r')}{4(t-t')}\Big)\ud \bs r' \ud t'.
\end{equation}
It is straightforward to confirm that the total number of diffusing species $Q_{\rm tot}(t)$ at time $t$ in a system with a diffusion distribution given by Eq.~\eqref{eq:general-pde-constant-d-solution} is equal to the total number of units deposited by the source from the activation time of the source to the current time, since;
\begin{equation}\label{eq:total_electrons}
    Q_{\rm tot}(t) \coloneqq  \int_{\bb R^{d}} \phi(\bs r, t)\, \ud \bs r =  \int_{0}^t\int_{\bb R^{d}}h(\bs r', t')\, \ud \bs r' \ud t',
\end{equation}
which agrees with principle of conservation of energy~\cite{feynman2006qed}.

\subsection{Special Cases in 2-D}\label{subsec:spatial-cases}
From Eq.\eqref{eq:general-pde-constant-d-solution}, we now develop the diffusion distribution formulation for specific cases of continuous sources in a 2-D medium. From ~\ref{remark:units_of_diffusion} we recall that the units of diffusion distribution in 2-D are $\diffunitnm$.

\begin{figure}[t]
    \centering
    \includegraphics[width=1\columnwidth]{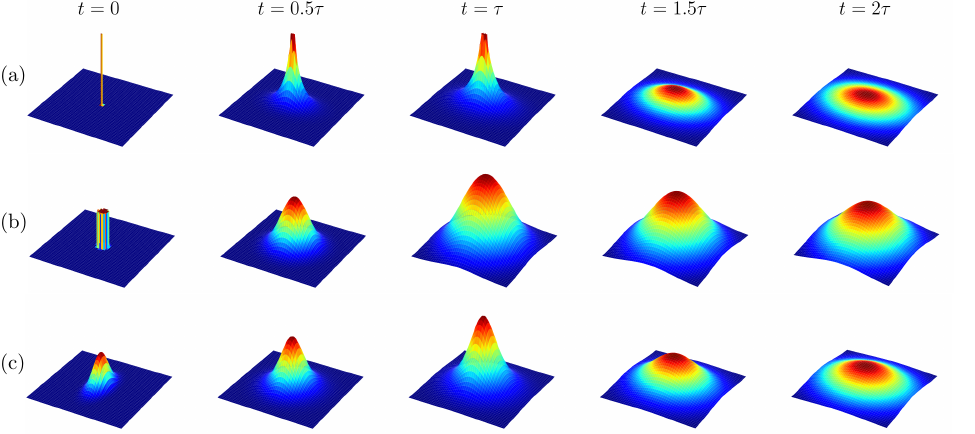}
    \caption{Diffusion distributions for three special cases in Section ~\ref{sec:math-diffusion}. (a) Point source: Eq.~\eqref{eq:general-pde-constant-d-solution-continued-point} with $\bs D = {\rm diag}(0.25, 0.5)$, (b) Circular disc source: Eq.~\ref{eq:general-pde-constant-d-solution-continued-circular} with $D = 0.25$ and $r_s = 0.2$, and (c) Gaussian-shape source: Eq.~\ref{eq:general-pde-constant-d-solution-continued-gaussian} with $\bs D_s = {\rm diag}(0.2,0.05)$ and $\bs D = {\rm diag}(0.25,0.5)$. For all cases $\tau = 1$, $Q_0 = 1$, $t_0 = 0$, and $\bs r_0 = \bs 0$. The range of vertical axes are identical intra-row but adjusted inter-row for better visualisation.}
    \label{fig:example_special_cases}
\end{figure}

\paragraph{Continuous point source activated at location $\bs r = \bs r_0$ during time $t \in [t_0, t_0 +\tau]$:} 
In this case, the corresponding source function can be decomposed into spatial and temporal components, \ie $h(\bs r, t) = Q_0 \cdot h_s(\bs r) \cdot h_t(t)$ with
\begin{equation}\label{eq:source-function-continued-point}
    h_s(\bs r) = \delta_{\bs r_0}(\bs r)\hspace{1cm}{\rm and} \hspace{1cm}
    h_t(t) = 
    \begin{cases}
        1, &{\rm if~} t_0 \le t \le t_0+\tau,\\
        0, &{\rm otherwise},
    \end{cases}
\end{equation}
where $Q_0$, in $\qunit$ and $\delta_{\bs r_0}(\bs r) = +\infty$, if $\bs r = \bs r_0$, and $\delta_{\bs r_0}(\bs r) = 0$, if $\bs r \ne \bs r_0$, is the delta Dirac function, whose integral over the entire domain is unity: $\int_{\bb R^{d}}\delta_{\bs r_0}(\bs r) \ud \bs r = 1$. Inserting Eq.~\eqref{eq:source-function-continued-point} in Eq.~\eqref{eq:general-pde-constant-d-solution} for $d = 2$ dimensions gives (Section ~\ref{subsec:developing-continued-point} gives details);
\begin{equation}
\label{eq:general-pde-constant-d-solution-continued-point}
    \phi(\bs r,t) = 
    \begin{cases}
        \frac{Q_0}{4\pi\sqrt{|\bs D|}}\eint{(\bs r - \bs r_0)^\top \bs D^{-1} (\bs r-\bs r_0)}{4(t-t_0)},
        & t_0\le t \le t_0+\tau,\\
        \frac{Q_0}{4\pi\sqrt{|\bs D|}}\Big(\eint{(\bs r - \bs r_0)^\top \bs D^{-1} (\bs r-\bs r_0)}{4(t-t_0)} - \eint{(\bs r - \bs r_0)^\top \bs D^{-1} (\bs r-\bs r_0)}{4(t-(t_0+\tau))}\Big),
        & t > t_0+\tau,
    \end{cases}
\end{equation}
at any spatial point other than the activation point $\bs r \ne \bs r_0$; and at the activation point $\bs r = \bs r_0$,
\begin{equation*}
\label{eq:general-pde-constant-d-solution-continued-point_2}
    \phi(\bs r_0,t) = 
    \begin{cases}
        +\infty,
        & t_0\le t \le t_0+\tau,\\
        \frac{Q_0}{4\pi \sqrt{|\bs D|}}\ln{\big(\frac{t-t_0}{t-(t_0+\tau)}\big)},
        & t > t_0+\tau.
    \end{cases}
\end{equation*}
In Eq.~\eqref{eq:general-pde-constant-d-solution-continued-point}, $E_1(v) \coloneqq \int_{v}^{+\infty} \frac{1}{u} e^{-u}\, \ud u$ for $v \in \bb{R}/\{0\}$ and $E_1(0) = +\infty$ is the Exponential integral of order one. 
For an isotropic medium with a diffusion coefficient $D$ or with a matrix of diffusion coefficients $\bs D = {\rm diag}(D,D)$  Eq.~\eqref{eq:general-pde-constant-d-solution-continued-point} reduces to Eq.~(1) in \cite{jannis2022reducingpart2}. However, the value of the diffusion distribution at the activation point $\phi(\bs r_0,t)$ when $t\ge \tau$ is not included in \cite{jannis2022reducingpart2} as illustrated in  Fig.~\ref{fig:example_special_cases}(a). 

This special case is relevant to imaging applications using STEM as studied in \cite{jannis2022reducingpart2}. However, experimentally the source is not a point . Therefore, in the following we provide solutions to \eqref{eq:general-pde-constant-d} for non-point sources. 

\paragraph{Continuous circular disc source with radius $r_s$ centered at $\bs r = \bs r_0$ and activated during time $t \in [t_0, t_0 +\tau]$:} In this case, the corresponding source function is $h(\bs r, t) = Q_0 \cdot h_s(\bs r) \cdot h_t(t)$ with
\begin{equation}\label{eq:source-function-continued-circular}
    h_s(\bs r) = 
    \begin{cases}
        \frac{1}{\pi r_s^2}, &{\rm if~} \|\bs r - \bs r_0\|_2 \le r_s,\\
        0, &{\rm otherwise},
    \end{cases}
    \hspace{1cm}{\rm and} \hspace{1cm}
    h_t(t) = 
    \begin{cases}
        1, &{\rm if~} t_0 \le t \le t_0+\tau,\\
        0, &{\rm otherwise},
    \end{cases}
\end{equation}
where the rate of charge $Q_0$ is $\qunit$. In Eq.~\eqref{eq:source-function-continued-circular}, the $\ell_2$-norm $\|\bs u\|_2 \coloneqq (\sum_i |u_i|^2)^{1/2}$, is the square root of the sum of the squared vector values, and is used to model the circular disc source. For simplicity, we assume isotropic diffusion coefficients, \ie $D_1 = D_2 = D$. Inserting \eqref{eq:source-function-continued-circular} in \eqref{eq:general-pde-constant-d-solution} for $d = 2$ gives (see Section ~\ref{subsec:developing-continued-circular} for details)
\begin{equation}\label{eq:general-pde-constant-d-solution-continued-circular}
    \phi(\bs r,t) \!= \!\!\int_{t_0}^{\min (t,t_0+\tau)} \!\!\!\!\frac{Q_0}{2\pi r_s^2 D(t-t')}\exp\Big(-\frac{\|\bs r - \bs r_0\|_2^2}{4D(t-t')}\Big)
    \!\!\int_{0}^{r_s} \!\!u
    \exp\Big(-\frac{u^2}{4D(t-t')}\Big) I_0\big(\frac{ u\|\bs r - \bs r_0\|_2}{2D(t-t')}\big) \ud u\, \ud t',
\end{equation}
where $I_0(v) \coloneqq \dfrac{1}{\pi}\int_0^{\pi} \exp{(v \cos{\theta})}\ud\theta$ is the modified Bessel function of the first kind of order zero. 

This case is pertinent to imaging using wide (parallel) beams as in conventional TEM or Fourier Ptychography~\cite{haigh2009atomic} and in certain X-ray applications ~\cite{batey2014information,claus2019diffraction}. For completeness in \ref{subsec:continued-square} we extend our analysis to continuous square disc source.
\paragraph{Continuous Gaussian source with a shape matrix $\bs D_s = {\rm diag}(D_{s,1}, D_{s,2})$ centered at $\bs r = \bs r_0$ and activated during time $t \in [t_0, t_0 +\tau]$:} in this case, the corresponding source function is $h(\bs r, t) = Q_0\cdot h_s(\bs r) \cdot h_t(t)$ with
\begin{equation}\label{eq:source-function-continued-gaussian}
    h_s(\bs r) = \frac{1}{\sqrt{|4\pi \bs D_s|}}\exp{{\big(-\frac{1}{2}(\bs r - \bs r_0)^\top \bs D_s^{-1}(\bs r - \bs r_0)\big)}}
    \hspace{1cm}{\rm and} \hspace{1cm}
    h_t(t) = 
    \begin{cases}
        1, &{\rm if~} t_0 \le t \le t_0+\tau,\\
        0, &{\rm otherwise},
    \end{cases}
\end{equation}
where the rate of charge $Q_0$ is $\qunit$. Inserting \eqref{eq:source-function-continued-gaussian} in \eqref{eq:general-pde-constant-d-solution} for $d = 2$ gives (see Section ~\ref{subsec:developing-continued-gaussian} for details)
\begin{equation}\label{eq:general-pde-constant-d-solution-continued-gaussian}
    \phi(\bs r,t) = \int_{t_0}^{\min (t,t_0+\tau)} 
    \frac{Q_0}{2\pi\sqrt{|\bs D_e|}}
    \exp{\big(-\frac{1}{2} (\bs r-\bs r_0)^\top \bs D_e^{-1} (\bs r- \bs r_0)\big)}\,\ud t',
\end{equation}
with $\bs D_e = \bs D_s + 2(t-t')\bs D$. Furthermore, assuming an isotropic source and diffusion coefficients, \ie $D_{s,1} = D_{s,2} = D_s$ and $D_1 = D_2 = D$,  \eqref{eq:general-pde-constant-d-solution-continued-gaussian} reduces to
\begin{equation}\label{eq:general-pde-constant-d-solution-continued-gaussian_2}
    \phi(\bs r,t) = 
    \begin{cases}
        \frac{Q_0}{4\pi D}\Big(\eint{\|\bs r - \bs r_0\|_2^2}{2D_s + 4D(t-t_0)}-\eint{\|\bs r - \bs r_0\|_2^2}{2D_s}\Big),
        & t_0\le t \le t_0+\tau,\\
        \frac{Q_0}{4\pi D}\Big(\eint{\|\bs r - \bs r_0\|_2^2}{2D_s +4D(t-t_0)} - \eint{\|\bs r - \bs r_0\|_2^2}{2D_s+4D(t-(t_0+\tau))}\Big),
        & t > t_0+\tau,
    \end{cases}
\end{equation}
for $\bs r \ne \bs r_0$ and
\begin{equation}\label{eq:general-pde-constant-d-solution-continued-gaussian_3}
        \phi(\bs r_0,t) = 
    \begin{cases}
        \frac{Q_0}{4\pi D}\ln{\big(\frac{D_s+2D(t-t_0)}{D_s}\big)},
        & t_0\le t \le t_0+\tau,\\
        \frac{Q_0}{4\pi D} \ln{\big(\frac{D_s+2D(t-t_0)}{D_s+2D(t-t_0-\tau)}\big)},
        & t > t_0+\tau.
        \end{cases}
\end{equation}

This special case, which forms the basis of the discussion in the remainder of this paper, is a good approximation for focused or defocused electron beams ~\cite{yang20154d} in the absence of any non-circular aberrations.
\begin{remark}\label{remark:total_electron}
By computing the integral in Eq.~\eqref{eq:total_electrons} for the above source functions described by Eqs.~\eqref{eq:source-function-continued-point},~\eqref{eq:source-function-continued-circular}, and ~\eqref{eq:source-function-continued-gaussian}, the total number of diffusing species $Q_{\rm tot}(t)$ at time $t$ in all systems is;
\begin{equation*}
    Q_{\rm tot}(t) = Q_0 \cdot \min(t-t_0,\tau).
\end{equation*}

We note that the normalisation factors in the spatial component of the activation functions in Eqs.~\eqref{eq:source-function-continued-circular} and  \eqref{eq:source-function-continued-gaussian} mean that the number of diffusing species in the system is constant when the size of the source changes. Since by definition, the function $\cl M$ maps every value of the beam current to a single value of the diffusing species, this normalisation of the electron beam is consistent with the optical conditions available in modern STEM instruments: for a fixed beam current, changing the electron probe size does not affect the number of electrons (and in turn, the number of diffusing species) in the system.
\end{remark}

%%%%%% Mathematical Model of Diffusion %%%%%%
%%%%%% Start of diffusion is STEM %%%%%%

\section{Diffusion in STEM} 
\label{sec:diffusion-in-stem}
We start by introducing a general formulation of diffusion processes in STEM. Subsequently, we restrict our analysis to the following assumptions:
\begin{itemize}
    \item [(A1)] The rate of initial diffusing species $Q_0$ is related to the beam current through a general function $\cl M$. It is not necessary to specify the function $\cl M$ and we here assume that the units of diffusion distribution is $\diffunit$ where ``$\rm u$'' is an arbitrary unit independent of the dwell time, diffusion coefficient, and probe size; 
    \item[(A2)] The sample is uniform and infinitesimally thin and is significantly larger than the probe;
    \item[(A3)] The diffusion coefficient within the sample is constant;
    \item[(A4)] An airy disk electron probe can be approximated by a Gaussian probe;
    \item[(A5)] The number of diffusing species in the system is constant with respect to changes in the size of the electron probe.
\end{itemize}
Assumption (A1) is a restatement of Remark~\ref{remark:units_of_diffusion}. Assumption (A2) implies that diffusion in STEM can be modeled by a 2D process in an infinite medium. Assumptions (A3) and (A4) together allow the use of the diffusion profile of a Gaussian-shaped source as formulated in Eq.~\eqref{eq:general-pde-constant-d-solution-continued-gaussian_2}. Assumption (A5) is made for convenience to ensure that the beam current is constant with changes in the probe conditions.

\subsection{Diffusion distribution of an electron probe activated at a single location}\label{subsec:stem_single_probe}
In this section, we study the behaviour of the diffusion profile of a single electron probe, (a probe activated at a single location). Following the assumptions above and from Eqs.\eqref{eq:general-pde-constant-d-solution-continued-gaussian_2} and \eqref{eq:general-pde-constant-d-solution-continued-gaussian_3}, the diffusion distribution of the $i$-th scanned electron probe positions with square width $D_s$, activated at location $\bs r_i$ and time $t_i$, and during a dwell-time $\tau_i$ is
\begin{equation}\label{eq:stem_single_probe_beamon}
    \phi_{i}^{\rm on}(\bs r,t) = \begin{cases}
        \frac{Q_0}{4\pi D}\Big(\eint{\|\bs r - \bs r_i\|_2^2}{2D_s + 4D(t-t_i)}-\eint{\|\bs r - \bs r_i\|_2^2}{2D_s}\Big), & {\rm if~} \bs r \ne \bs r_i,\\
        \frac{Q_0}{4\pi D}\ln{\big(\frac{D_s+2D(t-t_i)}{D_s}\big)}, & {\rm if~} \bs r = \bs r_i,
    \end{cases}
\end{equation}
When the beam is on, \ie, $t_i \le t\le t_i + \tau_i$, and 
\begin{equation}\label{eq:stem_single_probe_beamoff}
    \phi_i^{\rm off}(\bs r,t) = \begin{cases}
        \frac{Q_0}{4\pi D}\Big(\eint{\|\bs r - \bs r_i\|_2^2}{2D_s +4D(t-t_i)} - \eint{\|\bs r - \bs r_i\|_2^2}{2D_s+4D(t-(t_i+\tau_i))}\Big), & {\rm if~} \bs r \ne \bs r_i,\\
        \frac{Q_0}{4\pi D} \ln{\big(\frac{D_s+2D(t-t_i)}{D_s+2D(t-t_i-\tau_i)}\big)}, & {\rm if~} \bs r = \bs r_i,
         \end{cases}
\end{equation}
and when the  beam is off, \ie $t>t_i+\tau_i$. 
In this paper, for convenience, we also use the following compact form to denote the diffusion distribution:
\begin{equation}\label{eq:stem_single_probe}
    \phi_i(\bs r,t) = \begin{cases}
    0, & {\rm for~}  0\le t< t_i,\\
        \phi_i^{\rm on}(\bs r,t), & {\rm for~} t_i \le t\le t_i + \tau_i,\\
        \phi_i^{\rm off}(\bs r,t), & {\rm for~} t_i + \tau_i < t.
         \end{cases}
\end{equation}

From Eqs.~\eqref{eq:stem_single_probe_beamon} and \eqref{eq:stem_single_probe_beamoff} and using L'H\^{o}pital's rule we have
\begin{equation}\label{eq:stem_single_probe_limit_d_zero}
    \lim_{D\rightarrow 0^+} \phi_i^{\rm on}(\bs r,t) = \frac{Q_0(t-t_i)}{2\pi D_s}e^{-\frac{\|\bs r - \bs r_i\|^2}{2D_s}}\quad{\rm and}\quad
    \lim_{D\rightarrow 0^+} \phi_i^{\rm off}(\bs r,t) = \frac{Q_0\tau_i}{{2\pi D_s}} e^{-\frac{\|\bs r - \bs r_i\|^2}{2D_s}}
\end{equation}
and 
\begin{equation}\label{eq:stem_single_probe_limit_d_infty}
    \lim_{D\rightarrow \infty} \phi_i^{\rm on}(\bs r,t) = 
    \lim_{D\rightarrow \infty} \phi_i^{\rm off}(\bs r,t) = 0.
\end{equation}
Equation \eqref{eq:stem_single_probe_limit_d_zero} states that for an asymptotically small diffusion coefficient, the diffusion distribution follows the Gaussian shape of the electron probe with a magnitude that increases proportional to $t-t_i$ reaching a maximum at $t=t_i+\tau_i$. Equation \eqref{eq:stem_single_probe_limit_d_infty} shows that for a large diffusion coefficient, diffusion develops quickly as expected.

In Sections ~\ref{sec:diffusion_function_of_distance} and \ref{sec:diffusion_function_of_time} we highlight important properties of the diffusion distribution in Eq.~\eqref{eq:stem_single_probe}. For example, Lemmas \ref{lem:stem_single_probe_derivatives_beamon} and \ref{lem:stem_single_probe_derivatives_beamoff} summarise the behaviour of the diffusion distribution as a function of distance to the activation point and predict situations where the diffusion distribution is an increasing, decreasing, convex, or concave function of distance to the activation point. Lemmas \ref{lem:stem_single_probe_derivatives_t_beamon} and \ref{lem:stem_single_probe_derivatives_t_beamoff} provide similar analyses for diffusion distribution with respect to time. From those findings we report the following result.
\begin{corollary}\label{cor:stem_single_prob}
From Lemmas \ref{lem:stem_single_probe_derivatives_t_beamon} and \ref{lem:stem_single_probe_derivatives_t_beamoff}, for a single electron probe in STEM, at a given time instance, the maximum distribution of diffusing substances happens at the activation point. Moreover, the Maximum Beam Diffusion Distribution (M-BDD), denoted by $ A^{\max}_{\rm bdd}$, occurs at the activation point at the end of the activation time. Mathematically,
\begin{equation}\label{eq:cor_stem_single_probe}
        \phi_i(\bs r,t) \le  \phi_i(\bs r_i,t)\le \phi_i(\bs r_i,t_i+\tau_i) = A^{\max}_{\rm bdd} \coloneqq\frac{Q_0}{4\pi D}\ln(1+2\rho^{-1}\tau_i), \quad {\rm for~} \bs r \in \bb R^{2}{~\rm and~}  t \ge t_i,
\end{equation}
where $\rho \coloneqq \frac{D_s}{D}$ is the ratio between the square width of the Gaussian-shaped probe and the diffusion coefficient.
 \end{corollary}

\subsection{Diffusion distribution during STEM acquisition} \label{subsec:diffusion_in_stem}
Having formulated the diffusion distribution activated from an electron probe at a single location, we now focus on the full STEM acquisition. 

Our model is based on a STEM acquisition where $N$ spatial locations collected in a set of probe locations  $\cl R \coloneqq \{\bs r_1,\cdots,\bs r_N\}$ are sequentially scanned with an arbitrary trajectory. Let $\cl T_{\rm s} \coloneqq\{t_1,\cdots,t_N\}$ and $\cl T_{\tau}\coloneqq\{\tau_1,\cdots,\tau_N\}$ be, respectively, the set of activation and dwell times for these positions. Using Eq.~\eqref{eq:stem_single_probe}, the \textit{Cumulative Diffusion Distribution (CDD)} from $j$ scanned electron probe positions, for $j\in \{1,\cdots,N\}$, is
\begin{equation}\label{eq:stem_cumulative_simple}
    \psi_j(\bs r, t) \coloneqq \sum_{i=1}^{j} \phi_i(\bs r, t),
\end{equation}
or equivalently,
\begin{equation}\label{eq:stem_cumulative}
    \psi_j(\bs r, t) = \begin{cases}
        0, & t<t_1,\\
        \phi_1^{\rm on}(\bs r, t), & t_1\le t< t_1+\tau_1,\\
        \phi_1^{\rm off}(\bs r, t), & t_1+\tau_1 \le t< t_2,\\
        \phi_2^{\rm on}(\bs r, t)+ \phi_1^{\rm off}(\bs r, t), & t_2\le t< t_2+\tau_2,\\
        \vdots & \vdots \\
        \phi_j^{\rm on}(\bs r, t) + \sum_{i=1}^{j-1}\phi_i^{\rm off}(\bs r, t), & t_{j}\le t < t_j + \tau_j,\\
        \sum_{i=1}^{j}\phi_i^{\rm off}(\bs r, t), & t_{j}+\tau_{j} \le t \le t_{j+1}.
    \end{cases}
\end{equation}

Equation \eqref{eq:stem_cumulative} assumes a general STEM configuration with variable dwell and settling times for every electron probe position and an arbitrary scanning probe trajectory. In practice STEM often operates with a constant dwell time and close to zero settling time, \ie for all $i \in \{1,\cdots,N\}$,
\begin{align} \label{eq:stem_common_stem_config}
    \tau_i = \tau \quad {\rm and} \quad t_{i+1} = t_{i}+\tau_i. 
\end{align}
Equation \eqref{eq:stem_cumulative} is applicable to any space filling scanning probe trajectory such as raster, snake, Hilbert~\cite{velazco2020evaluation}, Z-order, alternating, spiral~\cite{sang2016dynamic, li2018compressed} and random, which can all be defined by simply considering an appropriate ordering of probe locations within the probe location set $\cl R$. 

In the numerical simulations in Section ~\ref{sec:numerical_results}, we show examples of raster, snake, random, and alternating scans.

Our analysis of the role of diffusion coefficient on the diffusion distribution caused by a single scanned electron probe position in Eq.~\eqref{eq:stem_single_probe_limit_d_zero} can be extended to the diffusion distribution in a full STEM acquisition. From Eqs.~\eqref{eq:stem_single_probe_limit_d_zero} and \eqref{eq:stem_cumulative_simple}, we can compute
\begin{equation}\label{eq:stem_role_of_d}
    \lim_{D\rightarrow 0^{+}} \psi_j(\bs r, t) = \lim_{D\rightarrow 0^{+}} \phi_j(\bs r, t) + \sum_{i = 1}^{j-1} \lim_{D\rightarrow 0^{+}} \phi_i(\bs r, t) = \frac{Q_0(t-t_j)}{2\pi D_s} e^{\frac{-\|\bs r - \bs r_j\|^2}{2D_s}} + \frac{Q_0}{{2\pi D_s}}\sum_{i=1}^{j-1} \tau_i e^{\frac{-\|\bs r - \bs r_i\|^2}{2D_s}},
\end{equation}
which shows that for an asymptotically small diffusion coefficient the CDD will be a sum of Gaussian functions centered at specific probe locations with time-invariant widths.

\section{Damage as a diffusion mechanism in STEM}\label{subsec:damage} 
The CDD in Eq.~\eqref{eq:stem_cumulative} does not model the physical processes underlying the damage induced by electron beam in a sample. 
We recall from Section ~\ref{sec:intro} that our hypothesis is that sample damage is a function of CDD, without specifying the exact physical mechanisms of damage. We subsequently refer to this generically as \textit{Diffusion Induced Damage (DID)}. In this section we show that the proposed diffusion model can be coupled to any physical damage mechanism that is induced by a diffusion process.

Let $\lambda \ge 0$ be a DID threshold and let $\Lambda_j(\bs r, t; \lambda)$ denote the DID w.r.t to the $j$-th activated probe at a given location $\bs r$ and time $t$. Therefore, $\Lambda_j(\bs r, t; \lambda)$ can be related to the CDD $\psi_j(\bs r,t)$ in Eq.~\eqref{eq:stem_cumulative} using a 
(temporal) integration, a non-linear activation function $g$, and a pupil function $p$ as
\begin{equation}\label{eq:stem_did}
    {\rm (Irreversible~DID~model)}\quad \Lambda_j(\bs r, t; \lambda) \coloneqq \int_{t_1}^{t} p(\bs r,t')\cdot g(\psi_j(\bs r, t') - \lambda)\, \ud t'.
\end{equation}
Eq.~\eqref{eq:stem_did} describes different models for DID. In Eq.~\eqref{eq:stem_did}, the non-linear function $g$ determines how damage is related to the CDD at a given location in space and time and the pupil function $p$ controls how damage at one location impacts damage at other locations. Moreover, integrating the damage quantity over time ensures that the damage is irreversible. 

Depending on the nature of the DID, a suitable activation function $g$ can be defined to represent the important features of the damage. As examples, we provide two illustrations:
\begin{itemize}
    \item  \textit{(DID frequency)} In cases where the frequency of the DID event during STEM acquisition is critical, a sign function can be used:
\begin{equation}\label{eq:stem_did_frequency}
    g(x) = {\rm sign}(x) \coloneqq \begin{cases}
        1, & x \ge 0,\\
        0, & x < 0.
    \end{cases}
\end{equation}
    \item  \textit{(DID intensity)} The intensity of the damage can alternatively be defined using a Rectified Linear Unit (ReLU) function:
\begin{equation}\label{eq:stem_did_intensity}
    g(x) = {\rm ReLU}(x) \coloneqq \begin{cases}
        x, & x \ge 0,\\
        0, & x < 0.
        \end{cases}
\end{equation}
\end{itemize}

There is also a sensitive feature in the damage model Eq.~\ref{eq:stem_did} that is accounted for by the use of the pupil function. As an illustration, we consider the following two specific cases for the observation of DID:
\begin{itemize}
    \item  \textit{(Offline DID)} This case accounts for damage that occurs in a location even after that location is scanned. In this case
\begin{equation}\label{eq:stem_did_offline}
    p(\bs r,t) = 1, \quad\forall \bs r {~\rm and~} \forall t.
\end{equation}
Hence, this type of damage can only be observed  by re-scanning the sample provided that re-scanning does not create further damage. This damage type is referred to as ``damage after scan'' in \cite{jannis2022reducingpart2}.

    \item  \textit{(Online DID)} This case, which is referred to as ``damage during scan'' in \cite{jannis2022reducingpart2}, considers damage that occurs during the irradiation of a given location by the probe. Any damage created in that position after moving probe is not further considered. However, this examples requires a careful definition of the pupil function. 
    For example, assuming  
\begin{equation}\label{eq:stem_did_online}
    p(\bs r,t) = \begin{cases}
        1, & {\rm if~} \exists j{\rm ~s.t.~} t_{j}\le t\le t_{j+1}, {\rm and~if~} \exists i>j {\rm ~s.t.~} \|\bs r - \bs r_i\|_2 \le r_p,\\
        0, & {\rm otherwise}.
        \end{cases}
\end{equation}
This pupil function states that when the $j^{\rm th}$ position is activated, the value of the damage will increase at a location $\bs r$ only if that location is within a radius $r_p$ of at least one of the probes positions that is activated in the remaining scan.
\end{itemize}

From Eq.~\eqref{eq:stem_did}, regardless of the choice of activation and pupil functions, the overall point-wise DID quantity can be computed at the end of the acquisition as:
\begin{equation}\label{eq:stem_did_overal}
    \Lambda(\bs r; \lambda) \coloneqq \Lambda_N(\bs r, t_N+\tau_N; \lambda),
\end{equation}
as well as the overall DID as $\Lambda(\lambda) \coloneqq \int \Lambda(\bs r; \lambda)\,\ud \bs r$.
\begin{remark}\label{remark:damage-assumption}
    We recall that our goal is to show that the proposed framework of diffusion distribution can be used in conjunction with a damage mechanism. However, the feasibility of Eqs.~\eqref{eq:stem_did}and~\eqref{eq:stem_did_overal} depends on the accuracy of the damage activation and pupil functions, as well as on the damage threshold. Hence, we require a quantity that is agnostic to any damage related assumptions so that different systems can be compared only based on variations on models of diffusion distribution.
\end{remark}
We note that the CDD $\psi_j(\bs r, t)$ does not incorporate sufficient information about DID, since the DID may happen at any time prior to the time the CDD is computed. Hence, we define 
a quantity, which can be used as an indicator of a DID event; the maximum value of the CDD at every spatial point taken over all previous time instances. More formally, we introduce the Point-wise Maximum Cumulative Diffusion Distribution (PM-CDD) w.r.t. to the $j^{\rm th}$ electron probe position, as $\chi_j(\bs r, t)$ and defined as 
\begin{equation}\label{eq:stem_mp_cdd}
    \chi_j(\bs r, t) \coloneqq \underset{\substack{ 1\le j' \le j \\t_{j'} \le t' \le t   }}{\rm max}~\psi_{j'}(\bs r, t'), \quad {\rm for~} t \ge t_j.
\end{equation}
The benefit of introducing PM-CDD is twofold. First, in the case of unknown damage threshold and activation and pupil functions, PM-CDD can be used to compare different STEM scans and provide information about how likely the DID can occur. Second, the design of DID-free STEM scans,  discussed in Section ~\ref{subsec:mask_design}, will involve a direct comparison of the PM-CDD with a damage threshold, which is computationally less demanding compared to the use of CDD.

We introduce a reduced and simplified notation, where if the index of the electron probe position is not specified, the term PM-CDD refers to the PM-CDD at the end of the scan as;
\begin{equation}
    \chi(\bs r) \coloneqq \chi_N(\bs r, t_N+\tau_N).
\end{equation}
Therefore, as mentioned above, the PM-CDD provides information about the likelihood of DID occurring during the full scan regardless of the damage mechanism. Subsequently in Section ~\ref{subsec:mask_design}, we will use the PM-CDD in a comparison of different STEM scans.

Furthermore, from the PM-CDD, we can define a Global Maximum of Cumulative Diffusion Distribution (GM-CDD) $\chi^{\max}$, as
\begin{equation}\label{eq:gm-cdd}
    \chi^{\max} \coloneqq \max_{\bs r} \chi(\bs r),
\end{equation}
which gives the maximum value of the CDD over all spatial and temporal points during an acquisition.
This value can be used to identify a DID-free STEM scan that is agnostic to the DID model. We define a STEM scan to be DID-free, if the overall DID quantity is zero \ie $\Lambda(\lambda)=0$. The following theorem, which is proved in Section ~\ref{subsec:proof-thm-did-free-stem}, relates the DID profile to the GM-CDD.
\begin{theorem}\label{thm:did-mpcdd}
    Consider a DID model characterised by a non-negative activation function $g$, \ie $g : \mathbb{R}\mapsto \mathbb{R}_{\ge 0}$, in either Eq.~\eqref{eq:stem_did_frequency} or Eq.~\eqref{eq:stem_did_intensity}, and a pupil function $p$, \ie $ p : \mathbb{R}\mapsto \mathbb{R}$, such that $\int_{t_1}^{t_N} p(\bs r,t')\, \ud t' >0, \forall \bs r$. A STEM scan is DID-free, iff, the GM-CDD is less than or equal to the DID threshold. Formally,
    \begin{equation}\label{eq:stem_did_free}
    {\rm (DID-free~STEM)}\quad \Lambda(\lambda) = 0, \quad \Longleftrightarrow \quad \chi^{\max} \le \lambda.
\end{equation}
\end{theorem}
Theorem~\ref{thm:did-mpcdd} shows that for a DID-free STEM scan, it is only necessary to examine the GM-CDD value. Hence, that quantity, or the PM-CDD, will be used as a measure in a comparison of different STEM scans in Section~\ref{subsec:simulations_cs_stem}.
%%%%%%%%%%%%%%%%%%%%%%%%%%%%%%%%

\section{Diffusion distribution in compressive STEM}\label{subsec:compressive_stem}
To reduce sample damage in STEM, several alternative (non-raster) scans~\cite{kovarik2016implementing, robinson2022sim, browning2023advantages} have been proposed based on CS~\cite{CStheory2006,candes2006robust} which have demonstrated that sample damage can be reduced or mitigated. A common theme in these studies is to activate the electron probe at a small subset of possible locations, \ie to subsample probe locations, and to recover a STEM image from those incomplete measurements by solving an inverse problem, known as inpainting.

The CDD and DID formulated in Eqs.~\eqref{eq:stem_cumulative} and \eqref{eq:stem_did}, respectively, can be simply adapted for compressive STEM scans. Let $\bs s\coloneqq [s_1,\cdots,s_N]^\top \in \{0,1\}^N$, with $\sum_j s_j = M \ll N$, be a binary vector associated with a subsampling of $M$ probe locations; $s_j = 1$
if the electron probe is activated at the $j$-th location $\bs r_j$ and $s_j = 0$, if it is not activated. From Eq.~\eqref{eq:stem_cumulative_simple}, the CDD in compressive STEM w.r.t. to the subsampling strategy $\bs s$ is given as
\begin{equation}\label{eq:stem_cumulative_compressive_stem}
    \psi_j^{\rm cs}(\bs r, t) \coloneqq \sum_{i=1}^{j} s_i\phi_i(\bs r, t).
\end{equation}
We note that Eq.~\eqref{eq:stem_cumulative_compressive_stem} supports any arbitrary subsampling strategy of  probe positions. By comparing the DID profile for full STEM, \ie $\Lambda(\bs r;\lambda)$, and compressive STEM, \ie,  $ \Lambda^{\rm cs}(\bs r;\lambda)$, it is clear that compressive STEM has the advantage of a lower DID. Alternatively, since the PM-CDD in compressive STEM is less than the PM-CDD in full STEM, 
\begin{equation*}
\Lambda^{\rm cs}(\bs r;\lambda) \le \Lambda(\bs r; \lambda).
\end{equation*}

Direct application of Eq.\eqref{eq:stem_cumulative_compressive_stem} allows the modelling of a compressive STEM scan using a beam blanker to perform probe subsampling~\cite{reed2022electrostatic}. Alternatively for the use of a programmable scan generator,  Eq.~\eqref{eq:stem_cumulative_simple} can be used to redefine the set of scanned probe locations $\cl R$ and associated activation times $\cl T_s$. We also recall that the acquisition time in compressive STEM with a beam blanker and a scan generator is proportional to, respectively, the number of total probe positions and subsampled probe positions.

From the findings in Corollary ~\ref{cor:stem_single_prob}, we derive the following theorem, which states the sufficient condition in compressive STEM for reducing DID.
\begin{theorem}\label{thm:cs_stem}
Let $\lambda>0$ be the DID threshold of a sample and $\tau^{\rm max} \coloneqq \max_{i} \tau_i$  the longest dwell time of the electron probe. Therefore, if
\begin{equation}\label{eq:thm_sc_stem}
A^{\max}_{\rm bdd} = \frac{Q_0\rho}{2}\ln(1+2\rho^{-1}\tau^{\rm max}) \ge \lambda \Longrightarrow \Lambda(\lambda) > 0.
\end{equation}
Hence, if DID is due to an individual electron probe, rather than the CDD, subsampling probe positions using any strategy will not result in a DID-free STEM.
\end{theorem}

The total number of diffusing species during the acquisition also needs to be considered when comparing conventional and compressive STEM. The application of Remark~\ref{remark:total_electron} to the CDD in Eqs.~\eqref{eq:stem_cumulative} and \eqref{eq:stem_cumulative_compressive_stem}, gives the total number of deposited electrons for conventional and compressive STEM, $Q_{\rm stem}$ and $Q_{\rm cstem}$, respectively, as
\begin{equation}
    Q_{\rm stem} = Q_0\sum_{j=1}^N \tau_j, \quad {\rm and} \quad    Q_{\rm cstem} = Q_0\sum_{j=1}^N s_j\tau_j,\label{eq:total_electrons_stem}
\end{equation}
From the above it is evident that $Q_{\rm cstem} \le Q_{\rm stem}$; fewer units are diffused in compressive STEM compared to conventional STEM as intuitively expected.
In a simplified model with a constant dwell time for all electron probe positions, \ie $\tau = \tau_j$ for all $j \in {1,\cdots,N}$, these quantities become
\begin{equation}
    Q_{\rm stem} = N Q_0 \tau , \quad {\rm and} \quad    Q_{\rm cstem} = M Q_0 \tau.\label{eq:total_electrons_stem_simplified}
\end{equation}

\subsection{Diffusion controlled sampling strategy for DID-free compressive STEM }\label{subsec:mask_design}
As described above, compressive STEM results in a lower DID compared to conventional STEM. However, this raises the following questions:
\begin{itemize}
    \item[\textit{(i)}] Given a DID threshold for constraining the PM-CDD, what is the optimal design for subsampling mask containing the maximum number of probe locations?
    \item[\textit{(ii)}] Given a fixed number of scanned probe locations, what is the optimal design of a mask minimising DID or PM-CDD?
\end{itemize}
Finding a general subsampling strategy for both cases will require advanced tools from optimisation theory and is deferred to a subsequent publication. Hence, in this paper we confine ourselves to providing a solution to the first question only. 

Our approach for designing a Diffusion Controlled Sampling (DCS) strategy of the probe positions that results in DID-free STEM is as follows; given a DID threshold $\lambda$ as in Eq.~\eqref{eq:stem_did_overal}, a probe location in the subsampling mask $\cl R$ is selected, only if the PM-CDD w.r.t. to that electron probe over all spatial locations is lower than that DID threshold as; 

\begin{equation}\label{eq:stem_did_free_csstem}
    j \in \cl R, \quad{\rm if} \quad \chi_j(\bs r, t_j+\tau_j) < \lambda, \quad {\rm for~all~} \bs r.
\end{equation}
Using Theorem~\ref{thm:did-mpcdd}, this approach guarantees that the GM-CDD $\chi^{\max} < \lambda$ and hence, $\Lambda^{\rm cs}(\bs r; \lambda) = \bs 0$ for all locations $\bs r$ and DID does not occur. We provide numerical examples of this approach in Section ~\ref{subsec:simulations_mask_design}. We note here that this problem does not necessarily have a unique solution and the approach described provides only one example of such a DCS strategy. However, other approaches, \eg based on constraining the minimum distance between two subsampled electron probes, can be simply incorporated within our approach.

%%%%%%%%%%% Simulations %%%%%
\section{Numerical Results}\label{sec:numerical_results}
In this section we describe numerical simulations in support of the analyses above , which highlight important properties of DID in STEM.

At the start of this Section we note that despite the availability of a closed-form solution for CDD in STEM, the associated numerical simulations are still computationally demanding. Therefore, in Section ~\ref{subsec:time_complexity}, we analyse the time complexity of computing  the CDD in Eqs.~\eqref{eq:stem_cumulative} and \eqref{eq:stem_cumulative_compressive_stem} for full STEM and compressive STEM, respectively. We also report approximation alternatives for accelerated simulation followed by numerical confirmations.

\begin{table}
    \centering
    \caption{\textbf{Parameter setting for a baseline STEM scan.} A $20\times 20$ square grid is considered for a raster scan of probe positions. These equate to a $4~{\rm ms}$ acquisition time and $0.1 {~\rm nm}$ probe radius.}
    \begin{tabu}{|[2pt]l|l|[2pt]l|l|[2pt]}
    \tabucline[2pt]{1-4}
        Parameter & Value & Parameter & Value\\
        \tabucline[2pt]{1-4}
        Dwell time & $\tau = 10~\mu{\rm s}$ & Scan step size & $\Delta_p = 0.05{~\rm nm}$\\
         Settling time & $\bar{\tau} = 0~{\rm s}$ & diffusion coefficient & $D = 10 {~\rm nm}^2\cdot {\rm s}^{-1}$\\
         Probe width parameter & $D_s = 0.01 {~\rm nm}^2$ & Initial deposited units & $Q_0 = 63.45 ~{\rm M}\qunit$
    \\
        \tabucline[2pt]{1-4}
    \end{tabu}

    \label{tab:baseline_parameters}
\end{table}
\noindent\textbf{Baseline STEM.} The parameters for the baseline simulation are given in Table ~\ref{tab:baseline_parameters}. Subsequently, for other simulations we report only those parameters that differ from this baseline setup. We consider a square array of probe positions giving $N = 20^2$ total probes. We further assume a conventional STEM scan described by Eq.~\eqref{eq:stem_common_stem_config} with a constant dwell time and zero settling time. This equates to a $4~{\rm ms}$ acquisition time and $0.1 {~\rm nm}$ probe radius. The diffusion coefficient in Table ~\ref{tab:baseline_parameters} is in the range of the values that was studied in~\cite{jannis2022reducingpart2,interleaveSTEM2022} for a specific zeolite sample. This setting is equivalent to M-BDD $A^{\max}_{\rm bdd} = 10^4~\diffunitnm$ and a total number of diffusing species $Q_{\rm stem} = 253.83$ Ku. We now let $\cl R_{\rm sim}$ be a set of spatial points defined as the simulation grid. We assume that $N_T$ temporal grids are defined during the scan of every two consecutive electron probe positions, \ie $t_i \le t \le t_{i+1}$. Each scan step length is simulated using 10 pixels, \ie $|\cl R_{\rm sim}| = 100N$. Furthermore, only time instances w.r.t to the probe dwell times are simulated, \ie $N_T =1$.
\begin{figure}[tb]
    \centering
    \includegraphics[width=1\columnwidth]{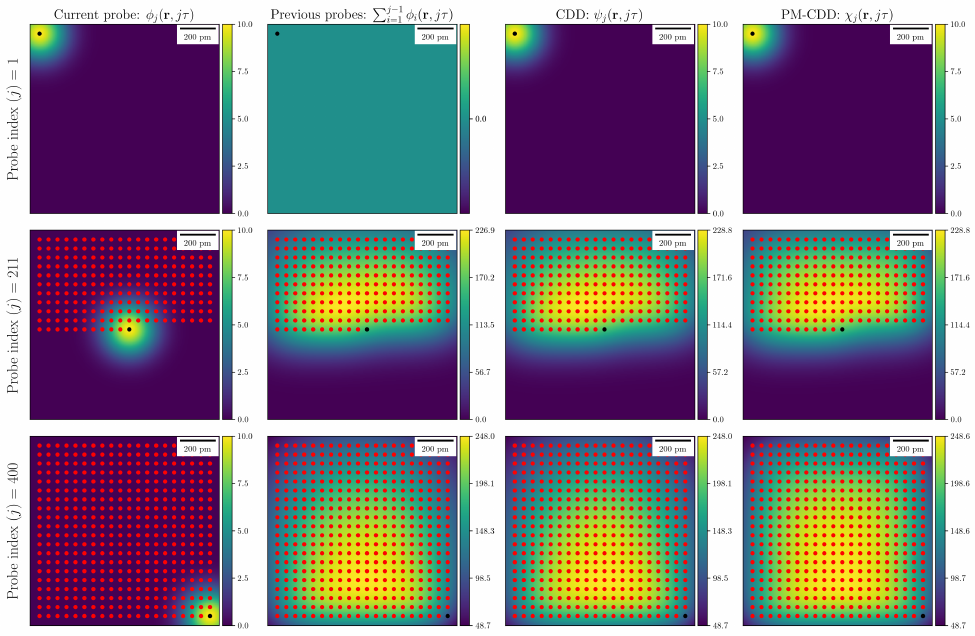}
    \caption{\textbf{Snapshots of diffusion distribution in a baseline STEM scan for a probe activated at three selected positions.} Top, middle, and bottom rows correspond to the first, middle, and last activated probe, respectively. The first column is the diffusion distribution caused by the electron probe activated at the current position. The second column shows the cumulative diffusion distribution caused by the electron probe activated at the previous location. The third column is the sum of the first two columns. The last column is the point-wise maximum of the cumulative diffusion distribution. Red and black points denote, respectively, the previously and currently activated electron probe positions. Every probe has an M-BDD $A^{\max}_{\rm bdd} = 10^4 ~\diffunitnm$, although the PM-CDD, which controls the DID, is 24.8 times greater. The units of color bars are $10^3~\diffunitnm$.}
\label{fig:baseline}
\end{figure}

\subsection{Diffusion distribution in a full STEM scan}\label{subsec:simulations_baseline}
We first simulate the CDD for a full STEM scan using the parameters in Table ~\ref{tab:baseline_parameters}. Figure ~\ref{fig:baseline} illustrates \textit{(i)} the diffusion distribution $\phi_j$ of the activated electron probe at the first, middle and last locations (first column), \textit{(ii)} the CDD of previously scanned electron probe positions $\sum_{i=1}^{j-1}\phi_i$ (second column), \textit{(iii)} CDD $\psi_j$ w.r.t to the $j^{\rm th}$ electron probe position (third column), and \textit{(iv)} the corresponding PM-CDDs $\chi_j$, during a full STEM scan, for three representative activated electron probe positions (last column). We use Eq.~\eqref{eq:stem_cumulative} for \textit{(i-iii)} and Eq.~\eqref{eq:stem_mp_cdd} for \textit{(iv)}. Fig.~\ref{fig:baseline} shows that although the M-BDD $A^{\max}_{\rm bdd} = 10^4~\diffunitnm$, the PM-CDD $\chi(\bs r)$ reaches a maximum value of $ 248.6\cdot 10^3~\diffunitnm$, which determines the DID.

\begin{figure}[t!bh]
    \centering
\includegraphics[width=1\columnwidth]{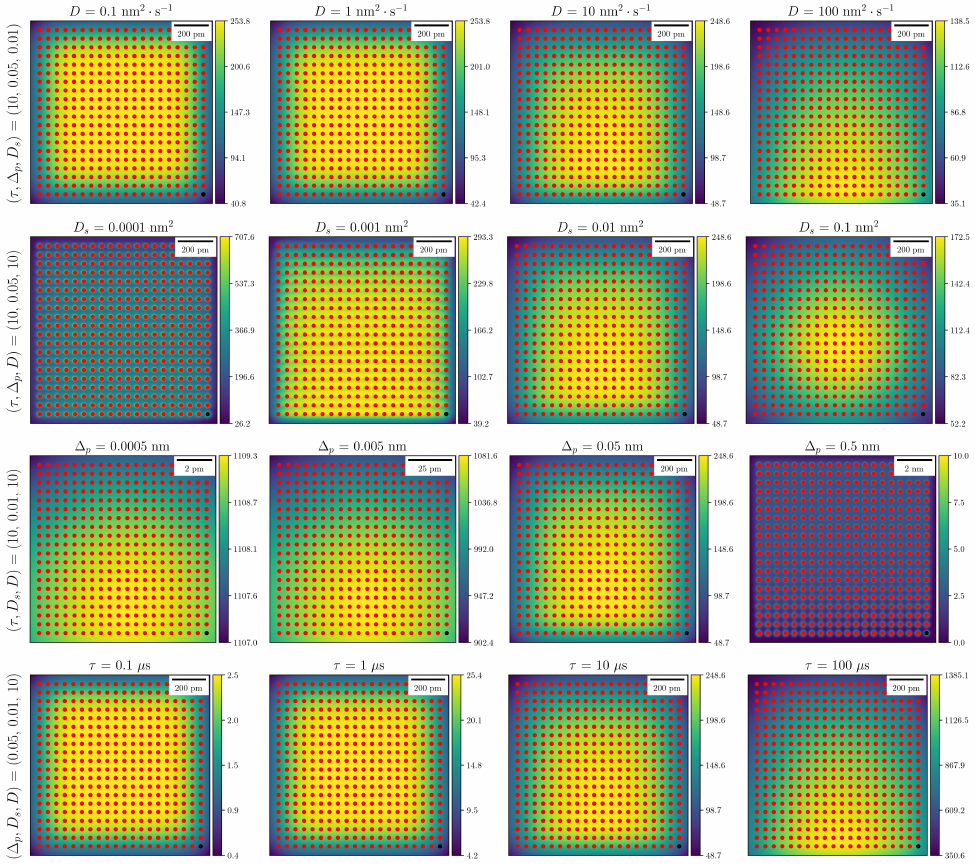}
    \caption{\textbf{Comparison of PM-CDD $\chi(\bs r)$ for different STEM scans.} Diffusion coefficient (first row), probe width parameter (second row), scan step (third row), and dwell time (fourth row) are varied. The PM-CDD of a STEM scan increases by either decreasing the diffusion coefficient, increasing the probe radius, decreasing the scan step size, or increasing the dwell time. Other parameters are the same as in Table ~\ref{tab:baseline_parameters}. The total number of deposited diffusing units is constant in the systems shown in the first three rows, \ie  $Q_{\rm stem} = 253.8$ Ku, whereas in the forth row, it is proportional to the dwell time. The units of color bars are $10^3~\diffunitnm$.}
\label{fig:fig_full_stem_system_examples}
\end{figure}

In Figure ~\ref{fig:fig_full_stem_system_examples}, we show an extended analysis with additional simulations  for multiple values of the diffusion coefficient $D \in \{0.1, 1, 10, 100\}$ ${\rm nm}^2\cdot {\rm s}^{-1}$, probe width parameter $D_s \in \{10^{-4}, 10^{-3}, 10^{-2},10^{-1}\}$ ${\rm nm}^2$, scan step size $\Delta_p \in \{0.0005, 0.005, 0.05, 0.5\}~{\rm nm}$, dwell time $ \tau \in \{0.1,1, 10,100\}~\mu{\rm s}$. Fig.~\ref{fig:fig_full_stem_system_examples} compares the PM-CDD of these systems. 

From Fig.~\ref{fig:fig_full_stem_system_examples} we observe that the PM-CDD increases if \textit{(i)} the diffusion coefficient decreases, \textit{(ii)} the probe radius  increases, \textit{(iii)} the scan step size decreases, or \textit{(iv)} the dwell time increases. Moreover as we showed in Eq.~\eqref{eq:stem_role_of_d} for a very small diffusion coefficient, the PM-CDD reduces to a sum of Gaussian functions with constant width parameter $D_s$ consistent with this behaviour. In the first row of Fig.~\ref{fig:fig_full_stem_system_examples}, it is clear that for a diffusion coefficient less than $D = 10 {~\rm nm}^2\cdot{\rm s}^{-1}$, the PM-CDD stays almost constant and the resulting heatmap is a sum of Gaussian functions. For a very small probe radius, the Gaussian-shaped electron probe approximates to a point like source, which is evident in the second row of Fig.~\ref{fig:fig_full_stem_system_examples}. However, a very small probe radius is functionally equivalent to very distant probe locations. Hence, the two heatmaps in Fig.~\ref{fig:fig_full_stem_system_examples} corresponding to $D_s = 0.0001 {~\rm nm}^2$ and $\Delta_p = 0.5 {~\rm nm}$ are similar in shape but are significantly different in their range of the values. We recall that in the work described here the total number of diffusing units in the system does not depend on the probe radius; hence, reducing the probe to a point source simply amounts to a very high value of the diffusion distribution at the scanned probe positions.

We emphasise that our aim in simulations is to illustrate the variability of the distribution of cumulative diffusion rather than to propose a specific experimental damage reduction strategy. In this regard we further note that \textit{(i)} the diffusion coefficient is sample-dependant; \textit{(ii)} increasing the probe radius reduces the image resolution in STEM; \textit{(iii)} increasing the electron probe step size also reduces the image resolution in STEM; and \textit{(iv)} reducing the dwell time can result in scan distortions and reduced signal-to-noise ratio~\cite{buban2009highresolution}.

\begin{figure}[!tbh]
    \centering
    \includegraphics[width=1\columnwidth]{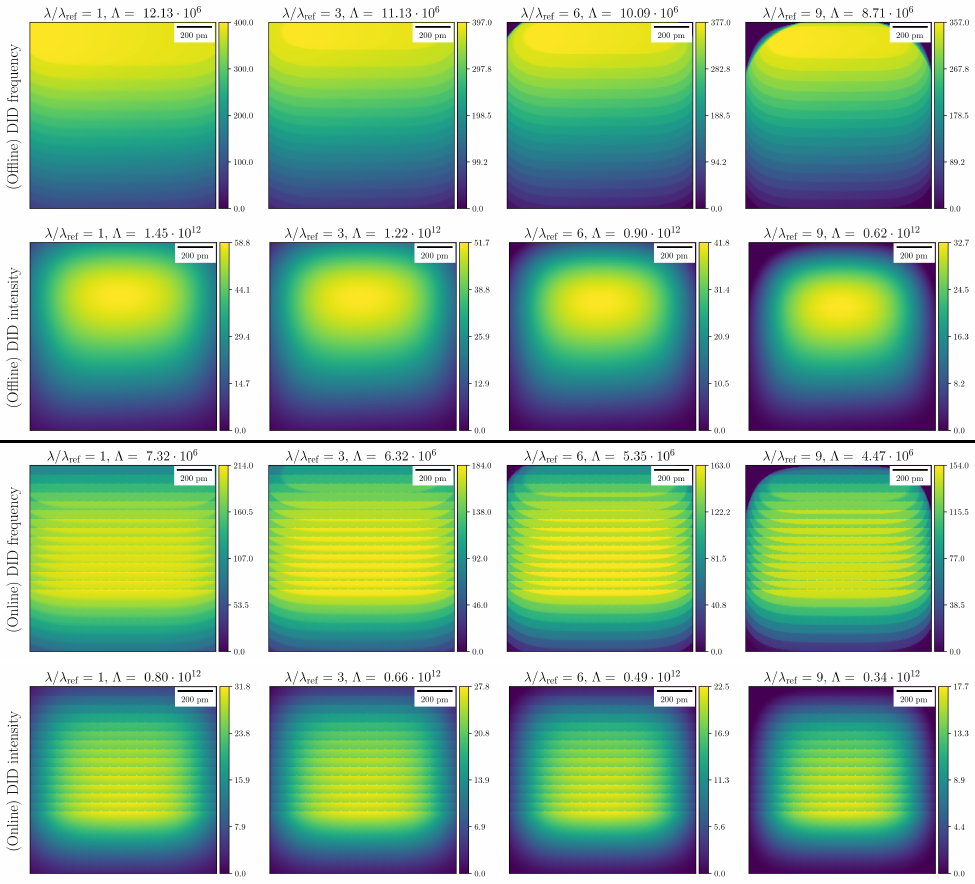}
    \caption{\textbf{DID profiles $\Lambda(\bs r;\lambda)$} with offline (top) and online (bottom) observations for different damage thresholds. The units of color bars and $\Lambda$ for DID are $\diffunitnm$, with color bars scaled by a factor $10^6$.}
\label{fig:stem_damage}
\end{figure}

Different DID models are illustrated in Fig.~\ref{fig:stem_damage} for the baseline STEM scan. We have chosen the models characterised by the sign and reLU activation functions in Eqs.~\eqref{eq:stem_did_frequency} and \eqref{eq:stem_did_intensity}, respectively. Both offline and online observations of DID with a pupil function given by Eqs.~\eqref{eq:stem_did_offline} and \eqref{eq:stem_did_online}, respectively, are considered. For the former we set $r_p = 3\cdot \Delta_p$. We define $\lambda_{\rm ref} \coloneqq A^{\max}_{\rm bdd} = \frac{Q_0\rho}{2}\ln(1+2\rho^{-1}\tau)$  \ie $\lambda_{\rm ref} =10^4~\diffunitnm$ in this example to be the reference quantity for the DID threshold. We subsequently simulated the DID profiles for different DID thresholds $\lambda$, such that $\lambda/\lambda_{\rm ref} \in \{1,3,6,9\}$. 

Fig.~\ref{fig:stem_damage} shows that the choice of both activation and pupil functions significantly changes the shape of the DID profile which emphasises the importance of accurate modeling of damage mechanisms. Offline DID would be observed if the sample was re-scanned and that damage is exacerbated in the upper part of the FoV, since this region has experienced more DID during the full acquisition. In contrast, online DID is observed during the scan and since damage will be recorded only at locations that will be scanned later on, the lower part of the FoV shows higher DID quantities. Together, these results imply that even if DID is not observed during a scan, it may have occurred at previously irradiated locations in the sample. The discontinuity observed in the online DID maps is the effect of the pupil function.
%%%%%%%%%%%%%%%%%%%%%%%%%%%%%
%%%%%%%%%%%%%%%%%%%%%%%%%%%%%%%%%%%
\subsection{Impact of a blanking time on diffusion distribution}\label{subsec:simulations_beamblanker}
As discussed in Section ~\ref{subsec:simulations_baseline}, decreasing the dwell time has a substantial impact in reducing the PM-CDD and DID, but at a potential cost of increased scan distortions. One way to mitigate DID is by allowing the diffusion process to dissipate (or equivalently, allowing the sample to relax) between scan positions. This can be achieved by blanking and unblanking the electron beam \cite{beche2016development, peters2023new} at every probe positions with a specific duty cycle, which can also reduce the scan distortions. Importantly, using this approach the number of deposited electrons in the system and hence, the signal-to-noise-ratio, remains unchanged. 

\begin{figure}[!tb]
    \centering
    \includegraphics[width=1\columnwidth]{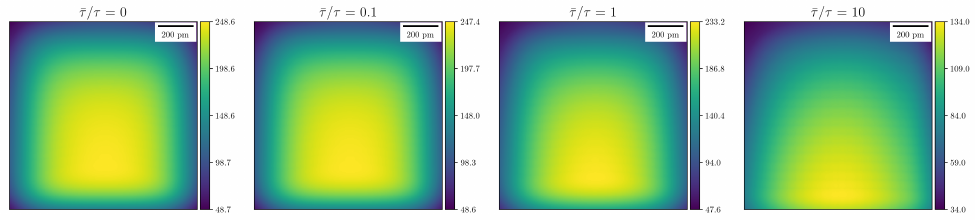}
    \caption{\textbf{STEM scans with identical total number of diffusing species but different blanking times.} PM-CDDs $\chi(\bs r)$ given by Eq.~\eqref{eq:stem_mp_cdd} are shown. Increasing the blanking time reduces the PM-CDD. The total number of diffusing units in all four systems is $Q_{\rm stem} = 253.83~{\rm Ku}$. The units of color bars are $10^3~\diffunitnm$.}
\label{fig:fig_full_stem_settling_time}
\end{figure}

Our framework can accommodate beam blanking through consideration of the settling time in Eq.~\eqref{eq:stem_cumulative}. To further explore this we let $\bar{\tau}$ be a constant blanking time for every probe position, \ie in Eq.~\eqref{eq:stem_cumulative} we set $\bar{\tau} = t_j - (t_{j-1} + \tau_{j-1})$ for all $j \in \{2,\cdots,N\}$. Figure ~\ref{fig:fig_full_stem_settling_time} shows the PM-CDD maps for STEM scans with different ratios of blanking to dwell time $\bar{\tau}/\tau \in \{0, 0.1, 1, 10\}$. We note that $\bar{\tau}/\tau = 0$ is baseline STEM. As expected, increasing the blanking time reduces the PM-CDD, for a constant total number of diffusing units.

\subsection{Impact of the scan trajectory on the diffusion distribution}
\begin{figure}[!t]
        \centering
\includegraphics[width=1\columnwidth]{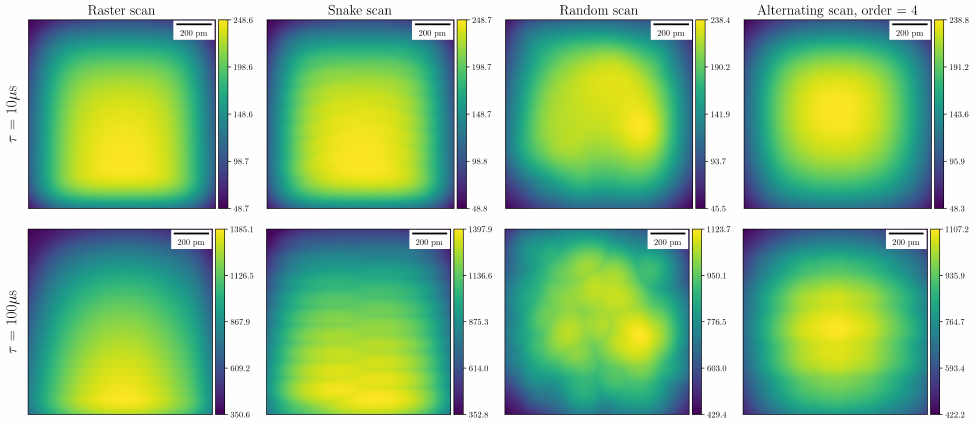}
    \caption{\textbf{PM-CDD $\chi(\bs r)$ for different scan trajectories.} Longer dwell times result in higher PM-CDD for all four scan trajectories. Random and alternating scans show a reduced PM-CDD compared to a raster scan, while a snake scan increases the PM-CDD. The impact of scan trajectory is more evident for longer dwell times. The units of color bars are $10^3~\diffunitnm$.}
\label{fig:full_stem_example_path}
\end{figure}
\begin{figure}[!ht]
    \centering
\includegraphics[width=1\columnwidth]{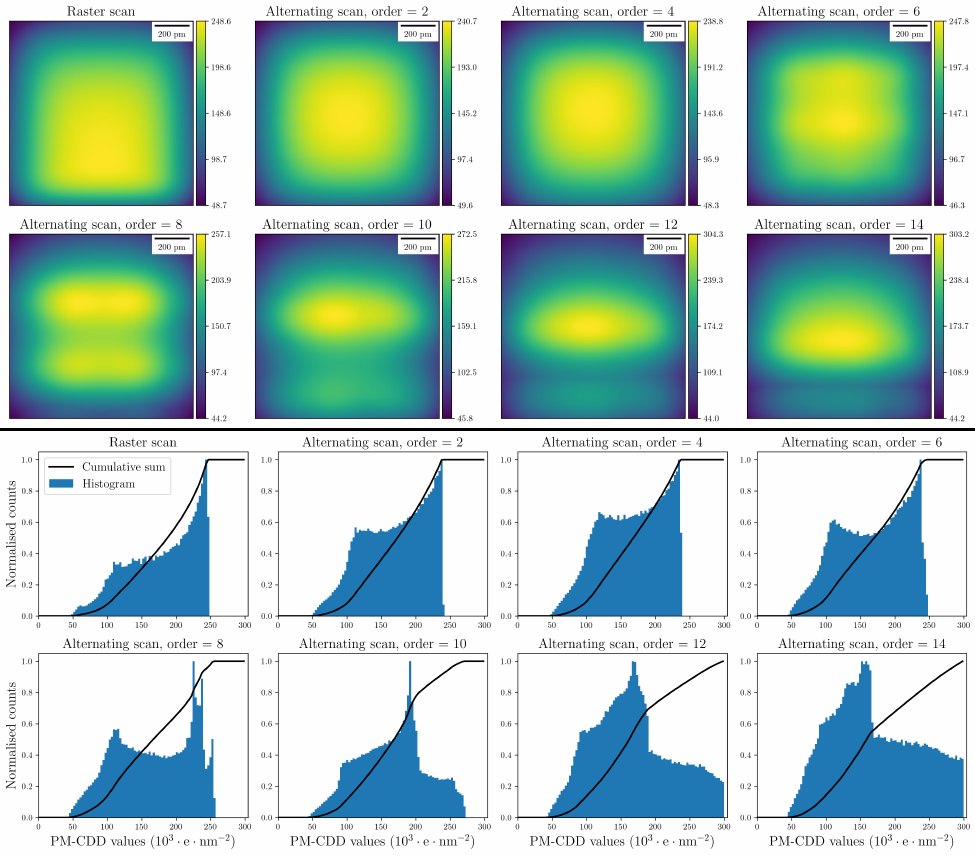}
    \caption{\textbf{PM-CDD $\chi(\bs r)$ for different orders of alternating scans.} Spatial distributions (top) and corresponding histograms (bottom) are plotted. The units of color bars are $10^3~\diffunitnm$.}
\label{fig:full_stem_example_alternating}
\end{figure}
The discussion in Section~\ref{sec:intro} showed that changing the sequence of scanned probe positions ~\cite{interleaveSTEM2022} is a useful approach for damage reduction.

To support this hypothesis we have simulated the PM-CDD for different scanning probe trajectories The values of $Q_0, D$, $ D_s$, and $\Delta_p$ are identical to those in Table ~\ref{tab:baseline_parameters}.

Fig.~\ref{fig:full_stem_example_path} compares the PM-CDD for raster, snake, random, and alternating scans as defined in Section ~\ref{subsec:diffusion_in_stem} for two values of dwell time $\tau \in \{10, 100\}\,\mu s$. As expected, the scan trajectory has a significant impact on the spatial distribution of the PM-CDD; random and alternating scans also show an overall reduction in the PM-CDD, \ie less DID, whereas in contrast the snake scan increases the DID.
\begin{figure}[tbh]
    \centering
\includegraphics[width=0.5\columnwidth]{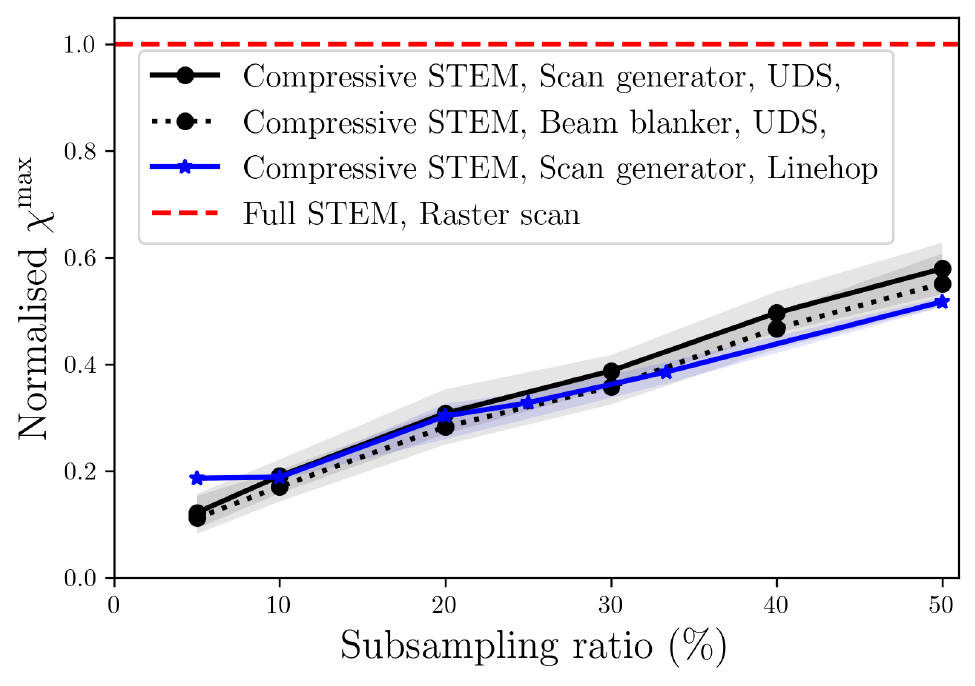}
    \caption{\textbf{Normalised GM-CDDs  $\chi^{\rm max}$ in compressive scans.} Subsampling probe positions using either a linehop or a UDS strategy reduces the GM-CDD. This reduction is not proportional to the subsampling ratio. The use of a beam blanker can further reduce the GM-CDD. A Linehop strategy shows a reduction of the GM-CDD for sampling ratios greater than $10\%$. Example maps taken from this figure are shown in Fig.~\ref{fig:cs_stem_maps}.}
\label{fig:cs_stem_curve}
\end{figure}

In Fig.~\ref{fig:full_stem_example_alternating}, alternating scans with different orders $\kappa \in \{1,2,4,6,8,10,12,14\}$ and with a dwell time $\tau = 10~\mu $s are shown. An alternating scan of order one is equivalent to a conventional raster scan. However, the PM-CDD $\chi(\bs r)$ is a non-monotonic function of alternating scan order $\kappa$ and hence, changing the order of an alternating scan does not necessarily reduce the DID. Fig.~\ref{fig:full_stem_example_alternating} shows the corresponding normalised histogram of the PM-CDD, where the shape of the histogram and the cumulative sum of the histogram is affected by the order of the alternating scan. Lower order alternating scans, \eg when $\kappa \in \{1,2,4\}$, have a sharp peak in their histograms, which indicates that a substantial portion of sample will be damaged as the PM-CDD crosses the DID threshold. However, for higher order alternating scans the histograms are more uniform. In these cases if a DID occurs it will affect a smaller portion of the sample. Accordingly the slope of the cumulative sum curve can be considered as an indicator of the risk of DID. For example, the slope of the cumulative sum, around $248.62~{\rm K}\diffunitnm$, is sharp for a raster scan, but is reduced by increasing the order of an alternating scan. Overall, in comparison to a raster scan, an alternating scan with an arbitrary order does not necessarily reduce the maximum value of PM-CDD, but can result in more uniform values of this quantity.

\begin{figure}[tb!h]
    \centering
\includegraphics[width=1\columnwidth]{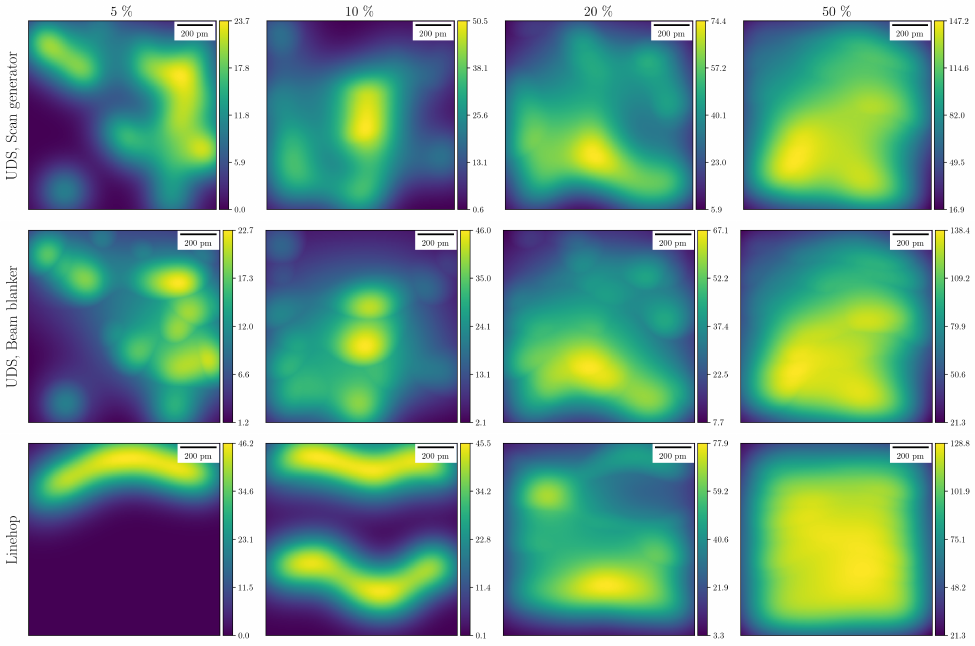}
    \caption{\textbf{ PM-CDDs $\chi(\bs r)$ in compressive STEM scans for different sampling ratios.} These are example maps taken from the simulations in Fig.~\ref{fig:cs_stem_curve}. The same subsampling mask of electron probe positions is used for both UDS strategies, whereas the use of a beam blanker yields lower GM-CDD value. The units of color bars are $10^3~\diffunitnm$.}
\label{fig:cs_stem_maps}
\end{figure}

\subsection{Diffusion distribution in compressive STEM}\label{subsec:simulations_cs_stem} 

Compressive STEM is a method for reducing the total fluence deposited over a region of interest without reducing the signal-to-noise ratio per measurement. In this section we simulate the PM-CDD for compressive STEM following Eq.~\eqref{eq:stem_cumulative_compressive_stem}, with two subsampling strategies; linehop~\cite{nicholls2022compressive} and Uniform Density Sampling (UDS), (subsampling probe positions uniformly at random). For the UDS strategy we consider subsampling of the electron probe positions with either a beam blanker or a scan generator, as described in Section ~\ref{sec:intro}. However, we simulate the linehop strategy using only a scan generator, since, for zero settling time, performing a linehop scan with a beam blanker and with a scan generator results in an identical CDD. Other parameters used are the same as those in Table ~\ref{tab:baseline_parameters}.

Our choice to study both a scan generator and beam blanker is motivated as these are the experimental, practical options for subsampling in STEM.  However, these two regimes exhibit two distinct timings. Using a beam blanker, time passes while the blanked beam is not scanning. In contrast using a scan generator, the beam is moved immediately and each exposure is followed by another exposure. In both cases considered here, the sample is exposed to the same number of total electrons and the acquisition is considered to be perfect in that there is no delay between the end of a sampling period and the activation of the beam blanker and that there is no hysteresis moving the probe from
one location to another.

Fig.~\ref{fig:cs_stem_curve} compares the GM-CDD  defined in Eq.~\eqref{eq:gm-cdd} for a raster scan with two compressive STEM scans, for multiple subsampling ratios. We consider $M/N = \{5, 10, 20, 30, 40, 50\}\%$ for the UDS strategy and $M/N = \{5,10, 20, 25, 33.3, 50\}\%$ for the linehop strategy. All compressive scans yield lower GM-CDD value compared to those of the raster scan. The GM-CDD of the UDS strategy is further reduced by the use of a beam blanker. An important observation in Fig.~\ref{fig:cs_stem_curve} is that the ratio between the GM-CDD for full STEM and compressive STEM is not equal to the subsampling ratio. This indicates that compressive STEM with $M/N (\%)$  probes does not reduce the GM-CDD of the full STEM by the same $M/N$ factor. As an example, a 10\% subsampling of the probes reduces the GM-CDD by only a factor of $\approx 5$. Example CDDs of the experiments in Fig.~\ref{fig:cs_stem_curve} are shown in Fig.~\ref{fig:cs_stem_maps}.

%%%%%%%%%%%%%%%%%%%%%%%%%%%%%%%%%%
\begin{figure}[!tbh]
\begin{minipage}{\linewidth}
    \centering
\includegraphics[width=0.5\columnwidth]{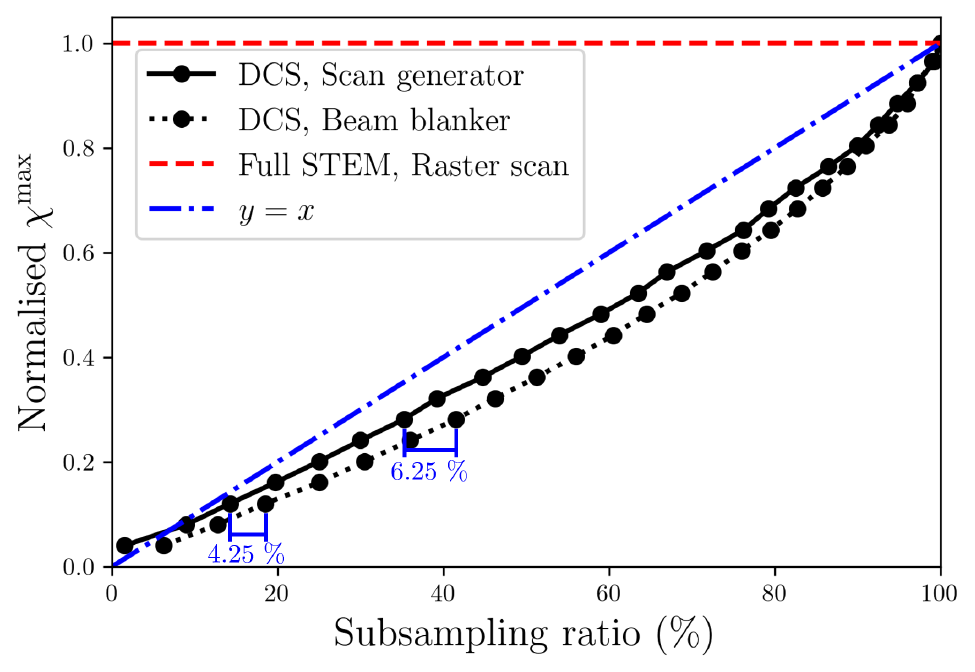}
    \caption{\textbf{Normalised GM-CDD values for \sss masks with respect to the damage threshold}. The use of a beam blanker improves subsampling efficiency in this experiment by 4-7\% for $\lambda/\lambda_{\rm ref}<10$, or equivalently, for subsampling ratios less than 50\%. Example maps taken from this figure are shown in Fig.~\ref{fig:designed_mask_maps}.}
\label{fig:designed_mask_curves}
\end{minipage}
\begin{minipage}{\linewidth}
    \centering
\includegraphics[width=1\columnwidth]{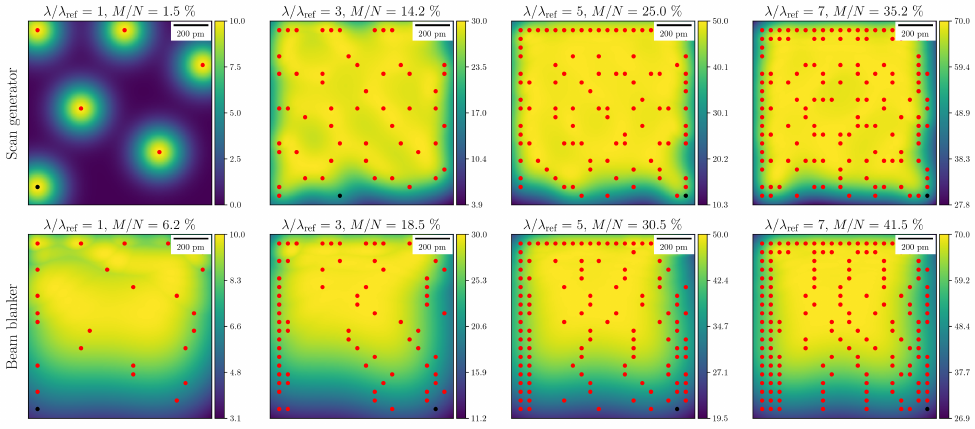}
    \caption{\textbf{PM-CDD $\chi(\bs r)$ from two \sss masks with different sampling ratios.} These are example maps taken from the simulations in Fig.~\ref{fig:designed_mask_curves}. The units of color bars are $10^3~\diffunitnm$.}
\label{fig:designed_mask_maps}
\end{minipage}
\end{figure}
\subsection{Designing \sss masks for DID-free STEM} 

\label{subsec:simulations_mask_design} 
In this section we use the previous results to design \sss subsampling strategies for compressive STEM applications. The STEM parameters in Table ~\ref{tab:baseline_parameters} are used and we consider two compressive STEM scans using either a scan generator or a beam blanker. 

As described in Section ~ \ref{subsec:mask_design}, the compressive framework allows the design of a subsampling mask of probe positions, ensuring that DID is eliminated. To design such a subsampling mask we follow the rule defined by Eq.~\eqref{eq:stem_did_free_csstem}; a probe location is selected, only if the corresponding PM-CDD is less than the DID threshold at every spatial location. Taking $\lambda_{\rm ref} =10^4~\diffunitnm$ from Section ~\ref{subsec:simulations_baseline} as the reference quantity for DID threshold, we have designed \sss masks for different DID thresholds $\lambda$, such that $\lambda/\lambda_{\rm ref} \in \{1,2,\cdots,25\}$. The resultant sampling ratio $M/N$ has been measured after each trial.

Fig.~\ref{fig:designed_mask_curves} shows the GM-CDD, which is constrained by the DID threshold ratio $\lambda/\lambda_{\rm ref}$, as a function of the subsampling ratio in the designed masks. Comparison of the curves in Fig.~\ref{fig:designed_mask_curves} with the $y=x$ line highlights the efficiency of the DCS strategy and shows that the reduction in relative GM-CDD is greater than the subsampling ratio. Moreover, we observe that for a DID threshold $\lambda$ equivalent to a horizontal line in Fig.~\ref{fig:designed_mask_curves}, the use of a beam blanker improves the efficiency of the sampling strategy by 4-7\% for $\lambda/\lambda_{\rm ref}<10$. This arises because, in the simplified  models used here a beam blanker allows the system to "cool" or relax between successive subsampled probe positions, while a scan generator instantaneously moves the electron probe to the next sampled location. Furthermore, when comparing the two masks, we note that the acquisition time for compressive STEM with a beam blanker and scan generator is, respectively, $N\tau$ and $M\tau$ seconds.

\section{Conclusions}
We have described a mathematical framework for understanding diffusion distribution in STEM scans. We have derived the first closed-form formulation of the diffusion distribution for a Gaussian-shaped electron source, used as a proxy to the airy disc function. Our analyses can be directly applied to 4-D STEM scans and can be used in various STEM acquisition modalities. 

Despite significantly improving both the simulation speed and accuracy of the diffusion process in STEM compared to previous work ~\cite{nicholls2020minimising,jannis2022reducingpart2}, numerical simulation within the framework described is limited by the size of the simulation space in turn determined by the size of the electron probe positions grid, and the dimensions of the temporal and spatial grids. Parallel programming or GPU implementation of the simulations can significantly boost the time performance given suitable storage requirements.

In Section ~\ref{sec:diffusion-in-stem}, we limited our analysis to the role of the diffusion coefficient in the asymptotic regime. A more in-depth study, \eg based on the first and second derivatives of both the diffusion distribution and the CDD, will the subject of future research.  
We have also limited our discussion in Section ~\ref{subsec:stem_single_probe} to studies of the first and second derivatives of a single electron probe diffusion distribution without considering the dynamics of the CDD and PM-CDD.

Our proposed damage models have yet to be validated using experimental data for which one of the required modifications to our model would be to support a secondary damage mechanism and possibly a healing effect. Moreover, we postulate that the units of the proposed damage model, \ie $\diffunitnm$, need to be converted to give a physically tractable unit of damage.

An important extension of this work would be to develop a framework for diffusion in a 3-D
medium, as for example represented by thick samples. This simply requires a derivation of the counterpart of Eq.~\eqref{eq:stem_single_probe} from Eq.~\ref{eq:general-pde-constant-d-solution} in 3-D.

Recent work into the effect of scan coil dynamics has shown that, when moving from traditional raster scanning, the behaviour of the scan coils can be difficult to predict \cite{nicholls2023scan}, particularly over large distances. It has been assumed in this work that the beam moves instantaneously and perfectly between scan positions, which can differ from the real case where scan coil hysteresis is present. Incorporating a scan dynamics model into the diffusion model presented in this work would allow a more detailed understanding of the usage of a scan generator and beam blanker for performing subsampling.
%%%%%% End of Discussion %%%%%%
\section*{Acknowledgments}                        
\label{sec:acknowledgment}
We thank Nafiseh Janatian and Mohsen Ghasimi for their help with the proofs. Amirafshar Moshtaghpour, Abner Velazco-Torrejon, and Angus I. Kirkland acknowledge the support of the Rosalind Franklin Institute, funded by the UK Research and Innovation, Engineering and Physical Sciences Research Council. Daniel Nicholls, Alex~W. Robinson, and Nigel D. Browning are partly funded by SenseAI Innovations Limited.

\bibliographystyle{IEEEtran} 
\bibliography{Main}

\input{Supplementary}

\end{document}

%% file: packages.tex
\usepackage{graphicx}
\usepackage[mode=buildnew]{standalone}
\usepackage{comment}
\graphicspath{{Figures/}}
\usepackage[sort&compress, numbers]{natbib}
\usepackage{amsmath}
\usepackage{array}
\usepackage{url}
\usepackage{amssymb}
\usepackage{mathtools}
\usepackage{amsthm}
\usepackage{tabu}
\usepackage{xcolor}
\usepackage{colortbl}
\usepackage{fullpage}
\usepackage[footnotesize]{caption}
\usepackage[caption=false,font=footnotesize]{subfig}
\usepackage{multirow}
\usepackage{booktabs}
\usepackage{hhline}
\usepackage{soul}
\usepackage{stmaryrd} \SetSymbolFont{stmry}{bold}{U}{stmry}{m}{n}
\usepackage{bm} 
\usepackage{bold-extra}
\usepackage{tikz}
\usetikzlibrary{arrows}

\usepackage{pgfplots}
\pgfplotsset{compat=1.18}
\usepackage{color}
\usepackage{xr-hyper}
\usepackage{hyperref}
\makeatletter
\newcommand*{\addFileDependency}[1]{% argument=file name and extension
  \typeout{(#1)}
  \@addtofilelist{#1}
  \IfFileExists{#1}{}{\typeout{No file #1.}}
}
\makeatother

\newtheorem{theorem}{Theorem}
\newtheorem{lemma}{Lemma}

\newtheorem{corollary}{Corollary}

\newtheorem{remark}{Remark}

\newcommand{\inv}[1]{\frac{1}{#1}}
\newcommand{\tinv}[1]{{\textstyle\frac{1}{#1}}}
\newcommand{\ud}{\mathrm{d}}
\newcommand{\eint}[2]{E_1\big(\frac{#1}{#2}\big)}

\newcommand{\bs}{\boldsymbol}
\newcommand{\bb}{\mathbb}
\newcommand{\cl}{\mathcal}

\newcommand{\ts}{\textstyle}
\newcommand{\ie}{\emph{i.e.}, }
\newcommand{\eg}{\emph{e.g.}, }
\newcommand{\etal}{\emph{et al.}}

\newcommand{\iid}{%
    \ifmmode% math mode
        \mathrm{iid}%
    \else%
        iid\xspace%
    \fi%
}

\newcommand{\sss}{DCS~}
\newcommand{\qunit}{{\rm u} \cdot {\rm s}^{-1}}
\newcommand{\diffunitnm}{{\rm u} \cdot {\rm nm}^{-2}}
\newcommand{\diffunit}{{\rm u} \cdot {\rm m}^{-2}}

%% file: Supplementary.tex
% \title{Supplementary Information: Diffusion Distribution for Damage Mitigation
% in Scanning Transmission Electron Microscopy}
% \author{
% Amirafshar~Moshtaghpour\footnotemark[1]\footnote{Corresponding author. Email:  amirafshar.moshtaghpour@rfi.ac.uk}
% \and 
% Abner~Velazco-Torrejon
% \and
% Daniel~Nicholls
% \and
% Alex~W.~Robinson
% \and
% Angus~I.~Kirkland
% \and
% Nigel~D.~Browning}
% \markboth{}%
% {}
% \date{}
% \maketitle
%%%%%% End of Title %%%%%%
\section{Supplementary Information}
% \tableofcontents
%%%%%% Start of Section 1 %%%%%%
\subsection{Derivation of Eq.~\texorpdfstring{\eqref{eq:general-pde-constant-d-solution-instantaneous-point}}{(3)}}
\label{subsec:developing-instantaneous-point}
     
We note that, firstly, that if $\phi(\bs r,t)$ is a solution to the PDE in Eq.~\eqref{eq:general-pde-constant-d} with an initial condition $\phi(\bs r,0) = \delta_{\bs 0}$, then $\phi(\bs r-\bs r_0,t-t_0)$ will be a solution to Eq.~\eqref{eq:general-pde-constant-d} with an initial condition $\phi(\bs r,t_0) = \delta_{\bs r_0}$. Therefore, we first solve Eq.~\eqref{eq:general-pde-constant-d} with an initial condition $\phi(\bs r,0) = \delta_{\bs 0}$. Secondly, it is evident from Eq.~\eqref{eq:general-pde-constant-d} that the diffusion distribution along the the $l$-th direction depends only on the diffusion distribution in the same $l$-th direction. Hence, the $d$-dimensional diffusion distribution can be decoupled into $d$ single dimensional terms, following the form
\begin{equation}\label{eq:aux-proof-instantaneous-point-0}
\phi(\bs r,t) = \prod_{l=1}^d \phi_l(r_l,t).
\end{equation}
Inserting Eq.~\eqref{eq:aux-proof-instantaneous-point-0} in Eq.~\eqref{eq:general-pde-constant-d},
\begin{equation*}
\sum_{l=1}^d \frac{\partial \phi_l(r_l, t)}{\partial t} \prod_{j\ne l} \phi_j(r_j,t) = \sum_{l=1}^d D_l \frac{\partial^2 \phi_l(r_l, t)}{\partial t^2} \prod_{j\ne l} \phi_j(r_j,t),
\end{equation*}
or equivalently,
\begin{equation*}
\sum_{l=1}^d \prod_{j\ne l} \phi_j(r_j,t) \left(\frac{\partial \phi_l(r_l, t)}{\partial t} - D_l \frac{\partial^2 \phi_l(r_l, t)}{\partial t^2} \right)=0,
\end{equation*}
whose trivial solution is $\phi_l(r_l,t) = 0$ for all $l$, while the non-trivial solution should satisfy
\begin{equation}\label{eq:aux-proof-instantaneous-point-00}
\frac{\partial \phi_l(r_l, t)}{\partial t} = D_l \frac{\partial^2 \phi_l(r_l, t)}{\partial t^2}, \quad \forall l.
\end{equation}
We note that if $\phi_l(r_l,t)$ is a solution to an individual PDE in Eq.~\eqref{eq:aux-proof-instantaneous-point-00},  $\phi_l(a r_l,a^2t)$ is also a solution for any $a\ne0$. To avoid that scaling ambiguity, let $z \coloneqq r_l^2/t$ and let $\psi(z) \coloneqq \phi_l(r_l,t)$ be a candidate solution to Eq.~\eqref{eq:aux-proof-instantaneous-point-00}. Hence, by noting $\partial z/\partial r_l = 2r_l/t$, we have
\begin{align}
\frac{\partial \phi_l(r_l, t)}{\partial t} &= \psi'\left(z\right) \cdot \frac{\partial z}{\partial t} = -\psi'\left(z\right) \cdot \frac{z}{t},\label{eq:aux-proof-instantaneous-point-1}\\
\frac{\partial \phi_l(r_l, t)}{\partial r_l} &= \psi'\left(z\right) \cdot \frac{\partial z}{\partial r_l} = \psi'\left(z\right) \cdot \frac{2r_l}{ t},\nonumber\\
\frac{\partial^2 \phi_l(r_l, t)}{\partial r_l^2} &= \psi''\left(z\right) \cdot \frac{4r_l^2}{ t^2} +  \psi'\left(z\right) \cdot \frac{2}{ t}.\label{eq:aux-proof-instantaneous-point-2}
\end{align}
Substituting Eqs.~\eqref{eq:aux-proof-instantaneous-point-1} and \eqref{eq:aux-proof-instantaneous-point-2} in Eq.~\eqref{eq:aux-proof-instantaneous-point-00} yields the following first order homogeneous linear differential equation:
\begin{equation}\label{eq:aux-proof-instantaneous-point-3}
\psi''(z) + \frac{z+2D_l}{4D_l z} \psi'(z) = 0, \quad{\rm or}\quad \frac{\psi''(z)}{\psi'(z)} = -\frac{z+2D_l}{4D_l z}.
\end{equation}
The solution to Eq.~\eqref{eq:aux-proof-instantaneous-point-3} is
\begin{equation*}
\psi(z) = c_1 \int_0^z q^{-\frac{1}{2}} e^{-\frac{q}{4D_l}} \ud q + c_2,
\end{equation*}
for constants $c_1, c_2 \in \bb{R}$.
However, this solution is still not a specific solution to Eq.~\eqref{eq:aux-proof-instantaneous-point-00}. Since Eq.~\eqref{eq:aux-proof-instantaneous-point-00} is linear, if $\phi_l$ is a solution, so is its partial derivatives with respect to $r_l$, \ie
\begin{align*}
\phi_l(r_l,t) &= \frac{\partial \psi(z)}{\partial r_l} = c_1 \frac{\partial z}{\partial r_l} z^{-\frac{1}{2}} e^{-\frac{z}{4D_l}}\\
& = \frac{2c_1}{\sqrt{t}} \cdot \exp{\left(-\frac{r_l^2}{4D_l t}\right)}.
\end{align*}
Finally, we set $c_1 = \bar{Q}_0/4\sqrt{\pi D}$ so that $\int_\bb{R} \phi_l(r_l,t) \ud r_l = \bar{Q}_0$. Therefore, the solution to all PDE terms in Eq.~\eqref{eq:aux-proof-instantaneous-point-00} becomes
\begin{equation}\label{eq:aux-proof-instantaneous-point-4}
\phi_l(r_l,t) =
\frac{\bar{Q}_0}{\sqrt{4\pi D_l t}} \cdot \exp{\left(-\frac{r_l^2}{4D_l t}\right)}.
\end{equation}
By substituting Eq.~\eqref{eq:aux-proof-instantaneous-point-4} in Eq.~\eqref{eq:aux-proof-instantaneous-point-0} and simplification of the formula, we get
\begin{equation}\label{eq:aux-proof-instantaneous-point-5}
\phi(\bs r,t) =
\frac{Q_0}{\sqrt{|4\pi\bs D| t^d}} \cdot \exp{\left(-\frac{\bs r^\top \bs D^{-1} \bs r}{4 t}\right)},
\end{equation}
where $Q_0 = \bar{Q}_0^d$. Recalling that Eq.~\eqref{eq:aux-proof-instantaneous-point-5} is a solution to Eq.~\eqref{eq:general-pde-constant-d} with an initial condition $\phi(\bs r,0) = \delta_{\bs 0}$, the desired solution is $\phi(\bs r - \bs r_0,t-t_0)$.
%%%%%% End of Section 1 %%%%%%
%%%%%% Start of Section 2 %%%%%%
\subsection{Derivation of Eq.~\texorpdfstring{\eqref{eq:general-pde-constant-d-solution-continued-point}}{(6)}}
\label{subsec:developing-continued-point}
Inserting Eq.~\eqref{eq:source-function-continued-point} in Eq.~\eqref{eq:general-pde-constant-d-solution} and setting $d = 2$,  we have
\begin{equation*}
    \phi(\bs r,t) = \frac{Q_0}{4\pi\sqrt{|\bs D|}}\int_{t_0}^{\min (t,t_0+\tau)} 
    \frac{1}{t-t'}
    \exp\Big(-\frac{(\bs r - \bs r_0)^\top \bs D^{-1} (\bs r-\bs r_0)}{4(t-t')}\Big) \ud t'.
\end{equation*}
An application of Lemma~\ref{lem:e1} below, \ie setting $a = 0$, $b = 1$, and $r = \frac{(\bs r - \bs r_0)^\top \bs D^{-1} (\bs r-\bs r_0)}{4}$, and noting $E_1(+\infty) = 0$ yields
\begin{equation*}
    \phi(\bs r,t) = 
    \begin{cases}
        \frac{Q_0}{4\pi\sqrt{|\bs D|}}\eint{(\bs r - \bs r_0)^\top \bs D^{-1} (\bs r-\bs r_0)}{4(t-t_0)},
        & t_0\le t \le t_0+\tau,\\
        \frac{Q_0}{4\pi\sqrt{|\bs D|}}\Big(\eint{(\bs r - \bs r_0)^\top \bs D^{-1} (\bs r-\bs r_0)}{4(t-t_0)} - \eint{(\bs r - \bs r_0)^\top \bs D^{-1} (\bs r-\bs r_0)}{4(t-(t_0+\tau))}\Big),
        & t > t_0+\tau,
    \end{cases}
\end{equation*}
for $\bs r \ne \bs r_0$ and 
\begin{equation*}
    \phi(\bs r_0,t) = 
    \begin{cases}
        +\infty,
        & t_0\le t \le t_0+\tau,\\
        \frac{Q_0}{4\pi \sqrt{|\bs D|}}\ln{\big(\frac{t-t_0}{t-(t_0+\tau)}\big)},
        & t > t_0+\tau.
    \end{cases}
\end{equation*}

\begin{lemma}
\label{lem:e1}
Let $f(r) \coloneqq \int_{t_1}^{t_2} \inv{a+b(t-t')} e^{-\frac{r}{a+b(t-t')}} \ud t'$ for $r \in \bb R_{\ge 0}$ with $a,b,t \in \bb R_{\ge 0}$ and $t_1 \le t\le t_2$. Then, $f(0) = \frac{1}{b}\ln{\big(\frac{a+b(t-t_1)}{a+b(t-t_2)}\big)}$ and, for $r \ne 0$, 
\begin{equation*}
    f(r) = \frac{1}{b}\Big(\eint{r}{a+b(t-t_1)} - \eint{r}{a+b(t-t_2)}\Big).
\end{equation*}
Moreover, if $a+b(t-t_1) \ne 0$ and $a+b(t-t_2) \ne 0$, $f(r)$ is continuous on $r \in (0,\infty)$ and is right-continuous at $r = 0$.
\end{lemma}
\begin{proof}
For $r = 0$, the integral becomes $\int_{t_1}^{t_2} \inv{a+b(t-t')} \ud t' = \frac{1}{b}\ln{\big(\frac{a+b(t-t_1)}{a+b(t-t_2)}\big)}$. For $r\ne 0$, by a change of variable $u = \frac{r}{a+b(t-t')}$, which gives $\ud t' = \frac{r}{b u^2}\ud u$, giving
\begin{align*}
    \int_{t_1}^{t_2} \inv{a+b(t-t')} e^{-\frac{r}{a+b(t-t')}} \ud t' = &
    \int_{\frac{r}{a+b(t-t_1)}}^{\frac{r}{a+b(t-t_2)}} \inv{b u} e^{-u} \ud u\\
    = & \int_{\frac{r}{a+b(t-t_1)}}^{+\infty} \inv{bu} e^{-u} \ud u
    - \int_{\frac{r}{a+b(t-t_2)}}^{+\infty} \inv{b u} e^{-u} \ud u\\
    = & \frac{1}{b}\Big(\eint{r}{a+b(t-t_1)} - \eint{r}{a+b(t-t_2)}\Big).
\end{align*}
It is straightforward to show that for $a+b(t-t_1) \ne 0$ and $a+b(t-t_2) \ne 0$, $f(r)$ is continuous on $(0,\infty)$. Moreover, using the above equations in reverse order we show that $f(r)$ is right continuous at $r=0$:
\begin{align*}
    \lim_{r\rightarrow 0^{+}} \frac{1}{b}\Big(\eint{r}{a+b(t-t_1)} - \eint{r}{a+b(t-t_2)}\Big) &= \lim_{r\rightarrow 0^{+}}
    \int_{t_1}^{t_2} \inv{a+b(t-t')} e^{-\frac{r}{a+b(t-t')}} \ud t'\\
    &= \frac{1}{b}\ln{\big(\frac{a+b(t-t_1)}{a+b(t-t_2)}\big)}.
\end{align*}
\end{proof}
%%%%%% End of Section 2 %%%%%%
%%%%%% Start of Section 3 %%%%
\subsection{Derivation of Eq.~\texorpdfstring{\eqref{eq:general-pde-constant-d-solution-continued-circular}}{(10)}}
\label{subsec:developing-continued-circular}
Inserting Eq.~\eqref{eq:source-function-continued-circular} in Eq.~\eqref{eq:general-pde-constant-d-solution} and setting $d = 2$ gives
\begin{equation}\label{eq:aux-proof-circular-source-1}
    \phi(\bs r,t) = \int_{t_0}^{\min (t,t_0+\tau)} \frac{Q_0}{4\pi^2 r_s^2 (t-t')\sqrt{|\bs D}|}\, I\,\ud t',
\end{equation}
where 
\begin{equation*}
    I \coloneqq \int_{\|\bs r' - \bs r_0\|_2 \le r_s}
    \exp\Big(-\frac{(\bs r - \bs r')^\top\bs D^{-1}(\bs r - \bs r')}{4(t-t')}\Big) \ud \bs r'.
\end{equation*}
By a change of variable $\bs r'' = \bs r' - \bs r_0$,
\begin{align}
    I \coloneqq &  \int_{\|\bs r''\|_2 \le r_s}
    \exp\Big(-\frac{(\bs r - (\bs r''+\bs r_0))^\top\bs D^{-1}(\bs r - (\bs r''+\bs r_0))}{4(t-t')}\Big) \ud \bs r''\nonumber \\
    = &
    \exp\Big(-\frac{(\bs r - \bs r_0)^\top \bs D^{-1}(\bs r - \bs r_0)}{4(t-t')}\Big)
    \cdot I_1,\label{eq:aux-proof-circular-source-2}
\end{align}
where 
\begin{equation}\label{eq:aux-proof-circular-source-3}
    I_1 \coloneqq \int_{\|\bs r''\|_2 \le r_s}
    \exp\Big(-\frac{\bs r''^\top \bs D^{-1} \bs r''}{4(t-t')}\Big)
    \exp\Big(\frac{(\bs r - \bs r_0)^\top\bs D^{-1} \bs r''}{2(t-t')}\Big) \ud \bs r''.
\end{equation}
To compute $I_1$ it is necessary to convert Cartesian to Polar coordinates by defining $\bs r'' = [r''_1, r''_2]^\top \coloneqq [u\cos{\theta}, u\sin{\theta}]^\top$ and $\bs r - \bs r_0 = [r_1 - r_{0,1}, r_2 - r_{0,2}]^\top \coloneqq [v\cos{\alpha}, v\sin{\alpha}]^\top$; hence 
\begin{align}
    \ud \bs r'' & = \ud r''_1 \,\ud r''_2 = u\, \ud u\, \ud \theta,\label{eq:aux-proof-circular-source-4}
    \\
    \bs r''^\top \bs D^{-1} \bs r'' & = u^2(D_1^{-1}\cos^2{\theta}+D_2^{-1}\sin^2{\theta}) = \frac{u^2}{2}\big((D_1^{-1} + D_2^{-1}) + (D_1^{-1} - D_2^{-1})\cos{2\theta}\big),\label{eq:aux-proof-circular-source-5}
    \\
    (\bs r - \bs r_0)^\top\bs D^{-1} \bs r'' & = u\,v \big(D_1^{-1} \cos{\theta}\cos{\alpha} + D_2^{-1} \sin{\theta}\sin{\alpha}\big) = u \, v\, a \cos{(\theta-\beta)},\label{eq:aux-proof-circular-source-6}
\end{align}
where in Eq.~\eqref{eq:aux-proof-circular-source-6}, $a \coloneqq \ts\sqrt{D_1^{-2}\cos^2{\alpha}+D_2^{-2}\sin^2{\alpha}}$ and $\beta \coloneqq \arctan{(\frac{D_1}{D_2}\tan{\alpha})}$. Substituting Eqs.~\eqref{eq:aux-proof-circular-source-4}, \eqref{eq:aux-proof-circular-source-5}, and \eqref{eq:aux-proof-circular-source-6} in Eq.~\eqref{eq:aux-proof-circular-source-3} yields
\begin{align*}
    I_1 = &
    \int_{0}^{r_s} u
    \exp\Big(-\frac{(D_1^{-1} + D_2^{-1}) u^2}{8(t-t')}\Big)
    \int_{0}^{2\pi}
    \exp\Big(\frac{\inv{2}u^2 (D_1^{-1} - D_2^{-1})\cos{2\theta}+ u\,v\,a\cos{(\theta-\beta)}}{2(t-t')}\Big) \ud \theta\,\ud u,
\end{align*}
which does not simplify further. 
However, by setting $D_1 = D_2=D$ the Integral $I_1$ reduces to
\begin{align}
    I_1 = &
    \int_{0}^{r_s} u
    \exp\Big(-\frac{u^2}{4D(t-t')}\Big)
    \int_{0}^{2\pi}
    \exp\Big(\frac{ u\,v\cos{(\theta-\alpha)}}{2D(t-t')}\Big) \ud \theta\,\ud u,\nonumber\\
    =&
    2\pi\int_{0}^{r_s} u
    \exp\Big(-\frac{u^2}{4D(t-t')}\Big) I_0\big(\frac{ u\,v}{2D(t-t')}\big) \ud u,\label{eq:aux-proof-circular-source-7}
\end{align}
where in the second line we use the fact that 
\begin{align*}
    \int_{0}^{2\pi}
    \exp\Big(\frac{ u\,v\cos{(\theta-\alpha)}}{2D(t-t')}\Big) \ud \theta = & 2 \int_{0}^{\pi}
    \exp\Big(\frac{ u\,v\cos{\theta}}{2D(t-t')}\Big) \ud \theta 
    = 2\pi I_0\big(\frac{ u\,v}{2D(t-t')}\big).
\end{align*}
Substituting Eq.~\eqref{eq:aux-proof-circular-source-7} in Eq.~\eqref{eq:aux-proof-circular-source-2} and combining the result with Eq.~\eqref{eq:aux-proof-circular-source-1} results in
\begin{equation*}
    \scalebox{1}{$\phi(\bs r,t) = \int_{t_0}^{\min (t,t_0+\tau)} \frac{Q_0}{2\pi r_s^2 D(t-t')}\exp\Big(-\frac{\|\bs r - \bs r_0\|_2^2}{4D(t-t')}\Big)
    \int_{0}^{r_s} u
    \exp\Big(-\frac{u^2}{4D(t-t')}\Big) I_0\big(\frac{ u\|\bs r - \bs r_0\|_2}{2D(t-t')}\big) \ud u\, \ud t'.$}
\end{equation*}
%%%%%% End of Section 3 %%%%%%
%%%%%% Start of Section 4 %%%%%%
\subsection{Diffusion distribution for a continuous square disc source}
\label{subsec:continued-square}
In this section we study the diffusion distribution for a continuous square disc source with half-width $r_s$ centered at $\bs r = \bs{r}_0$ and activated during time $t \in [t_0, t_0 +\tau]$. In this case, the corresponding source function is $h(\bs r, t) = Q_0\cdot h_s(\bs r) \cdot h_t(t)$ with
\begin{equation}\label{eq:source-function-continued-square}
    h_s(\bs r) = 
    \begin{cases}
        \frac{1}{4r_s^2}, &{\rm if~} \|\bs r - \bs r_0\|_\infty \le r_s,\\
        0, &{\rm otherwise},
    \end{cases}
    \hspace{1cm}{\rm and} \hspace{1cm}
    h_t(t) = 
    \begin{cases}
        1, &{\rm if~} t_0 \le t \le t_0+\tau,\\
        0, &{\rm otherwise},
    \end{cases}
\end{equation}
with $Q_0$ in $\qunit$. In Eq.~\eqref{eq:source-function-continued-square}, the $\ell_\infty$-norm $\|\bs u\|_\infty \coloneqq \max_i |u_i|$, which returns the maximum absolute value entry of the input vector, is used to model the square disc source. Inserting \eqref{eq:source-function-continued-square} in \eqref{eq:general-pde-constant-d-solution} for $d = 2$ gives (see Section ~\ref{subsec:developing-continued-square} for the details)
\begin{align}
    \phi(\bs r,t) = \frac{Q_0}{16 r_s^2}\int_{t_0}^{\min (t,t_0+\tau)}
    & \Big({\rm erf}\big(\frac{r_1-r_{0,1}+r_s}{\sqrt{4D_1 (t-t')}}\big)-{\rm erf} \big(\frac{r_1-r_{0,1}-r_s}{\sqrt{4D_1 (t-t')}}\big)\Big)\nonumber
    \\
    \cdot & \Big({\rm erf}\big(\frac{r_2-r_{0,2}+r_s}{\sqrt{4D_2 (t-t')}}\big)-{\rm erf} \big(\frac{r_2-r_{0,2}-r_s}{\sqrt{4D_2 (t-t')}}\big)\Big)
    \,\ud t', \label{eq:general-pde-constant-d-solution-continued-square}
\end{align}
where ${\rm erf}(v) \coloneqq \dfrac{2}{\sqrt{\pi}}\int_0^{v} \exp{(-u^2)}\,\ud u$ is the error function. This special case is pertinent to imaging applications using wide (or parallel) beams as in some applications of X-ray ptychography , \eg with coherently illuminated condensed aperture~\cite{batey2014information,claus2019diffraction}.
\begin{figure}[t]
    \centering
    \includegraphics[width=1\columnwidth]{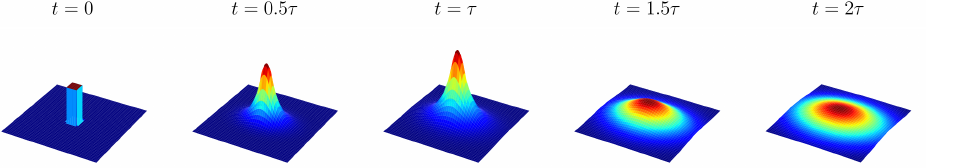}
    \caption{Diffusion distributions for a square disc source following Eq.~\eqref{eq:general-pde-constant-d-solution-continued-square} with $\bs D = {\rm diag}(0.25, 0.5)$, $r_s = 0.2$, $\tau = 1$, $Q_0 = 1$, $t_0 = 0$, and $\bs{r}_0 = \bs 0$. The range of vertical axes are identical.}
    \label{fig:example_special_cases_square}
\end{figure}
\subsubsection{Derivation of Eq.~\texorpdfstring{\eqref{eq:general-pde-constant-d-solution-continued-square}}{(8)}}
\label{subsec:developing-continued-square}
First, we note that 
\begin{equation*}
    \{\bs{r}' : \|\bs{r}'-\bs r_0\|_\infty \le r_s\} = \{\bs r' : |r'_1-r_{0,1}| \le r_s {\rm ~and~} |r'_2-r_{0,2}| \le r_s\}.
\end{equation*}
Inserting Eq.~\eqref{eq:source-function-continued-square} in Eq.~\eqref{eq:general-pde-constant-d-solution}, setting $d = 2$, and since $(\bs r - \bs{r}')^\top \bs D^{-1} (\bs r-\bs r') = \tinv{D_1} (r_1 - r'_1)^2+\tinv{D_2} (r_2 - r'_2)^2$, gives
\begin{equation}\label{eq:aux-proof-square-source-1}
    \phi(\bs r,t) = \frac{Q_0}{16\pi r_s^2 \sqrt{|\bs D|}}\int_{t_0}^{\min (t,t_0+\tau)} I_1 \cdot I_2\,\ud t',
\end{equation}
where 
\begin{equation*}
    I_i \coloneqq \frac{1}{\sqrt{t-t'}}\int_{r_{0,i}-r_s}^{r_{0,i}+r_s}
    \exp\Big(-\frac{(r_i - r'_i)^2}{4D_i(t-t')}\Big) \ud r'_i, {~~\rm for~} i \in \{1,2\}.
\end{equation*}

An application of Lemma~\ref{lem:erf} below, \ie setting $T = 4D_i (t-t')$, $r = r_i$, $r'= r'_i$, $a_1 = r_{0,i}-r_s$, and $a_2 = r_{0,i}+r_s$, yields
\begin{equation}\label{eq:aux-proof-square-source-2}
    I_i = \sqrt{\pi D_i}\Big({\rm erf} \big(\frac{r_i-r_{0,i}+r_s}{\sqrt{4D_i (t-t')}}\big)-{\rm erf} \big(\frac{r_i-r_{0,i}-r_s}{\sqrt{4D_i (t-t')}}\big)\Big). 
\end{equation}
Substituting Eq.~\eqref{eq:aux-proof-square-source-2} in Eq.~\eqref{eq:aux-proof-square-source-1} for $i \in \{1,2\}$ and noting $|\bs D| = D_1 D_2$ gives
\begin{align*}
    \phi(\bs r,t) = \frac{Q_0}{16 r_s^2}\int_{t_0}^{\min (t,t_0+\tau)}
    & \Big({\rm erf}\big(\frac{r_1-r_{0,1}+r_s}{\sqrt{4D_1 (t-t')}}\big)-{\rm erf} \big(\frac{r_1-r_{0,1}-r_s}{\sqrt{4D_1 (t-t')}}\big)\Big)
    \\
    \cdot & \Big({\rm erf}\big(\frac{r_2-r_{0,2}+r_s}{\sqrt{4D_2 (t-t')}}\big)-{\rm erf} \big(\frac{r_2-r_{0,2}-r_s}{\sqrt{4D_2 (t-t')}}\big)\Big)
    \,\ud t'.
\end{align*}
\begin{lemma}
\label{lem:erf}
\begin{equation*}
    \frac{1}{\sqrt{T}}\int_{a_1}^{a_2} e^{-\frac{(r-r')^2}{T}} \ud r' =\frac{\sqrt{\pi}}{2}\Big( {\rm erf}\big(\frac{r-a_1}{\sqrt{T}}\big)-{\rm erf}\big(\frac{r-a_2}{\sqrt{T}}\big)\Big).
\end{equation*}
\end{lemma}
\begin{proof}
A change of a variable $u = \frac{(r-r')}{\sqrt{T}}$, gives $\ud r' = -\sqrt{T}\ud u$, giving
\begin{align*}
    \int_{a_1}^{a_2} e^{-\frac{(r-r')^2}{T}} \ud r' = &
    \sqrt{T}\int_{\frac{r-a_2}{\sqrt{T}}}^{\frac{r-a_1}{\sqrt{T}}}  e^{-u^2} \ud u
    = \frac{\sqrt{\pi T}}{2}\Big({\rm erf}\big(\frac{r-a_1}{\sqrt{T}}\big)-{\rm erf}\big(\frac{r-a_2}{\sqrt{T}}\big)\Big).
\end{align*}

\end{proof}
%%%%%% End of Section 4 %%%%%%
%%%%%% Start of Section 5 %%%%%%
\subsection{Derivation of Eqs.~\texorpdfstring{\eqref{eq:general-pde-constant-d-solution-continued-gaussian}}{(12)} and \texorpdfstring{\eqref{eq:general-pde-constant-d-solution-continued-gaussian_2}}{13}}
\label{subsec:developing-continued-gaussian}
Inserting Eq.~\eqref{eq:source-function-continued-gaussian} in Eq.~\eqref{eq:general-pde-constant-d-solution} and setting $d = 2$ gives
\begin{equation}\label{eq:aux-proof-gaussian-source-1}
    \phi(\bs r,t) = Q_0\int_{t_0}^{\min (t,t_0+\tau)} \int_{\bb R^2} \, f_1(\bs r') \cdot f_2(\bs r') \,\ud \bs r'\,\ud t',
\end{equation}
where 
\begin{align*}
    f_1(\bs r') \coloneqq & \frac{1}{2\pi \sqrt{|\bs D_s}|}
    \exp{\big(-\frac{1}{2} (\bs r'-\bs r_0)^\top \bs D_s^{-1} (\bs r'- \bs r_0)\big)},\\
    f_2(\bs r') \coloneqq & \frac{1}{4\pi (t-t')\sqrt{|\bs D}|}
    \exp{\big(-\frac{(\bs r'-\bs r)^\top \bs D^{-1} (\bs r'- \bs r)}{4(t-t')}\big)}.
\end{align*}
Setting $\bs m_1 = \bs r_0$, $\bs m_2 = \bs r$, $\bm \Sigma_1 = \bs D_s$, $\bm \Sigma_2 = 2(t-t')\bs D$ in Lemma \ref{lem:gaussian} and solving Eq.~\eqref{eq:aux-proof-gaussian-source-1} yields
\begin{equation}\label{eq:aux-proof-gaussian-source-2}
    \phi(\bs r,t) = \int_{t_0}^{\min (t,t_0+\tau)} 
    \frac{Q_0}{2\pi\sqrt{|\bs D_e|}}
    \exp{\big(-\frac{1}{2} (\bs r-\bs r_0)^\top \bs D_e^{-1} (\bs r- \bs r_0)\big)}\,\ud t',
\end{equation}
with $\bs D_e = \bs D_s + 2(t-t')\bs D$. 
Furthermore, assuming isotropic source and diffusion coefficients, \ie $D_{s,1} = D_{s,2} = D_s$ and $D_1 = D_2 = D$, Eq.~\eqref{eq:aux-proof-gaussian-source-2} becomes
\begin{equation}\label{eq:aux-proof-gaussian-source-3}
    \phi(\bs r,t) = \int_{t_0}^{\min (t,t_0+\tau)} 
    \frac{Q_0}{2\pi(D_s + 2D(t-t'))}
    \exp{\big(-\frac{\|\bs r-\bs r_0\|_2^2}{2D_s + 4D(t-t')}\big)}\,\ud t'.
\end{equation}
Using Lemma \ref{lem:e1} and setting $a = 2D_s$, $b = 4D$, and $r = \|\bs r - \bs r_0\|_2^2$ gives
\begin{equation*}
    \phi(\bs r,t) = 
    \begin{cases}
        \frac{Q_0}{4\pi D}\Big(\eint{\|\bs r - \bs r_0\|_2^2}{2D_s + 4D(t-t_0)}-\eint{\|\bs r - \bs r_0\|_2^2}{2D_s}\Big),
        & t_0\le t \le t_0+\tau,\\
        \frac{Q_0}{4\pi D}\Big(\eint{\|\bs r - \bs r_0\|_2^2}{2D_s +4D(t-t_0)} - \eint{\|\bs r - \bs r_0\|_2^2}{2D_s+4D(t-(t_0+\tau))}\Big),
        & t > t_0+\tau,
    \end{cases}
\end{equation*}
and for $\bs r \ne \bs r_0$ and 
\begin{equation*}
        \phi(\bs r_0,t) = 
    \begin{cases}
        \frac{Q_0}{4\pi D}\ln{\big(\frac{D_s+2D(t-t_0)}{D_s}\big)},
        & t_0\le t \le t_0+\tau,\\
        \frac{Q_0}{4\pi D} \ln{\big(\frac{D_s+2D(t-t_0)}{D_s+2D(t-t_0-\tau)}\big)},
        & t > t_0+\tau.
        \end{cases}
\end{equation*}
\begin{lemma}
\label{lem:gaussian}
Let $\cl N({\bs u} | \bs m, \bm \Sigma) \coloneqq \frac{1}{\sqrt{|2\pi \bm \Sigma|}} \exp{\big(-\frac{1}{2} (\bs u - \bs m)^\top \bm \Sigma^{-1} (\bs u - \bs m)\big)}$. Then
\begin{equation*}
    \int_{\bb R^2} \cl N({\bs u} | \bs m_1, \bm \Sigma_1) \cdot \cl N( {\bs u} | \bs m_2, \bm \Sigma_2)\, \ud \bs u =
    \cl N( {\bs m_1} | \bs m_2, \bm \Sigma_c),
\end{equation*}
where $\bm \Sigma_c \coloneqq \bm \Sigma_1 + \bm \Sigma_2$.
\end{lemma}
\begin{proof}
An application of Lemma \ref{lem:gaussian_product} and recalling that $\int_{\bb R^2} \cl N( {\bs u} | \bs m', \bm \Sigma')\, \ud \bs u = 1$, for arbitrary $\bs m'$ and $\bm \Sigma'$, completes the proof.
\end{proof}

\begin{lemma}
\label{lem:gaussian_product}
\textsc{(\cite[p.~41]{petersen2008matrix})} Let $\cl N({\bs u} | \bs m, \bm \Sigma) \coloneqq \frac{1}{\sqrt{|2\pi \bm \Sigma|}} \exp{\big(-\frac{1}{2} (\bs u - \bs m)^\top \bm \Sigma^{-1} (\bs u - \bs m)\big)}$. Then
\begin{equation*}
     \cl N({\bs u} | \bs m_1, \bm \Sigma_1) \cdot \cl N({\bs u} \ \bs m_2, \bm \Sigma_2)= \cl N({\bs m_1}| \bs m_2, \bm \Sigma_c) \cdot \cl N({\bs u}| \bs m_3, \bm \Sigma_3),
\end{equation*}
where
\begin{align*}
    \bm \Sigma_c &\coloneqq \bm \Sigma_1 + \bm \Sigma_2\\
    \bm \Sigma_3 & \coloneqq (\bm \Sigma_1^{-1} + \bm \Sigma_2^{-1})^{-1}\\
    \bs m_3 &\coloneqq \bm \Sigma_3 (\bm \Sigma_1^{-1} \bm m_1 + \bm \Sigma_2^{-1} \bs m_2).
\end{align*}
\end{lemma}
%%%%%% End of Section 5 %%%%%%
%%%%%% Start of Section 6 %%%%%%
\subsection{Diffusion distribution as a function of distance to the activation point} \label{sec:diffusion_function_of_distance}
Let $d_i \coloneqq \|\bs r - \bs r_i\|_2$ be the Euclidean distance from the activation location of the electron beam. We compute the first and second derivatives of $\phi_i^{\rm on}$ in the following Lemma proved in Sec.~\ref{subsec:proof-stem-single-probe-derivatives-d-beamon}.
\begin{lemma}\label{lem:stem_single_probe_derivatives_beamon}
    For $\phi_i^{\rm on}$ defined in Eq.~\eqref{eq:stem_single_probe_beamon} and for $d_i \coloneqq \|\bs r - \bs r_i\|_2 \ne 0$ we have
    \begin{align}
     \frac{\partial}{\partial d_i} \phi_i^{\rm on} &= -\frac{Q_0}{2\pi D d_i}\big(e^{-\frac{d_i^2}{2D_s + 4D(t-t_i)}}-e^{-\frac{d_i^2}{2D_s}}\big),\label{eq:stem-single-probe-first-derivative-d-beamon}\\
    \frac{\partial^2}{\partial d_i^2} \phi_i^{\rm on} &= \frac{Q_0}{2\pi D d_i^2}\Big(\big(1+\frac{2d_i^2}{2D_s+4D(t-t_i)}\big)e^{-\frac{d_i^2}{2D_s + 4D(t-t_i)}}-\big(1+\frac{2d_i^2}{2D_s}\big)e^{-\frac{d_i^2}{2D_s}}\Big).\label{eq:stem-single-probe-second-derivative-d-beamon}
    \end{align}
    Moreover, 
    \begin{itemize}
        \item[(i)] $\frac{\partial}{\partial d_i} \phi_i^{\rm on} <0$, hence, $\phi_i^{\rm on}$ is a decreasing function of $d_i$, for all $d_i \ne 0$,
        \item[(ii)] $\frac{\partial^2}{\partial d_i^2} \phi_i^{\rm on} <0$, hence, $\phi_i^{\rm on}$ is a strictly concave function of $d_i$, if $\frac{d_i^2}{2D_s} <\frac{1}{2}$,
        \item[(iii)] $\frac{\partial^2}{\partial d_i^2} \phi_i^{\rm on} >0$, hence, $\phi_i^{\rm on}$ is a strictly convex function of $d_i$, if $\frac{d_i^2}{2D_s+4D(t-t_i)} >\frac{1}{2}$.
    \end{itemize}
\end{lemma}

Similarly, we can compute the first and second derivatives of $\phi_i^{\rm off}$ using the following Lemma. See Sec.~\ref{subsec:proof-stem-single-probe-derivatives-d-beamoff} for the proof.
\begin{lemma}\label{lem:stem_single_probe_derivatives_beamoff}
    For $\phi_i^{\rm off}$ defined in Eq.~\eqref{eq:stem_single_probe_beamoff} and for $d_i \coloneqq \|\bs r - \bs r_i\|_2 \ne 0$ we have
    \begin{align}
    \frac{\partial}{\partial d_i} \phi_i^{\rm off} &= -\frac{Q_0}{2\pi D d_i}\big(e^{-\frac{d_i^2}{2D_s + 4D(t-t_i)}}-e^{-\frac{d_i^2}{2D_s + 4D(t-t_i-\tau_i)}}\big),\label{eq:stem-single-probe-first-derivative-d-beamoff}\\
    \frac{\partial^2}{\partial d_i^2} \phi_i^{\rm off} &= \frac{Q_0}{2\pi D d_i^2}\Big(\big(1+\frac{2d_i^2}{2D_s+4D(t-t_i)}\big)e^{-\frac{d_i^2}{2D_s + 4D(t-t_i)}}\nonumber\\
    &\hphantom{= \frac{Q_0 D_s}{2D d_i^2}\Big(}-\big(1+\frac{2d_i^2}{2D_s+4D(t-t_i-\tau_i)}\big)e^{-\frac{d_i^2}{2D_s+4D(t-t_i-\tau_i)}}\Big).\label{eq:stem-single-probe-second-derivative-d-beamoff}
    \end{align}
    Moreover, 
    \begin{itemize}
        \item[(i)] $\frac{\partial}{\partial d_i} \phi_i^{\rm off} <0$, hence, $\phi_i^{\rm off}$ is a decreasing function of $d_i$, for all $d_i \ne 0$,
        \item[(ii)] $\frac{\partial^2}{\partial d_i^2} \phi_i^{\rm off} <0$, hence, $\phi_i^{\rm off}$ is a strictly concave function of $d_i$, if $\frac{d_i^2}{2D_s+4D(t-t_i-\tau_i)} <\frac{1}{2}$,
        \item[(iii)] $\frac{\partial^2}{\partial d_i^2} \phi_i^{\rm off} >0$, hence, $\phi_i^{\rm off}$ is a strictly convex function of $d_i$, if $\frac{d_i^2}{2D_s+4D(t-t_i)} >\frac{1}{2}$.
    \end{itemize}
\end{lemma}
%%%%%%%%%%%%%%%%%%%%%%
%%%%%%%%%% Figure 
\begin{figure}[th]
    \centering
    \includegraphics[width=1\columnwidth]{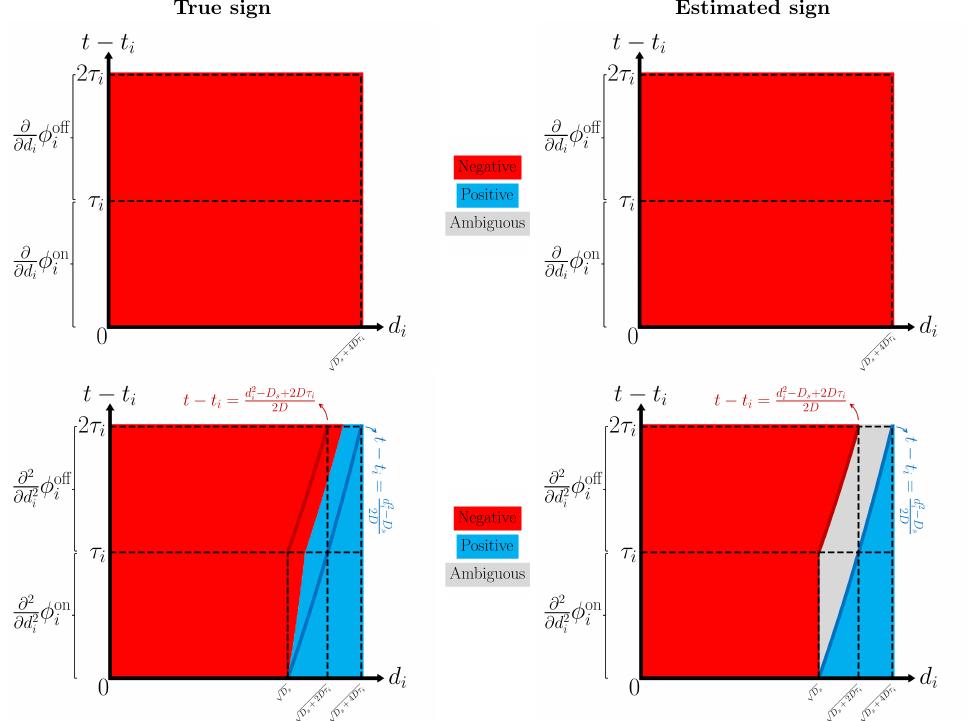}
    \caption{\textbf{Sign of the first (top) and second (bottom) derivative of the diffusion distribution from a single electron STEM probe with respect to distance from the activation point.} (Left) true sign of diffusion distribution  numerically computed from Eqs.~\eqref{eq:stem-single-probe-first-derivative-d-beamon}, \eqref{eq:stem-single-probe-second-derivative-d-beamon}, \eqref{eq:stem-single-probe-first-derivative-d-beamoff}, and \eqref{eq:stem-single-probe-second-derivative-d-beamoff}.
    (Right) Estimated sign following the sufficient conditions given by Cases \textit{(ii)} and \textit{(iii)} of Lemmas \ref{lem:stem_single_probe_derivatives_beamon} and \ref{lem:stem_single_probe_derivatives_beamoff}. The upper plots are shown for visual comparison with the plots in Fig.~\ref{fig:stem-single-beam-derivatives-t}(top)}
    \label{fig:stem-single-beam-derivatives-d}
\end{figure}

%%%%%%%% End of figure 
Lemmas \ref{lem:stem_single_probe_derivatives_beamon} and \ref{lem:stem_single_probe_derivatives_beamoff} state that, regardless of the status of the electron beam, the diffusion distribution is always a decreasing function of the distance to the activation point, hence, the maximum value of the diffusion distribution is always at the activation point. Moreover, cases \textit{(ii)} and \textit{(iii)} in Lemmas \ref{lem:stem_single_probe_derivatives_beamon} and \ref{lem:stem_single_probe_derivatives_beamoff} provide, respectively, sufficient conditions for the diffusion distribution to be a concave or convex function of distance to the activation point. These results provide an understanding of when a diffusion distribution has a global minimum or maximum, as a strictly convex (or concave) function has no more than one minimum (respectively, maximum). For example, when the condition of case \textit{(ii)} in Lemma~\ref{lem:stem_single_probe_derivatives_beamoff} is met, there is a point in space at a distance $d_i$ from the activation point with the maximum number of diffusing substances compared to the other spatial points. The conditions in Lemmas \ref{lem:stem_single_probe_derivatives_beamon} and \ref{lem:stem_single_probe_derivatives_beamoff}, as illustrated in Fig.~\ref{fig:stem-single-beam-derivatives-d} (left), correspond to regions identified by a vertical line and two parabolas. As an example, from the blue region in Fig.~\ref{fig:stem-single-beam-derivatives-d}(bottom), for points $(d_i,t)$ where $2D_s(t-t_i) < d_i^2$, the diffusion distribution is a convex function of $d_i$. However, those lemmas do not cover the points where $D_s < d_i^2 < D_s + 2D(t-t_i)$, \ie an ambiguous region, as highlighted in gray in Fig.~\ref{fig:stem-single-beam-derivatives-d} (right). We note that when $D\tau_i \rightarrow 0$, the area of ambiguous region reduces to zero. Fig.~\ref{fig:stem-single-beam-derivatives-d}(left) illustrates the true sign of the derivatives, numerically computed from Eqs.~\eqref{eq:stem-single-probe-first-derivative-d-beamon} and \eqref{eq:stem-single-probe-second-derivative-d-beamon} for $0< t-t_i < \tau_i$ and from Eqs.~\eqref{eq:stem-single-probe-first-derivative-d-beamoff} and \eqref{eq:stem-single-probe-second-derivative-d-beamoff} for $\tau_i< t-t_i$.

\subsubsection{Proof of Lemma \texorpdfstring{\ref{lem:stem_single_probe_derivatives_beamon}}{(1)}}
\label{subsec:proof-stem-single-probe-derivatives-d-beamon}
An application of Lemma \ref{lem:e1_delta-cos-derivatives} by setting $v \leftarrow 0$, and $u \leftarrow d_i$, for two settings of $\alpha$, \ie $\alpha \leftarrow \frac{1}{2D_s + 4D(t-t_i)}$ and  $\alpha \leftarrow \frac{1}{2D_s}$, gives \eqref{eq:stem-single-probe-first-derivative-d-beamon} and \eqref{eq:stem-single-probe-second-derivative-d-beamon}. Since $2D_s < 2D_s + 4D(t-t_i)$, it is evident that 
\begin{equation*}
     e^{-\frac{d_i^2}{2D_s + 4D(t-t_i)}}>e^{-\frac{d_i^2}{2D_s}},
\end{equation*}
hence, $\frac{\partial}{\partial d_i} \phi_i^{\rm on}< 0$ and $\phi_i^{\rm on}$ is a decreasing function of $d_i$, for all $d_i \ne 0$.

We now prove the cases \textit{(ii)} and \textit{(iii)}. Since $\frac{d_i^2}{2D_s + 4D(t-t_i)} < \frac{d_i^2}{2D_s}$, an application of Lemma~\ref{lem:aux:(1+2u)e(-u)-(1+v)e(-v)} by setting $u \leftarrow \frac{d_i^2}{2D_s + 4D(t-t_i)}$ and $v \leftarrow \frac{d_i^2}{2D_s}$ completes the proof.
\begin{lemma}\label{lem:e1_delta-cos-derivatives}
    Let $f(u) \coloneqq E_1\left(\alpha\left(v - u \right)^2\right)$ for $u \ge 0$ with $\alpha, v\ge 0$. The first and second derivatives of $f$ reads, for $u\ne v$,
    \begin{align*}
        f'(u) &=\frac{2}{v - u} e^{-\alpha\left(v-u
        \right)^2}, \\
        f''(u) &= \frac{2}{\left(v - u \right)^2}\left(1+2\alpha\left(v-u\right)^2\right)e^{-\alpha\left(v-u\right)^2}.
    \end{align*}
    Moreover,
    \begin{equation*}
        \begin{cases}
            f'(u) < 0, & {\rm if~} u>v,\\
            f'(u) > 0, & {\rm if~} u<v,
        \end{cases}
        \quad {\rm and}\quad 
        f''(u) >0,
    \end{equation*}
\end{lemma}
\begin{proof}
    We define $g(u) \coloneqq \alpha\left(v - u \right)^2$. Then, $g'(u) = -2\alpha(v-u)$. Using a chain rule and Lemma~\ref{lem:e1_derivative},
    \begin{equation*}
        f'(u) = g'(u)\left. E'_1(v)\right|_{v = g(u)} = \frac{2}{v - u}e^{-\alpha\left(v - u \right)^2},
    \end{equation*}
for $u\ne v$ and
    \begin{align*}
        f''(u) &= \frac{\partial}{\partial u} \left( \frac{2}{v - u}\right) e^{-\alpha\left(v - u \right)^2} + \frac{2}{v - u} \frac{\partial}{\partial u} \left(e^{-\alpha\left(v - u \right)^2}\right)\\
        &= \frac{2}{\left(v - u \right)^2} e^{-\alpha\left(v - u \right)^2} + \frac{2}{v - u} (2\alpha)(v - u)e^{-\alpha\left(v - u \right)^2}\\
        &=\frac{2}{\left(v - u \right)^2}\left(1+2\alpha\left(v-u\right)^2\right)e^{-\alpha\left(v-u\right)^2}.
    \end{align*}
    We now prove the second part of the Lemma. We initially note that $\exp(-w)>0$ for $w\in\bb{R}$ and
    \begin{equation*}
        \begin{cases}
            f'(u) < 0, & {\rm if~} \frac{1}{v - u}<0, {~\rm or~} v - u<0,\\
            f'(u) > 0, & {\rm if~} \frac{1}{v - u}>0, {~\rm or~} v - u>0.
        \end{cases}
    \end{equation*}
    Moreover, it is evident that $f''(u) >0$ for $u\ge0$, which completes the proof.
\end{proof}
\begin{lemma}\label{lem:e1_derivative}
    Let $E_1(u) \coloneqq \int_u^{+\infty}\inv{v} \exp{(-v)}\,\ud v$ be the Exponential integral of order one. Then,
    \begin{equation*}
        E_1'(u) =-\frac{e^{-u}}{u}, \quad {\rm for} \quad u>0.
    \end{equation*}
\end{lemma}
\begin{proof}
    The proof is a direct application of Thm.~\ref{thm:leibniz} by setting $g(u) \leftarrow u$, $h(u)\leftarrow +\infty$, $t\leftarrow v$, and $f(u,t) \leftarrow \inv{v}\exp{(-v)}$.
\end{proof}
\begin{theorem}\textsc{(\textbf{Leibniz integral rule} \cite[p.~156]{amazigo1980advanced})}\label{thm:leibniz}
Let $g(u)$ and $h(u)$ be continuously differentiable functions of $u$ for $u_0<u<u_1$. Let $f(u,t)$ and $\partial f(u,t)/\partial u$ be continuous for $g(u) < t< h(u)$ and $u_0 < u < u_1$. Then, for $u_0< u< u_1$,
\begin{equation*}
    \frac{\partial}{\partial u} \int_{g(u)}^{h(u)} f(u,t)\, \ud t = h'(u) f(u, h(u)) - g'(u) f(u,g(u)) + \int_{g(u)}^{h(u)} \frac{\partial}{\partial u} f(u,t)\,\ud t.
\end{equation*}    
\end{theorem}
\begin{lemma}\label{lem:aux:(1+2u)e(-u)-(1+v)e(-v)}
    Let $f(u)\coloneqq (1+2u)e^{-u}$ for $u\in \bb{R}$ and $g(u,v) \coloneqq f(u) - f(v)$ with $u<v$. Then, 
    \begin{equation*}
        \begin{cases}
            g(u,v) > 0, & {\rm if}\quad u> \frac{1}{2},\\
            g(u,v) < 0, & {\rm if}\quad v< \frac{1}{2},
        \end{cases}
    \end{equation*}
\end{lemma}
\begin{proof}
    From Lemma~\ref{lem:aux:(1+2u)e(-u)}, $f(u)$ is a decreasing function of $u$, if $u>\inv{2}$. Therefore, for $u<v$ and $u>\inv{2}$, $f(u)-f(v) > 0$. Similarly, $f(v)$ is an increasing function of $v$, if $v<\inv{2}$. Therefore, for $u<v$ and $v<\inv{2}$, $f(u)-f(v) < 0$.
\end{proof}
\begin{lemma}\label{lem:aux:(1+2u)e(-u)}
    Let $f(u) \coloneqq (1+2u)e^{-u}$. Then, 
    \begin{equation*}
        \begin{cases}
            f'(u) > 0, & {\rm for}\quad u< \frac{1}{2},\\
            f'(u) = 0, & {\rm for}\quad u= \frac{1}{2},\\
            f'(u) < 0, & {\rm for}\quad u> \frac{1}{2}.
        \end{cases}
    \end{equation*}
\end{lemma}
\begin{proof}
    The proof is straightforward by showing that $f'(u) = (1-2u)e^{-u}$ and noting that $e^{-u} >0$.
\end{proof}
%%%%%%%%%% End of subsection 1
\subsubsection{Proof of Lemma \texorpdfstring{\ref{lem:stem_single_probe_derivatives_beamoff}}{(2)}}
\label{subsec:proof-stem-single-probe-derivatives-d-beamoff}
We follow the same approach to the proof of Lemma~\ref{lem:stem_single_probe_derivatives_beamoff} in Sec.~\ref{subsec:proof-stem-single-probe-derivatives-d-beamon}. An application of Lemma \ref{lem:e1_delta-cos-derivatives} by setting $v \leftarrow 0$, and $u \leftarrow d_i$, for two settings of $\alpha$, \ie $\alpha \leftarrow \frac{1}{2D_s + 4D(t-t_i)}$ and  $\alpha \leftarrow \frac{1}{2D_s + 4D(t-t_i-\tau_i)}$, gives Eqs.~\eqref{eq:stem-single-probe-first-derivative-d-beamoff} and \eqref{eq:stem-single-probe-second-derivative-d-beamoff}. Since $2D_s + 4D(t-t_i-\tau_i) < 2D_s + 4D(t-t_i)$, it is evident that 
\begin{equation*}
     e^{-\frac{d_i^2}{2D_s + 4D(t-t_i)}}>e^{-\frac{d_i^2}{2D_s + 4D(t-t_i-\tau_i)}},
\end{equation*}
hence, $\frac{\partial}{\partial d_i} \phi_i^{\rm off}< 0$ and $\phi_i^{\rm off}$ is a decreasing function of $d_i$, for all $d_i \ne 0$.

We now prove the cases \textit{(ii)} and \textit{(iii)}. Since $\frac{d_i^2}{2D_s + 4D(t-t_i)} < \frac{d_i^2}{2D_s + 4D(t-t_i-\tau_i)}$, an application of Lemma~\ref{lem:aux:(1+2u)e(-u)-(1+v)e(-v)} by setting $u \leftarrow \frac{d_i^2}{2D_s + D(t-t_i)}$ and $v \leftarrow \frac{d_i^2}{2D_s + 4D(t-t_i-\tau_i)}$ completes the proof.
%%%%%% End of Section 6 %%%%%%
%%%%%% Start of Section 7 %%%%%%
\subsection{Diffusion distribution as a function of time}\label{sec:diffusion_function_of_time}
To understand the behaviour of the diffusion distribution as a function of time, we compute the first and second derivatives of $\phi_i^{\rm on}$ with respect to $t$ in the following Lemma. See the proof in Sec.~\ref{subsec:proof-stem-single-probe-derivatives-t-beamon}
\begin{lemma}\label{lem:stem_single_probe_derivatives_t_beamon}
    For $\phi_i^{\rm on}$ defined in Eq.~\eqref{eq:stem_single_probe_beamon} and for $d_i \coloneqq \|\bs r - \bs r_i\|_2 \ne 0$ we have
\begin{align}
    \frac{\partial}{\partial t}\phi_{i}^{\rm on} = &\begin{cases}
        \frac{Q_0}{\pi(2D_s + 4D(t-t_i))}e^{-\frac{d_i^2}{2D_s + 4D(t-t_i)}}, & {\rm if~} d_i \ne 0,\\
        \frac{Q_0}{\pi(2D_s + 4D(t-t_i))}, & {\rm if~} d_i =0.
    \end{cases}\label{eq:stem-single-probe-first-derivate-beamon-t}\\
    \frac{\partial^2}{\partial t^2}\phi_{i}^{\rm on} = &\begin{cases}
        \frac{-4Q_0 D}{\pi(2D_s + 4D(t-t_i))^2}e^{-\frac{d_i^2}{2D_s + 4D(t-t_i)}}\left(1-\frac{d_i^2}{2D_s + 4D(t-t_i)}\right), & {\rm if~} d_i \ne 0,\\
        \frac{-4Q_0 D}{\pi(2D_s+4D(t-t_i))^2}, & {\rm if~} d_i =0.
    \end{cases}\label{eq:stem-single-probe-second-derivate-beamon-t}
\end{align}
    Moreover, 
    \begin{itemize}
        \item[(i)] $\frac{\partial}{\partial t} \phi_i^{\rm on} >0$, hence, $\phi_i^{\rm on}$ is an increasing function of $t$, for all $t_i \le t\le\tau_i$,
        \item[(ii)] $\frac{\partial^2}{\partial t^2} \phi_i^{\rm on} =0$, hence, given $d_i$, $\phi_i^{\rm on}$ has an inflection point at time $t$, if $\frac{d_i^2}{2D_s+4D(t-t_i)} =1$,
        \item[(iii)] $\frac{\partial^2}{\partial t^2} \phi_i^{\rm on} <0$, hence, $\phi_i^{\rm on}$ is a strictly concave function of $t$, if $\frac{d_i^2}{2D_s+4D(t-t_i)} <1$,
        \item[(iv)] $\frac{\partial^2}{\partial t^2} \phi_i^{\rm on} >0$, hence, $\phi_i^{\rm on}$ is a strictly convex function of $t$, if $\frac{d_i^2}{2D_s+4D(t-t_i)} >1$.
    \end{itemize}
\end{lemma}
Similarly, following Lemma, proved in Sec.~\ref{subsec:proof-stem-single-probe-derivatives-t-beamoff}, which reports the first and second derivatives of $\phi_i^{\rm off}$ with respect to $t$.
\begin{lemma}\label{lem:stem_single_probe_derivatives_t_beamoff}
    For $\phi_i^{\rm off}$ defined in Eq.~\eqref{eq:stem_single_probe_beamoff} and for $d_i \coloneqq \|\bs r - \bs r_i\|_2 \ne 0$ we have
\begin{align}
    \frac{\partial}{\partial t}\phi_{i}^{\rm off} = &\begin{cases}
        \frac{Q_0}{\pi d_i^2}\big(A e^{-A}-Be^{-B}\big), & {\rm if~} d_i \ne 0,\\
        \frac{-4Q_0 D \tau_i}{\pi\left(2D_s+4D(t-t_i)\right)\left(2D_s+4D(t-t_i-\tau_i)\right)}, & {\rm if~} d_i =0.
    \end{cases}\label{eq:stem-single-probe-first-derivate-beamoff-t}\\
    \frac{\partial^2}{\partial t^2}\phi_{i}^{\rm off} = &\begin{cases}
        \frac{-4Q_0 }{\pi d_i^4}\Big(A^2 (1-A) e^{-A}- B^2 (1-B) e^{-B}\Big), & {\rm if~} d_i \ne 0,\\
        \frac{8Q_0 D^2\left(2D_s +4D(t-t_i-\tau_i)\right)}{\pi\left(2D_s+4D(t-t_i)\right)^2\left(2D_s+4D(t-t_i-\tau_i)\right)^2}, & {\rm if~} d_i =0.
    \end{cases}\label{eq:stem-single-probe-second-derivate-beamoff-t}
\end{align}
where
\begin{equation*}
    A \coloneqq\frac{d_i^2}{2D_s + 4D(t-t_i)} \quad {\rm and}\quad  B \coloneqq\frac{d_i^2}{2D_s + 4D(t-t_i-\tau_i)}.
\end{equation*}
    Moreover, 
    \begin{itemize}
        \item[(i)] $\frac{\partial}{\partial t} \phi_i^{\rm off} <0$, hence, $\phi_i^{\rm off}$ is a decreasing function of $t$, if $\frac{d_i^2}{2D_s + 4D(t-t_i-\tau_i)} < 1$,
        \item[(ii)] $\frac{\partial}{\partial t} \phi_i^{\rm off} >0$, hence, $\phi_i^{\rm off}$ is an increasing function of $t$, if $\frac{d_i^2}{2D_s + 4D(t-t_i)} > 1$,
        \item[(iii)] $\frac{\partial^2}{\partial t^2} \phi_i^{\rm off} <0$, hence, $\phi_i^{\rm off}$ is a strictly concave function of $t$, if $\frac{d_i^2}{2D_s+4D(t-t_i-\tau_i)} <2+\sqrt{2}$ and $\frac{d_i^2}{2D_s+4D(t-t_i)} >2-\sqrt{2}$,
        \item[(iv)] $\frac{\partial^2}{\partial t^2} \phi_i^{\rm off} >0$, hence, $\phi_i^{\rm off}$ is a strictly convex function of $t$, if $\frac{d_i^2}{2D_s+4D(t-t_i-\tau_i)} <2-\sqrt{2}$ or $\frac{d_i^2}{2D_s+4D(t-t_i)} >2+\sqrt{2}$,
    \end{itemize}
\end{lemma}
%%%%%%%%%% Figure 
\begin{figure}[h!t]
    \centering
    \includegraphics[width=1\columnwidth]{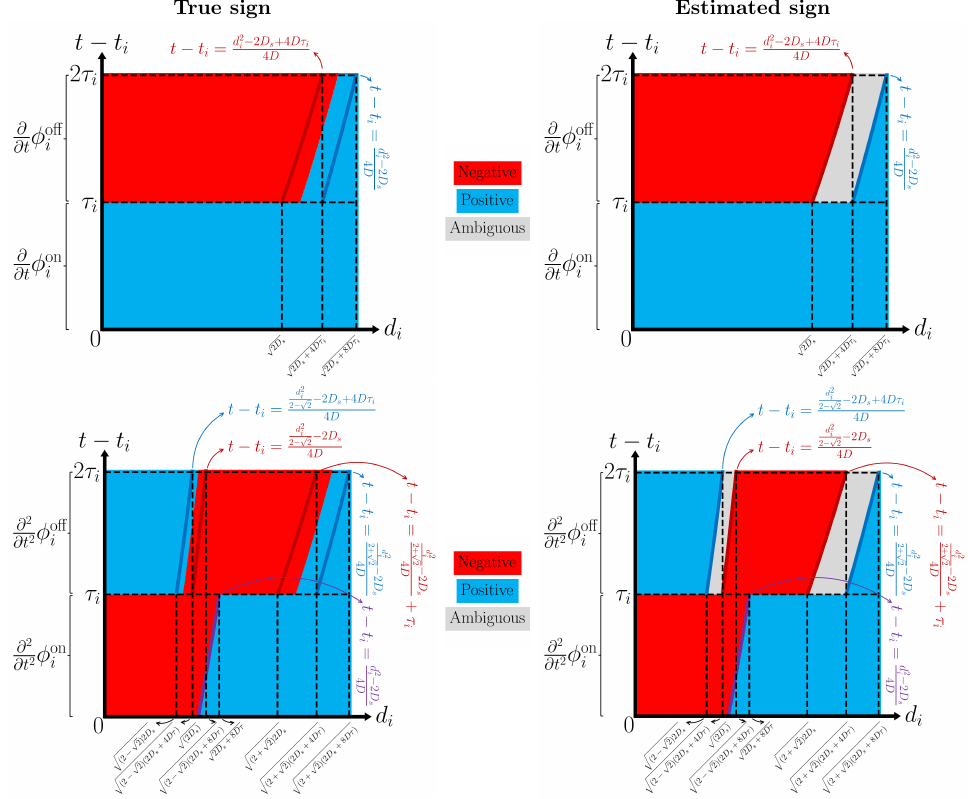}
    \caption{\textbf{Sign of the first (top) and second (bottom) derivatives of the diffusion distribution from a single STEM electron probe with respect to time.} Points with $0<t-t_i<\tau_i$ and $\tau_i<t-t_i$ correspond to the electron beam being, respectively, on and off. True sign of the first (top-left) and second (bottom-left) derivatives are numerically computed from the equations provided by Lemmas~\ref{lem:stem_single_probe_derivatives_t_beamon} and \ref{lem:stem_single_probe_derivatives_t_beamoff}.
    (Right) Estimated sign following the sufficient conditions given by cases \textit{(ii)} and \textit{(iii)} of Lemmas~\ref{lem:stem_single_probe_derivatives_t_beamon} and \ref{lem:stem_single_probe_derivatives_t_beamoff}. Ambiguous regions are the pairs of $(d_i, t-t_i)$ that are not covered by the sufficient conditions in Lemmas~\ref{lem:stem_single_probe_derivatives_t_beamon} and \ref{lem:stem_single_probe_derivatives_t_beamoff}.}
    \label{fig:stem-single-beam-derivatives-t}
\end{figure}
%%%%%%%% End of figure
Lemma~\ref{lem:stem_single_probe_derivatives_t_beamon} states that when the electron beam is on, diffusion distribution is always an increasing function of time, \ie the concentration of diffusing substances increases, and reaches a maximum value at the end of the dwell time. In contrast, when the electron beam is off depending on the ratios given in \textit{(i)} and \textit{(ii)} in Lemma~\ref{lem:stem_single_probe_derivatives_t_beamoff}, diffusion distribution can increase or decrease as a function of time. Cases \textit{(iii)} and \textit{(iv)} in Lemmas \ref{lem:stem_single_probe_derivatives_t_beamon} and \ref{lem:stem_single_probe_derivatives_t_beamoff} give sufficient conditions for convexity or concavity of the diffusion distribution. These conditions, when illustrated, as in Fig.~\ref{fig:stem-single-beam-derivatives-t} (left), correspond to regions identified by parabolas. Regions in time and space that are not covered by Lemma ~\ref{lem:stem_single_probe_derivatives_t_beamoff}, \ie ambiguous regions, are highlighted in gray in Fig.~\ref{fig:stem-single-beam-derivatives-t} (left). 
% We note that when $D\tau_i \rightarrow$, the area of inconclusive region goes to zero.
Fig.~\ref{fig:stem-single-beam-derivatives-t}(top-right) illustrates the true sign of the first derivative and Fig.~\ref{fig:stem-single-beam-derivatives-t}(bottom-right) illustrates the true sign of the second derivative, both numerically computed from Eq.~\eqref{eq:stem-single-probe-first-derivate-beamon-t} for $0< t-t_i < \tau_i$ and from \eqref{eq:stem-single-probe-second-derivate-beamoff-t} for $\tau_i< t-t_i$.
%% subsection 1
\subsubsection{Proof of Lemma \texorpdfstring{\ref{lem:stem_single_probe_derivatives_t_beamon}}{(3)}}
\label{subsec:proof-stem-single-probe-derivatives-t-beamon}
Using Lemma~\ref{lem:e1_derivative}, we first note that
\begin{align}
    &\frac{\partial}{\partial t}\eint{d_i^2}{2D_s+4D(t-t_i)} = \frac{4D}{2D_s+4D(t-t_i)}e^{-\frac{d_i^2}{2D_s+4D(t-t_i)}}\label{eq:aux:derivate-e1-t}\\
    &\frac{\partial}{\partial t}\ln\big(\frac{2D_s+4D(t-t_i)}{2D_s}\big) = \frac{4D}{2D_s+4D(t-t_i)}\label{eq:aux:derivate-ln1-t}.
\end{align}
Using Eqs.~\eqref{eq:aux:derivate-e1-t} and \eqref{eq:aux:derivate-ln1-t} gives
\begin{equation}\label{eq:aux:stem-single-probe-first-derivate-beamon-t}
    \frac{\partial}{\partial t}\phi_{i}^{\rm on} = \begin{cases}
        \frac{2Q_0 D_s}{2D_s + 4D(t-t_i)}e^{-\frac{d_i^2}{2D_s + 4D(t-t_i)}}, & {\rm if~} d_i \ne 0,\\
        \frac{2Q_0 D_s}{2D_s+4D(t-ti)}, & {\rm if~} d_i =0.
    \end{cases}
\end{equation}
It is evident that $\frac{\partial}{\partial t}\phi_{i}^{\rm on} > 0$. From Eq.~\eqref{eq:aux:stem-single-probe-first-derivate-beamon-t}, the second derivative of $\phi_{i}^{\rm on}$ with respect to $t$ reads
\begin{equation}\label{eq:aux:stem-single-probe-second-derivate-beamon-t}
    \frac{\partial^2}{\partial t^2}\phi_{i}^{\rm on} = \begin{cases}
        \frac{-8Q_0 D_sD}{(2D_s + 4D(t-t_i))^2}e^{-\frac{d_i^2}{2D_s + 4D(t-t_i)}}\left(1-\frac{d_i^2}{2D_s + 4D(t-t_i)}\right), & {\rm if~} d_i \ne 0,\\
        \frac{-8Q_0 D_s D}{(2D_s+4D(t-ti))^2}, & {\rm if~} d_i =0.
    \end{cases}
\end{equation}
It is also evident that $\frac{\partial^2}{\partial t^2}\phi_{i}^{\rm on} <0$ for $\frac{d_i^2}{2D_s + 4D(t-t_i)} < 1$ and $\frac{\partial^2}{\partial t^2}\phi_{i}^{\rm on} >0$ for $\frac{d_i^2}{2D_s + 4D(t-t_i)} > 1$. Moreover, $\frac{\partial^2}{\partial t^2}\phi_{i}^{\rm on} =0$, if $\frac{d_i^2}{2D_s + 4D(t-t_i)} = 1$.
%%%%%%%%%%%%%%%%%%%%%%%%%%%%%%%%%%%%%%%%%
\subsubsection{Proof of Lemma \texorpdfstring{\ref{lem:stem_single_probe_derivatives_t_beamoff}}{(4)}}
\label{subsec:proof-stem-single-probe-derivatives-t-beamoff}
We first note that
\begin{align}
    \frac{\partial}{\partial t}\ln\big(\frac{2D_s+4D(t-t_i)}{2D_s+4D(t-t_i-\tau_i)}\big) = \frac{-16D^2\tau_i}{\left(2D_s+4D(t-t_i)\right)\left(2D_s+4D(t-t_i-\tau_i)\right)}\label{eq:aux:derivate-ln2-t}.
\end{align}
Using Eqs.~\eqref{eq:aux:derivate-e1-t} and \eqref{eq:aux:derivate-ln2-t} gives
\begin{equation}\label{eq:aux:stem-single-probe-first-derivate-beamoff-t}
    \frac{\partial}{\partial t}\phi_{i}^{\rm off} = \begin{cases}
        \frac{2Q_0 D_s}{d_i^2}\big(Ae^{-A}-Be^{-B}\big), & {\rm if~} d_i \ne 0,\\
        \frac{-8Q_0 D_s D \tau_i}{\left(2D_s+4D(t-t_i)\right)\left(2D_s+4D(t-t_i-\tau_i)\right)}, & {\rm if~} d_i =0,
    \end{cases}
\end{equation}
where
\begin{equation*}
    A \coloneqq\frac{d_i^2}{2D_s + 4D(t-t_i)} \quad {\rm and}\quad  B \coloneqq\frac{d_i^2}{2D_s + 4D(t-t_i-\tau_i)}.
\end{equation*}
Since $2D_s + 4D(t-t_i-\tau_i) < 2D_s + 4D(t-t_i)$, then $A < B$. An application of Lemma~\ref{lem:aux:ue(-u)-ve(-v)} by setting $u \leftarrow A$ and $v \leftarrow B$ proves cases \textit{(i)} and \textit{(ii)}.

From Eq.~\eqref{eq:aux:stem-single-probe-first-derivate-beamoff-t}, the second derivative of $\phi_{i}^{\rm off}$ with respect to $t$ reads
\begin{equation}\label{eq:aux:stem-single-probe-second-derivate-beamoff-t}
    \frac{\partial^2}{\partial t^2}\phi_{i}^{\rm on} = \begin{cases}
        \frac{-8Q_0 D_s}{d_i^4}\Big(A^2 (1-A) e^{-A}- B^2 (1-B) e^{-B}\Big), & {\rm if~} d_i \ne 0,\\
        \frac{16Q_0 D_s D^2\left(2D_s +4D(t-t_i-\tau_i)\right)}{\left(2D_s+4D(t-t_i)\right)^2\left(2D_s+4D(t-t_i-\tau_i)\right)^2}, & {\rm if~} d_i =0.
    \end{cases}
\end{equation}
Since $A<B$, an application of Lemma~\ref{lem:aux:u^2(1-u)e(-u)-v^2(1-v)e(-v)} by setting $u\leftarrow A$ and $v \leftarrow B$ proves the cases \textit{(iii)} and \textit{(iv)}.
\begin{lemma}\label{lem:aux:ue(-u)-ve(-v)}
    Let $f(u)\coloneqq ue^{-u}$ for $u\in \bb{R}$ and $g(u,v) \coloneqq f(u) - f(v)$ with $u<v$. Then, 
    \begin{equation*}
        \begin{cases}
            g(u,v) > 0, & {\rm if}\quad u> 1,\\
            g(u,v) < 0, & {\rm if}\quad v< 1,
        \end{cases}
    \end{equation*}
\end{lemma}
\begin{proof}
    From Lemma~\ref{lem:aux:ue(-u)}, $f(u)$ is a decreasing function of $u$, if $u>1$. Therefore, for $u<v$ and $u>1$, $f(u)-f(v) > 0$. Similarly, $f(v)$ is an increasing function of $v$, if $v<1$. Therefore, for $u<v$ and $v<1$, $f(u)-f(v) < 0$.
\end{proof}
\begin{lemma}\label{lem:aux:ue(-u)}
    Let $f(u) \coloneqq ue^{-u}$. Then, 
    \begin{equation*}
        \begin{cases}
            f'(u) > 0, & {\rm for}\quad u< 1,\\
            f'(u) = 0, & {\rm for}\quad u= 1,\\
            f'(u) < 0, & {\rm for}\quad u> 1.
        \end{cases}
    \end{equation*}
\end{lemma}
\begin{proof}
    The proof is straightforward by showing that $f'(u) = (1-u)e^{-u}$ and noting that $e^{-u} >0$.
\end{proof}
\begin{lemma}\label{lem:aux:u^2(1-u)e(-u)-v^2(1-v)e(-v)}
    Let $f(u)\coloneqq u^2(1-u)e^{-u}$ for $u\in \bb{R}$ and $g(u,v) \coloneqq f(u) - f(v)$ with $u<v$. Then, 
    \begin{equation*}
        \begin{cases}
            g(u,v) > 0, & {\rm if}\quad v<2-\sqrt{2} {~\rm or~} u>2+\sqrt{2},\\
            g(u,v) < 0, & {\rm if}\quad v<2+\sqrt{2} {~\rm and~} u>2-\sqrt{2},
        \end{cases}
    \end{equation*}
\end{lemma}
\begin{proof}
    From Lemma~\ref{lem:aux:u^2(1-u)e(-u)}, $f(u)$ is a decreasing function of $u$, if $u\in (2-\sqrt{2},2+\sqrt{2})$. Therefore, for $u<v$, $v<2+\sqrt{2}$ and $u>2-\sqrt{2}$, $f(u)-f(v) > 0$. Similarly, $f(v)$ is an increasing function of $v$, if $v \in (0,2-\sqrt{2}) \cup (2+\sqrt{2},+\infty)$. Therefore, for $u<v$, $f(u)-f(v) < 0$, if $v<2-\sqrt{2}$ or $u>2+\sqrt{2}$.
\end{proof}
\begin{lemma}\label{lem:aux:u^2(1-u)e(-u)}
    Let $f(u) \coloneqq u^2(1-u)e^{-u}$. Then, 
    \begin{equation*}
        \begin{cases}
            f'(u) > 0, & {\rm for}\quad u \in (-\infty,2-\sqrt{2}) \cup (2+\sqrt{2},+\infty),\\
            f'(u) = 0, & {\rm for}\quad u\in \{0,2+\sqrt{2},2+\sqrt{2}\},\\
            f'(u) < 0, & {\rm for}\quad u\in (2-\sqrt{2},2+\sqrt{2}).
        \end{cases}
    \end{equation*}
\end{lemma}
\begin{proof}
    We note that $f'(u) = u(u^2-4u+2)e^{-u}$ and $f'(u) =0$ for $u\in\{0,2-\sqrt{2},2+\sqrt{2}\}$.
\end{proof}
%%%%%% End of Section 7 %%%%%%
%%%%%% Start of Section 8 %%%%%%
\subsection{Proof of Theorem \texorpdfstring{\ref{thm:did-mpcdd}}{(1)}}
\label{subsec:proof-thm-did-free-stem}
Since $\forall \bs r, \Lambda(\bs r;\lambda) \ge 0,$ we note that $ \Lambda(\lambda) = \int \Lambda(\bs r;\lambda)\,\ud \bs r =  0$ if and only $\forall \bs r, \Lambda(\bs r;\lambda) = 0$.

We first prove $ \Lambda(\lambda) = 0 \Longrightarrow \chi^{\max} \le \lambda$ by contradiction. Assume that $\chi^{\max} > \lambda$, then, from Eq.~\ref{eq:gm-cdd}, there exists a point $\bs r^*$ in space, such that $\chi (\bs r^*) > \lambda$. Eq.~\eqref{eq:stem_mp_cdd} implies that there also exists a point $t^*$ and a probe position $j^*$ such that $\psi_{j^*}(\bs r^*, t^*) > \lambda$. Therefore, since the function $g$ is non-negative and $g(u) > 0$ for $u > 0$, $g(\psi_{j^*}(\bs r^*, t^*) - \lambda) > 0$. Moreover, since $\int_{t_1}^{t_N} p(\bs r,t)\, \ud t'>0$ for every location $r$, this implies that $\int_{t_1}^{t_N} p(\bs r^*,t)\, \ud t'>0$. Therefore, $\Lambda(\bs r^*;\lambda) > 0$; hence, $ \Lambda(\lambda) > 0$.

We finally prove $\chi^{\max} \le \lambda  \Longrightarrow \Lambda(\lambda) = 0$. If $\chi^{\max} \le \lambda$, by definition in Eq.~\eqref{eq:gm-cdd}, for all spatial locations $\bs r$, we have $\chi(\bs r) \le \lambda$. Therefore, from the definition in Eq.~\eqref{eq:stem_mp_cdd},  for all spatial and temporal points $\bs r$ and $t$ and for all electron probes $j$, $\psi_j(\bs r,t)\le\lambda$; hence, $g(\psi_j(\bs r,t)-\lambda) =0$. This implies that $\Lambda(\bs r;\lambda) = 0$ for all $\bs r$.
%%%%%% End of Section 8 %%%%%%
\subsection{Computational complexity of diffusion distribution in STEM}\label{subsec:time_complexity}
In this section, we analyse the time complexity of computing (or simulating) the CDD in Eqs.~\eqref{eq:stem_cumulative} and \eqref{eq:stem_cumulative_compressive_stem} for full STEM and compressive STEM, respectively. Numerical supports are performed in Sec.~\ref{subsec:simulations_time_complexity}.
We emphasise that without the explicit formulation of a diffusion distribution in Eq.~\eqref{eq:stem_cumulative}, it is necessary to use numerical solvers for the PDE in Eq.~\eqref{eq:general-pde-constant-d}, for instance, as in~\cite{nicholls2020minimising}, for which the time complexity is greater than computing the exponential integral in Eq.~\eqref{eq:stem_single_probe}. 

\paragraph{Full STEM.} Let $\cl R_{\rm sim}$ be a set of spatial points defined as the simulation grid. We assume that $N_T$ temporal grids are defined during every period of two consecutive electron probes, \ie $t_i \le t \le t_{i+1}$. For a given time instance, the time complexity of every diffusion distribution $\phi_i(\bs r,t)$ in Eq.~\eqref{eq:stem_cumulative_simple} scales as $ O(|\cl R_{\rm sim}|\cdot T_{E_1})$, where $T_{E_1}$ is the time complexity required to compute the exponential integral function $E_1$ in \eqref{eq:stem_single_probe_beamon}. Hence, the time complexity of every CDD is $\psi_j(\bs r,t)$ is $O(j \cdot N_T\cdot |\cl R_{\rm sim}|\cdot T_{E_1})$. Since $\sum_{j=1}^N j = N(N+1)/2$, the time complexity of diffusion in a full STEM scan, with $N$ probe positions, is $O(N^2\cdot N_T \cdot|\cl R_{\rm sim}|\cdot T_{E_1})$. Since $|\cl R_{\rm sim}|$ scales with the number of probe positions $N$, the time complexity of diffusion with respect to only the number of probe positions will be $O(N^3)$. Notice that the irreversible DID modeled in Eq.~\eqref{eq:stem_did} shares the same time complexity. 

\paragraph{Compressive STEM.} Following the same steps as above for compressive STEM with $M$ subsampled probe results in a time complexity of $O(M^2\cdot N_T \cdot|\cl R_{\rm sim}|\cdot T_{E_1})$. Therefore, the simulation of CDD of compressive STEM is computationally more tractable.

\paragraph{Methods for reducing time complexity of diffusion distribution in STEM.} Although our explicit formulation of diffusion has orders of magnitude lower time complexity compared to approaches using PDE solvers, its application for large-scale, high-resolution, and accurate numerical analysis is limited. In the following, we provide two approaches to reduce computational time, albeit at the cost of reduced accuracy.
Therefore in a subsequent section we describe approximations of the CDD in STEM. 

\subparagraph{$K$-Nearest Probes ($K$-NP).} We assume that when an electron probe is activated, only a subset of previous probes have contribution to the CDD and that the diffusion distributions of the other electron probe positions are negligible. From Eq.~\eqref{eq:stem_cumulative}, let $\cl N_j$ with cardinality $|\cl N_j| = K$ be the set of $K$-nearest neighbor probes to the $j$-th electron probe position. Note that definition of a distance function identifying the nearest neighbor is arbitrary  and can be based on, Euclidean, Manhattan, or Chebyshev distances \cite{rodrigues2018combining}. Therefore, the CDD can be written as
\begin{equation}
    \psi_j(\bs r, t) = \psi^{K{\rm -NP}}_j(\bs r, t) + \epsilon^{K{\rm -NP}}_j(\bs r, t),
\end{equation}
where
\begin{equation}\label{eq:stem_cumulative_knnp}
\psi^{K{\rm -NP}}_j(\bs r, t) \coloneqq
        \phi_j(\bs r, t) + \sum_{i\in \cl N_j}\phi_i(\bs r, t)
\end{equation}
is the $K$-NP approximation of the CDD and $\epsilon^{K{\rm -NP}}_j$ is the corresponding approximation error. Following the same steps as above, the time complexity of every $\psi_j(\bs r,t)$ with a $K$-NP approximation is $O({\rm min}(j, K) \cdot N_T\cdot |\cl R_{\rm sim}|\cdot T_{E_1})$. Since $\sum_{j=1}^N {\rm min}(j, K) = K(K+1)/2 + K(N-K)$, the full STEM scan with an $K$-NP approximation will have a time complexity of $O(K(N-K/2) \cdot N_T\cdot |\cl R_{\rm sim}|\cdot T_{E_1})$ which is more scalable compared to that of $\psi_N(\bs r, t)$, \ie $O(N^2\cdot N_T \cdot|\cl R_{\rm sim}|\cdot T_{E_1})$. The number of nearest probes hence $K$ controls the trade-off between computational time  and accuracy.

\subparagraph{Upper bound.} For cases where the upper bound or the maximum value of the diffusion distribution is important, an upper bound on the diffusion distribution in Eq.~\eqref{eq:stem_cumulative} can be understood by considering Cor.~\ref{cor:stem_single_prob}. Let $\psi^{\rm UB}_j(\bs r, t)$ be an upper bound to the CDD $\psi_j(\bs r, t)$. Using Cor.~\ref{cor:stem_single_prob} and at a given time instance, the maximum value of the diffusion distribution happens at the activation point, we replace the value of $\phi_i^{\rm off}(\bs r, t)$ in Eq.~\eqref{eq:stem_cumulative} for $1\le i\le j-1$ with its maximum value at the activation point; Hence,
\begin{equation}\label{eq:stem_cumulative_ub}
    \psi_j(\bs r, t) \le \psi^{\rm UB}_j(\bs r, t) \coloneqq
        \phi_j^{\rm on}(\bs r, t) + \sum_{i=1}^{j-1}\phi_i^{\rm off}(\bs r_i, t).
\end{equation}

For a given time instance, the time complexity of every $\phi_i^{\rm off}(\bs r_i,t)$ in Eq.~\eqref{eq:stem_cumulative_ub} is $ O(T_{E_1})$. By taking into account the $j$ number of terms in Eq.~\eqref{eq:stem_cumulative_ub} and $N_T$ the number of time instances, the time complexity of the diffusion distribution $\psi^{\rm UB}_j(\bs r, t)$ is $O( (j + |\cl R_{\rm sim}|) \cdot N_T\cdot T_{E_1})$. Therefore, the time complexity of diffusion in a full STEM scan using the upper bound $\psi^{\rm UB}_N(\bs r, t)$ is $O(N(|\cl R_{\rm sim}| + N/2) \cdot N_T\cdot T_{E_1})$, which is lower than that of the diffusion distribution $\psi_N(\bs r, t)$ by order $O(N)$.
%%% Subsection 1
%%%%%%%%%%%%

\subsubsection{Numerical experiments for time complexity of diffusion distribution}\label{subsec:simulations_time_complexity}

To verify the time complexity analyses above, we have carried out several numerical tests using the same values as for the baseline STEM scan, reported in Tab.~\ref{tab:baseline_parameters}, for $\tau, Q_0, D, D_s, \Delta_p$. The remaining parameters are given below.

Fig.~\ref{fig:time_complexity_full} shows the empirical and theoretical time complexities, with the latter given by $O(N^2 \cdot N_T \cdot |\cl R_{\rm sim}| \cdot T_{E_1})$ in Sec.~\ref{subsec:time_complexity}. The elapsed times are averaged over 50 Monte-Carlo trials. In Fig.~\ref{fig:time_complexity_full}-left, $|\cl R_{\rm sim}|$ is set as $N$, \ie every scan step length is simulated with one pixel, $N_T = 1$, and we have simulated systems with $N\in\{1, 2^2,\cdots,20^2\}$. We observe that the elapsed time is proportional to $N^3$. In Fig.~\ref{fig:time_complexity_full}-middle, we set $N = 2^2$ and $N_T = 1$ and simulate systems with $|\cl R_{\rm sim}| \in\{1,10^2,\cdots,100^2\}$. In this case we observe that the elapsed time is proportional to $|\cl R_{\rm sim}|$. Finally, in Fig.~\ref{fig:time_complexity_full}-right, we set $N = 2^2$ and $|\cl R_{\rm sim}| = N$ and simulate systems with $N_T \in \{1,10,\cdots,90\}$, where we observe that the elapsed time is proportional to $N_T$. These simulations confirm our theoretical bound $O(N^2 \cdot N_T \cdot |\cl R_{\rm sim}| \cdot T_{E_1})$.

The simulations have also been extended to two acceleration methods for simulating CDD. Initially we set $N = 20^2$, $N_T=1$, and $|\cl R_{\rm sim}| = N$ with results averaged over 50 Monte-Carlo trials. From Sec.~\ref{subsec:time_complexity} the time complexity bounds of the $K-$NP approximation and upper bound approaches are respectively, $O(K(N-K/2)\cdot N_T \cdot |\cl R_{\rm sim}|\cdot T_{E_1})$ and $O(N(|\cl R_{\rm sim}| + N/2)\cdot N_T\cdot T_{E_1})$.

\begin{figure}[!tbh]
    \centering
    \includegraphics[width=1\columnwidth]{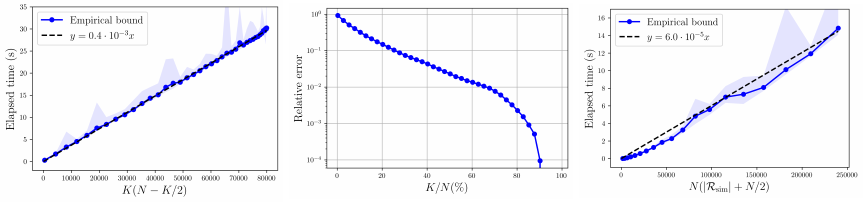}
    \caption{\textbf{Time complexity of CDD in STEM using acceleration methods.}
  Elapsed time for $K$-nearest neighbours approximation (left) with corresponding relative error (middle); Simulated elapsed time for the upper bound of CDD using \eqref{eq:stem_cumulative_ub}.} 
\label{fig:time_complexity_acceleration}
\end{figure}

Fig.~\ref{fig:time_complexity_acceleration} shows the time complexity of the $K$-NP approximation with $K\in \{10j+1\}_{j=0}^{N^2/10-1}$. It is clear that the elapsed time is proportional to $K(N-K/2)$, as expected. The relative errors of this $K$-NP approach are plotted in Fig.~\ref{fig:time_complexity_acceleration}. As a specific example approximating the CDD using only  $K = 50\%$ of the previous probes resulted in approximately a $2\%$ relative error. The elapsed time for simulating CDD in STEM using the upper bound approach is shown in Fig.~\ref{fig:time_complexity_acceleration}, which is proportional to $N(|\cl R_{\rm sim}| + N/2)$.
\begin{figure}[!tbh]
    \centering
    \includegraphics[width=1\columnwidth]{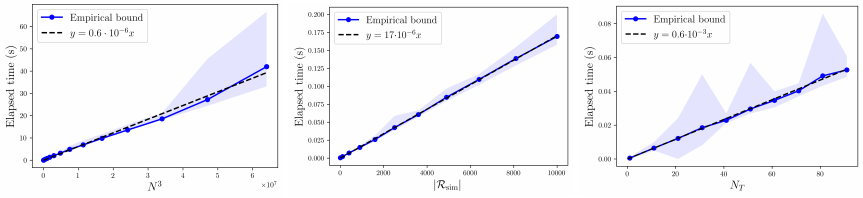}
    \caption{\textbf{Time complexity of CDD in STEM.}
  Number of probe positions (left); spatial resolution (middle); and temporal resolution (right).}
\label{fig:time_complexity_full}
\end{figure}

%%%%%% Start of Section 9 %%%%%%
% \bibliographystyle{IEEEtran} 
% \bibliography{Main}

%% file: Main.bbl
\begin{thebibliography}{10}
\providecommand{\url}[1]{#1}
\csname url@samestyle\endcsname
\providecommand{\newblock}{\relax}
\providecommand{\bibinfo}[2]{#2}
\providecommand{\BIBentrySTDinterwordspacing}{\spaceskip=0pt\relax}
\providecommand{\BIBentryALTinterwordstretchfactor}{4}
\providecommand{\BIBentryALTinterwordspacing}{\spaceskip=\fontdimen2\font plus
\BIBentryALTinterwordstretchfactor\fontdimen3\font minus \fontdimen4\font\relax}
\providecommand{\BIBforeignlanguage}[2]{{%
\expandafter\ifx\csname l@#1\endcsname\relax
\typeout{** WARNING: IEEEtran.bst: No hyphenation pattern has been}%
\typeout{** loaded for the language `#1'. Using the pattern for}%
\typeout{** the default language instead.}%
\else
\language=\csname l@#1\endcsname
\fi
#2}}
\providecommand{\BIBdecl}{\relax}
\BIBdecl

\bibitem{nellist1995resolution}
P.~Nellist, B.~McCallum, and J.~M. Rodenburg, ``Resolution beyond the information limit in transmission electron microscopy,'' \emph{nature}, vol. 374, no. 6523, pp. 630--632, 1995.

\bibitem{james1999practical}
E.~James and N.~Browning, ``Practical aspects of atomic resolution imaging and analysis in {STEM},'' \emph{Ultramicroscopy}, vol.~78, no. 1-4, pp. 125--139, 1999.

\bibitem{zhang2018atomic}
D.~Zhang, Y.~Zhu, L.~Liu, X.~Ying, C.-E. Hsiung, R.~Sougrat, K.~Li, and Y.~Han, ``Atomic-resolution transmission electron microscopy of electron beam--sensitive crystalline materials,'' \emph{Science}, vol. 359, no. 6376, pp. 675--679, 2018.

\bibitem{Krivanek1999subA}
O.~L. Krivanek, N.~Delby, and A.~R. Lupini, ``Towards sub-\r{A} electron beams,'' \emph{Ultramicroscopy}, vol.~78, pp. 1--11, 1999.

\bibitem{Sawada47pmSTEM}
H.~Sawada, Y.~Tanishiro, N.~Ohashi, T.~Tomita, F.~Hosokawa, T.~Kaneyama, Y.~Kondo, and K.~Takayanagi, ``Stem imaging of 47-pm-separated atomic columns by a spherical aberration-corrected electron microscope with a 300-kv cold field emission gun,'' \emph{Journal of Electron Microscopy}, vol.~58, pp. 357--361, 2009.

\bibitem{haider1998electron}
M.~Haider, S.~Uhlemann, E.~Schwan, H.~Rose, B.~Kabius, and K.~Urban, ``Electron microscopy image enhanced,'' \emph{Nature}, vol. 392, no. 6678, pp. 768--769, 1998.

\bibitem{johnson2022near}
C.~W. Johnson, A.~K. Schmid, M.~Mankos, R.~R{\"o}pke, N.~Kerker, E.~K. Wong, D.~F. Ogletree, A.~M. Minor, and A.~Stibor, ``Near-monochromatic tuneable cryogenic niobium electron field emitter,'' \emph{Physical Review Letters}, vol. 129, no.~24, p. 244802, 2022.

\bibitem{pooch2018coherent}
A.~Pooch, M.~Seidling, N.~Kerker, R.~R{\"o}pke, A.~Rembold, W.-T. Chang, I.-S. Hwang, and A.~Stibor, ``Coherent properties of a tunable low-energy electron-matter-wave source,'' \emph{Physical Review A}, vol.~97, no.~1, p. 013611, 2018.

\bibitem{kuo2006noble}
H.-S. Kuo, S.~Hwang, T.-Y. Fu, Y.-C. Lin, C.-C. Chang, and T.~T. Tsong, ``Noble metal/w (111) single-atom tips and their field electron and ion emission characteristics,'' \emph{Japanese journal of applied physics}, vol.~45, no. 11R, p. 8972, 2006.

\bibitem{faruqi2007direct}
A.~Faruqi, ``Direct electron detectors for electron microscopy,'' \emph{Advances in Imaging and Electron Physics}, vol. 145, pp. 55--93, 2007.

\bibitem{faruqi2015progress}
A.~Faruqi, R.~Henderson, and G.~McMullan, ``Progress and development of direct detectors for electron cryomicroscopy,'' \emph{Advances in Imaging and Electron Physics}, vol. 190, pp. 103--141, 2015.

\bibitem{pennycook2011scanning}
S.~J. Pennycook and P.~D. Nellist, \emph{Scanning transmission electron microscopy: imaging and analysis}.\hskip 1em plus 0.5em minus 0.4em\relax Springer Science \& Business Media, 2011.

\bibitem{zuo2017advanced}
J.~M. Zuo and J.~C. Spence, \emph{Advanced transmission electron microscopy}.\hskip 1em plus 0.5em minus 0.4em\relax Springer, 2017.

\bibitem{egerton2004radiation}
R.~Egerton, P.~Li, and M.~Malac, ``Radiation damage in the {TEM} and {SEM},'' \emph{Micron}, vol.~35, no.~6, pp. 399--409, 2004.

\bibitem{jiang2015electron}
N.~Jiang, ``Electron beam damage in oxides: a review,'' \emph{Reports on Progress in Physics}, vol.~79, no.~1, p. 016501, 2015.

\bibitem{jenkins1982characterization}
M.~Jenkins and C.~English, ``Characterization of displacement cascade damage in ordered alloys using transmission electron microscopy,'' \emph{Journal of Nuclear Materials}, vol. 108, pp. 46--61, 1982.

\bibitem{egerton2019radiation}
R.~Egerton, ``Radiation damage to organic and inorganic specimens in the {TEM},'' \emph{Micron}, vol. 119, pp. 72--87, 2019.

\bibitem{secondary_polys1999}
K.~Siangchaew and M.~Libera, ``The influence of fast secondary electrons on the aromatic structure of polystyrene,'' \emph{Philosophical Magazine A}, vol.~80, pp. 1001--1016, 1999.

\bibitem{secondary_litho2001}
B.~Wu and A.~R. Neureuther, ``Energy deposition and transfer in electron-beam lithography,'' \emph{Journal of Vacuum Science \& Technology B}, vol.~19, pp. 2508--–2511, 2001.

\bibitem{EgerOutrundamage2015}
R.~F. Egerton, ``Outrun radiation damage with electrons?'' \emph{Advanced Structural and Chemical Imaging}, vol.~1, pp. 1--5, 2015.

\bibitem{interleaveSTEM2022}
A.~Velazco, A.~Béché, D.~Jannis, and J.~Verbeeck, ``Reducing electron beam damage through alternative {STEM} scanning strategies, part i: Experimental findings,'' \emph{Ultramicroscopy}, vol. 232, p. 113398, 2022.

\bibitem{randomSTEM2020}
A.~Zobelli, S.~Y. Woo, A.~Tararan, L.~H.~G. Tizei, N.~Brun, X.~Li, O.~Stéphan, M.~Kociak, and M.~Tencé, ``Spatial and spectral dynamics in {STEM} hyperspectral imaging using random scan patterns,'' \emph{Ultramicroscopy}, vol. 212, p. 112912, 2020.

\bibitem{CStheory2006}
D.~L. Donoho, ``Compressed sensing,'' \emph{IEEE Transactions on Information Theory}, vol.~52, pp. 1289--1306, 2006.

\bibitem{candes2006robust}
E.~Cand\'es, J.~Romberg, and T.~Tao, ``Robust uncertainty principles: Exact signal reconstruction from highly incomplete frequency information,'' \emph{IEEE Transactions on information theory}, vol.~52, no.~2, pp. 489--509, 2006.

\bibitem{kovarik2016implementing}
L.~Kovarik, A.~Stevens, A.~Liyu, and N.~D. Browning, ``Implementing an accurate and rapid sparse sampling approach for low-dose atomic resolution {STEM} imaging,'' \emph{Applied Physics Letters}, vol. 109, no.~16, 2016.

\bibitem{beche2016development}
A.~B{\'e}ch{\'e}, B.~Goris, B.~Freitag, and J.~Verbeeck, ``Development of a fast electromagnetic beam blanker for compressed sensing in scanning transmission electron microscopy,'' \emph{Applied Physics Letters}, vol. 108, no.~9, 2016.

\bibitem{multiframeSTEM2018}
L.~Jones, A.~Varambhia, R.~Beanland, D.~Kepaptsoglou, I.~Griffiths, A.~Ishizuka, F.~Azough, R.~Freer, K.~Ishizuka, D.~Cherns, Q.~M. Ramasse, S.~Lozano-Perez, and P.~D. Nellist, ``Managing dose-, damage- and data-rates in multi-frame spectrum-imaging,'' \emph{Microscopy}, vol.~67, p. i98–i113, 2018.

\bibitem{knockonmodels2018}
K.~Nordlund, S.~J. Zinkle, A.~E. Sand, F.~Granberg, R.~S. Averback, R.~E. Stoller, S.~Tomoaki, L.~Malerba, F.~Banhart, W.~J. Weber, F.~Willaime, S.~L. Dudarev, and S.~David, ``Primary radiation damage: A review of current understanding and models,'' \emph{Journal of Nuclear Materials}, vol. 512, pp. 450--479, 2018.

\bibitem{EgerDose-rate1999}
R.~F. Egerton and I.~Rauf, ``Dose-rate dependence of electron-induced mass lossfrom organic specimens,'' \emph{Ultramicroscopy}, vol.~80, pp. 247--254, 1999.

\bibitem{fick1855v}
A.~Fick, ``V. on liquid diffusion,'' \emph{The London, Edinburgh, and Dublin Philosophical Magazine and Journal of Science}, vol.~10, no.~63, pp. 30--39, 1855.

\bibitem{crank1979mathematics}
J.~Crank, \emph{The mathematics of diffusion}.\hskip 1em plus 0.5em minus 0.4em\relax Oxford university press, 1979.

\bibitem{nicholls2020minimising}
D.~Nicholls, J.~Lee, H.~Amari, A.~J. Stevens, B.~L. Mehdi, and N.~D. Browning, ``Minimising damage in high resolution scanning transmission electron microscope images of nanoscale structures and processes,'' \emph{Nanoscale}, vol.~12, no.~41, pp. 21\,248--21\,254, 2020.

\bibitem{jannis2022reducingpart2}
D.~Jannis, A.~Velazco, A.~Béché, and J.~Verbeeck, ``Reducing electron beam damage through alternative {STEM} scanning strategies, part ii: Attempt towards an empirical model describing the damage process,'' \emph{Ultramicroscopy}, vol. 240, p. 113568, 2022.

\bibitem{moshtaghpour2023exploring}
A.~Moshtaghpour, A.~Velazco-Torrejon, A.~Robinson, A.~Kirkland, and N.~Browning, ``Exploring low-dose and fast electron ptychography using l0 regularisation of extended ptychographical iterative engine.'' \emph{Microscopy and Microanalysis}, vol.~29, no. Supplement\_1, pp. 344--345, 2023.

\bibitem{robinson2023simultaneous}
A.~W. Robinson, A.~Moshtaghpour, J.~Wells, D.~Nicholls, M.~Chi, I.~MacLaren, A.~I. Kirkland, and N.~D. Browning, ``Simultaneous high-speed and low-dose {4-D STEM} using compressive sensing techniques,'' \emph{arXiv preprint arXiv:2309.14055}, 2023.

\bibitem{moshtaghpour2022towards}
A.~Moshtaghpour, A.~Velazco-Torrejon, A.~Robinson, E.~Liberti, J.~S. Kim, N.~D. Browning, and A.~I. Kirkland, ``Towards low-dose and fast {4-D} scanning transmission electron microscopy: New sampling and reconstruction approaches,'' \emph{Microscopy and Microanalysis}, vol.~28, no.~S1, pp. 372--373, 2022.

\bibitem{nicholls2022compressive}
D.~Nicholls, A.~Robinson, J.~Wells, A.~Moshtaghpour, M.~Bahri, A.~Kirkland, and N.~Browning, ``Compressive scanning transmission electron microscopy,'' in \emph{proceedings of 2022 IEEE International Conference on Acoustics, Speech and Signal Processing (ICASSP)}.\hskip 1em plus 0.5em minus 0.4em\relax IEEE, 2022, pp. 1586--1590.

\bibitem{nicholls2023potential}
D.~Nicholls, M.~Kobylysnka, J.~Wells, Z.~Broad, D.~McGrouther, A.~Moshtaghpour, , A.~I. Kirkland, R.~A. Fleck, and N.~D. Browning, ``The potential of subsampling and inpainting for fast low-dose cryo {FIB-SEM} imaging and tomography,'' \emph{arXiv preprint arXiv:2309.09617}, 2023.

\bibitem{carslaw1906introduction}
H.~Carslaw, \emph{Introduction to the Mathematical Theory of the Conduction of Heat in Solids}, ser. Dover Scientific Reference Books.\hskip 1em plus 0.5em minus 0.4em\relax Macmillan and Company, limited, 1906.

\bibitem{pattle1959diffusion}
R.~Pattle, ``Diffusion from an instantaneous point source with a concentration-dependent coefficient,'' \emph{The Quarterly Journal of Mechanics and Applied Mathematics}, vol.~12, no.~4, pp. 407--409, 1959.

\bibitem{feynman2006qed}
R.~P. Feynman, \emph{QED: The strange theory of light and matter}.\hskip 1em plus 0.5em minus 0.4em\relax Princeton University Press, 2006, vol.~90.

\bibitem{haigh2009atomic}
S.~J. Haigh, H.~Sawada, and A.~I. Kirkland, ``Atomic structure imaging beyond conventional resolution limits in the transmission electron microscope,'' \emph{Phys. Rev. Lett.}, vol. 103, p. 126101, Sep 2009.

\bibitem{batey2014information}
D.~J. Batey, D.~Claus, and J.~M. Rodenburg, ``Information multiplexing in ptychography,'' \emph{Ultramicroscopy}, vol. 138, pp. 13--21, 2014.

\bibitem{claus2019diffraction}
D.~Claus and J.~M. Rodenburg, ``Diffraction-limited superresolution ptychography in the rayleigh--sommerfeld regime,'' \emph{JOSA A}, vol.~36, no.~2, pp. A12--A19, 2019.

\bibitem{yang20154d}
H.~Yang, L.~Jones, H.~Ryll, M.~Simson, H.~Soltau, Y.~Kondo, R.~Sagawa, H.~Banba, I.~MacLaren, and P.~Nellist, ``4{D} {STEM}: High efficiency phase contrast imaging using a fast pixelated detector,'' in \emph{Journal of Physics: Conference Series}, vol. 644, no.~1.\hskip 1em plus 0.5em minus 0.4em\relax IOP Publishing, 2015, p. 012032.

\bibitem{velazco2020evaluation}
A.~Velazco, M.~Nord, A.~B{\'e}ch{\'e}, and J.~Verbeeck, ``Evaluation of different rectangular scan strategies for {STEM} imaging,'' \emph{Ultramicroscopy}, vol. 215, p. 113021, 2020.

\bibitem{sang2016dynamic}
X.~Sang, A.~R. Lupini, R.~R. Unocic, M.~Chi, A.~Y. Borisevich, S.~V. Kalinin, E.~Endeve, R.~K. Archibald, and S.~Jesse, ``Dynamic scan control in {STEM}: Spiral scans,'' \emph{Advanced Structural and Chemical Imaging}, vol.~2, pp. 1--8, 2016.

\bibitem{li2018compressed}
X.~Li, O.~Dyck, S.~V. Kalinin, and S.~Jesse, ``Compressed sensing of scanning transmission electron microscopy ({STEM}) with nonrectangular scans,'' \emph{Microscopy and Microanalysis}, vol.~24, no.~6, pp. 623--633, 2018.

\bibitem{robinson2022sim}
A.~W. Robinson, D.~Nicholls, J.~Wells, A.~Moshtaghpour, A.~Kirkland, and N.~D. Browning, ``{SIM-STEM Lab}: Incorporating compressed sensing theory for fast {STEM} simulation,'' \emph{Ultramicroscopy}, vol. 242, p. 113625, 2022.

\bibitem{browning2023advantages}
N.~D. Browning, J.~Castagna, A.~I. Kirkland, A.~Moshtaghpour, D.~Nicholls, A.~W. Robinson, J.~Wells, and Y.~Zheng, ``The advantages of sub-sampling and inpainting for scanning transmission electron microscopy,'' \emph{Applied Physics Letters}, vol. 122, no.~5, 2023.

\bibitem{reed2022electrostatic}
B.~Reed, R.~Bloom, G.~Eyzaguirre, C.~Henrichs, A.~Moghadam, and D.~Masiel, ``Electrostatic switching for spatiotemporal dose control in a transmission electron microscope,'' \emph{Microscopy and Microanalysis}, vol.~28, no.~S1, pp. 2230--2231, 2022.

\bibitem{buban2009highresolution}
\BIBentryALTinterwordspacing
J.~P. Buban, Q.~Ramasse, B.~Gipson, N.~D. Browning, and H.~Stahlberg, ``{High-resolution low-dose scanning transmission electron microscopy},'' \emph{Journal of Electron Microscopy}, vol.~59, no.~2, pp. 103--112, 11 2009. [Online]. Available: \url{https://doi.org/10.1093/jmicro/dfp052}
\BIBentrySTDinterwordspacing

\bibitem{peters2023new}
J.~J. Peters, B.~W. Reed, Y.~Jimbo, A.~Porter, D.~Masiel, and L.~Jones, ``A new low-dose {STEM} imaging mode with probability driven intra-pixel beam blanking,'' 2023.

\bibitem{nicholls2023scan}
D.~Nicholls, J.~Wells, A.~W. Robinson, A.~Moshtaghpour, A.~I. Kirkland, and N.~D. Browning, ``Scan coil dynamics simulation for subsampled scanning transmission electron microscopy,'' \emph{arXiv preprint arXiv:2307.08441}, 2023.

\bibitem{petersen2008matrix}
K.~B. Petersen, M.~S. Pedersen \emph{et~al.}, ``The matrix cookbook,'' \emph{Technical University of Denmark}, vol.~7, no.~15, p. 510, 2008.

\bibitem{amazigo1980advanced}
J.~C. Amazigo and L.~A. Rubenfeld, \emph{Advanced calculus and its applications to the engineering and physical sciences}.\hskip 1em plus 0.5em minus 0.4em\relax Wiley, 1980.

\bibitem{rodrigues2018combining}
{\'E}.~O. Rodrigues, ``Combining minkowski and chebyshev: New distance proposal and survey of distance metrics using k-nearest neighbours classifier,'' \emph{Pattern Recognition Letters}, vol. 110, pp. 66--71, 2018.

\end{thebibliography}
